\documentclass{iopart}
\usepackage{iopams}
\usepackage{graphicx}
\expandafter\let\csname equation*\endcsname\relax
\expandafter\let\csname endequation*\endcsname\relax
\usepackage{amsmath}

\usepackage{bm}
\renewcommand{\theenumi}{\arabic{enumi}}

\newcommand{\upT}{\sf{T}}
\newcommand{\downT}{
\mathchoice
{\raisebox{\depth}{\rotatebox{180}{\mbox{$\displaystyle \sf{T}$}}}}
{\raisebox{\depth}{\rotatebox{180}{\mbox{$\textstyle \sf{T}$}}}}
{\raisebox{\depth}{\rotatebox{180}{\mbox{$\scriptstyle \sf{T}$}}}}
{\raisebox{\depth}{\rotatebox{180}{\mbox{$\scriptscriptstyle \sf{T}$}}}}
}

\usepackage[utf8]{inputenc}		
\usepackage{hyperref}
\usepackage[backend=biber,style=numeric,citestyle=numeric-comp,backref=true,sorting=none,natbib=true,autopunct=true,uniquename=false,uniquelist=false,giveninits]{biblatex}
\usepackage[all]{hypcap}		

\DeclareMathOperator{\wedgie}{\wedge}

\begin{document} 

\newcommand{\ee}{e}	
\newcommand{\im}{i}	
\newcommand{\transpose}{\top}

\newcommand{\fback}{\mkern-1.8mu f}	

\newcommand{\dd}{d}			
\newcommand{\DD}{D}			
\newcommand{\bDD}{\bm{\DD}}
\newcommand{\ddi}[1]{\dd^{#1}\mkern-1.5mu}	

\newcommand{\binomial}[2]{\bigl(\!\begin{smallmatrix}#1\\#2\end{smallmatrix}\!\bigr)}

\newcommand{\Cp}{C}			
\newcommand{\conj}[1]{\bar{#1}}		
\newcommand{\reverse}[1]{\tilde{#1}}	
\newcommand\longconj[1]{\ThisStyle{%
  \setbox0=\hbox{$\SavedStyle#1$}%
  \stackengine{1.3\LMpt}{$\SavedStyle#1$}{\rule{\dimexpr\wd0-2\LMpt\relax}{.3\LMpt}}{O}{c}{F}{t}{S}%
}}

\newcommand{\upTup}{{\upT}{\uparrow}}
\newcommand{\upTdown}{{\upT}{\downarrow}}
\newcommand{\downTup}{{\downT}{\uparrow}}
\newcommand{\downTdown}{{\downT}{\downarrow}}
\newcommand{\Upup}{{{\Uparrow}{\uparrow}}}
\newcommand{\Downdown}{{{\Downarrow}{\downarrow}}}
\newcommand{\Downup}{{{\Downarrow}{\uparrow}}}
\newcommand{\Updown}{{{\Uparrow}{\downarrow}}}
\newcommand{\upup}{{{\uparrow}{\uparrow}}}
\newcommand{\downdown}{{{\downarrow}{\downarrow}}}
\newcommand{\downup}{{{\downarrow}{\uparrow}}}
\newcommand{\updown}{{{\uparrow}{\downarrow}}}

\newcommand{\lchiral}{l}
\newcommand{\Lchiral}{L}
\newcommand{\rchiral}{r}
\newcommand{\Rchiral}{R}
\newcommand{\Xchiral}{X}

\newcommand{\spinordot}{{\mkern2mu \cdot}}

\newcommand{\ba}{\bm{a}}
\newcommand{\bb}{\bm{b}}
\newcommand{\bB}{\bm{B}}
\newcommand{\bC}{\bm{C}}
\newcommand{\be}{\bm{e}}
\newcommand{\bE}{\bm{E}}
\newcommand{\bF}{\bm{F}}
\newcommand{\bH}{\bm{H}}
\newcommand{\bL}{\bm{L}}
\newcommand{\bmm}{\bm{m}}
\newcommand{\bM}{\bm{M}}
\newcommand{\bp}{\bm{p}}
\newcommand{\bP}{\bm{P}}
\newcommand{\bR}{\bm{R}}
\newcommand{\bS}{\bm{S}}
\newcommand{\bT}{\bm{T}}
\newcommand{\bV}{\bm{V}}
\newcommand{\bW}{\bm{W}}
\newcommand{\bx}{\bm{x}}
\newcommand{\bX}{\bm{X}}

\newcommand{\G}{{\rm G}}
\newcommand{\SL}{{\rm SL}}
\newcommand{\SO}{{\rm SO}}
\newcommand{\Sp}{{\rm Sp}}
\newcommand{\Spin}{{\rm Spin}}
\newcommand{\SU}{{\rm SU}}
\newcommand{\U}{{\rm U}}

\newcommand{\Ityzrgb}{J}		

\newcommand{\unit}[1]{\, {\rm #1}}		

\newcommand\Ttab{\rule{0pt}{2.6ex}}		
\newcommand\Btab{\rule[-1.2ex]{0pt}{0pt}}	

\newcommand{\spintenchartfig}{
    \begin{figure*}[t!]
    \begin{center}
    \leavevmode
    \includegraphics[scale=.8]{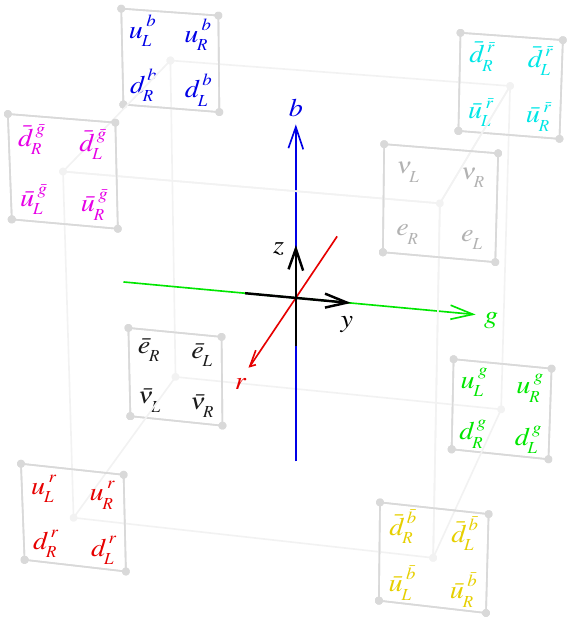}
    \caption[Chart of Spin(10) fermions]{
    \label{spin10chart}
A generation (the electron generation) of 32 fermions
arranged according to their $\Spin(10)$ $yzrgb$ charges.
The eight squares are distinguished by their colour $rgb$ bits,
with $r$-bit-up directed out of the page,
$g$-bit-up directed rightward,
and $b$-bit-up directed upward.
The fermions in each of the eight squares
are distinguished by their weak $yz$ bits,
with $y$-bit-up directed rightward,
and $z$-bit-up directed upward.
A $\Spin(11,1)$ version of this figure is Figure~\ref{spin111chart}.
    }
    \end{center}
    \end{figure*}
}

\newcommand{\spinelevenonechartfig}{
    \begin{figure*}[t!]
    \begin{center}
    \leavevmode
    \includegraphics[scale=.8]{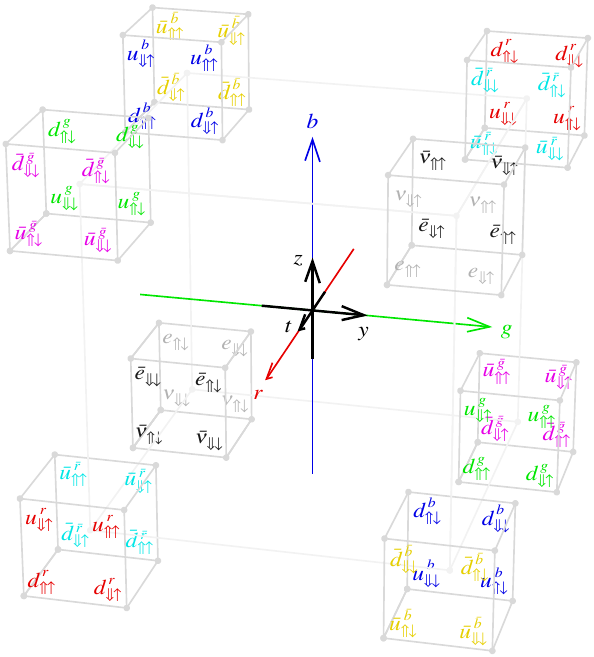}
    \caption[Chart of Spin(10) fermions]{
    \label{spin111chart}
A generation (the electron generation) of 64 fermions
arranged according to their $\Spin(11,1)$ $tyzrgb$ charges.
This is similar to Figure~\ref{spin10chart},
but with the addition of the $t$-bit,
with $t$-bit-up directed out of the page.
Whereas in the $\Spin(10)$ Figure~\ref{spin10chart}
each fermion was a 2-component Weyl fermion,
here the Weyl components are distinguished by their Dirac boost
($\Uparrow$ or $\Downarrow$) and spin ($\uparrow$ or $\downarrow$) bits.
A right-handed Weyl fermion has Dirac bits $\Upup$ or $\Downdown$,
while a left-handed Weyl fermion has Dirac bits $\Downup$ or $\Updown$.
Flipping all 6 $tyzrgb$ bits of a fermion flips its Dirac boost and spin bits,
thereby transforming the fermion to its Weyl companion.
Flipping the $t$-bit of a fermion flips the fermion to its antifermionic partner
of opposite boost and the same spin.
    }
    \end{center}
    \end{figure*}
}

\newcommand{\couplingfig}{
    \begin{figure*}[t!]
    \begin{center}
    \leavevmode
    \includegraphics[scale=.65]{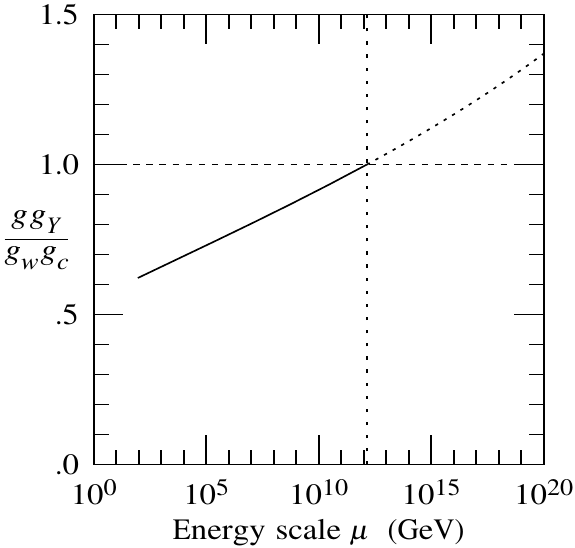}
    \hspace{1em}
    \includegraphics[scale=.65]{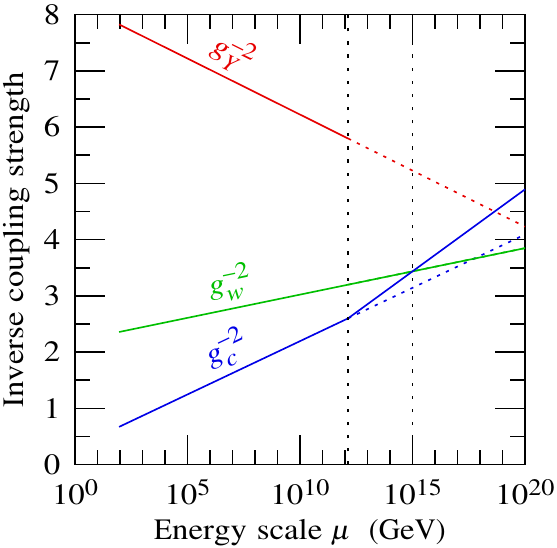}
    \caption[Running of coupling parameters with energy]{
    \label{coupling}
(Left)
Symmetry breaking of the Pati-Salam group $\Spin(4)_w \times \Spin(6)_c$
to the standard model should occur where $g g_Y / ( g_w g_c ) = 1$,
which happens at energy scale $\mu = 10^{12} \unit{GeV}$,
equation~(\ref{mubreak}).
(Right)
Running of the standard-model coupling parameters $g_Y$, $g_w$, and $g_c$
with renormalization energy scale $\mu$, equation~(\ref{grunning}).
The solid blue line ($g_c^{-2}$) tracks the colour force $\SU(3)_c$
below the $\Spin(4)_w \times \Spin(6)_c$ transition energy,
then $\Spin(6)_c$ above the transition energy.
The solid green line ($g_w^{-2}$) tracks the weak force $\SU(2)_\Lchiral$
below the transition energy,
then $\Spin(4)_w$ above the transition energy.
Grand unification,
in the sense that the weak and colour couplings $g_w$ and $g_c$ are equal,
occurs at $\mu = 10^{15} \unit{GeV}$,
equation~(\ref{mubreak}).
The lines are straight lines when $g^{-2}$
is plotted versus log energy.
    }
    \end{center}
    \end{figure*}
}

\title[Unification of the four forces in the Spin(11,1) geometric algebra]{Unification of the four forces in the Spin(11,1) geometric algebra}

\author{Andrew J. S. Hamilton$^{1,2}$ and Tyler McMaken$^{1,3}$}
\address{$^1$ JILA, Box 440, U.\ Colorado, Boulder, CO 80309, USA, $^2$ Dept.\ Astrophysical \& Planetary Sciences, U.\ Colorado, Boulder, $^3$ Dept.\ Physics, U.\ Colorado, Boulder}
\eads{\mailto{Andrew.Hamilton@colorado.edu}, \mailto{Tyler.McMaken@colorado.edu}}


\begin{abstract}
$\SO(10)$, or equivalently its covering group $\Spin(10)$,
is a well-known promising grand unified group that contains
the standard-model group.
The spinors of the group $\Spin(N)$ of rotations in $N$ spacetime dimensions
are indexed by a bitcode with $[N/2]$ bits.
Fermions in $\Spin(10)$
are described by five bits $yzrgb$,
consisting of two weak bits $y$ and $z$, and three colour bits $r$, $g$, $b$.
If a sixth bit $t$ is added, necessary to accommodate a time dimension,
then the enlarged $\Spin(11,1)$ algebra contains the
standard-model and Dirac algebras as commuting subalgebras,
unifying the four forces.
The minimal symmetry breaking chain that breaks $\Spin(11,1)$
to the standard model is unique,
proceeding via the Pati-Salam group.
The minimal Higgs sector is similarly unique,
consisting of the dimension~66 adjoint representation of $\Spin(11,1)$;
in effect, the scalar Higgs sector matches the vector gauge sector.
Although the unified algebra is that of $\Spin(11,1)$,
the persistence of the electroweak Higgs field after grand symmetry breaking
suggests that the gauge group before grand symmetry breaking is
$\Spin(10,1)$, not the full group $\Spin(11,1)$.
The running of coupling parameters predicts that
the standard model should unify to the Pati-Salam group
$\Spin(4)_w \times \Spin(6)_c$
at $10^{12} \unit{GeV}$,
and thence to $\Spin(10,1)$
at $10^{15} \unit{GeV}$.
The grand Higgs field
breaks $t$-symmetry,
can drive cosmological inflation,
and generates a large Majorana mass for the right-handed neutrino
by flipping its $t$-bit.
The electroweak Higgs field breaks $y$-symmetry,
and generates masses for fermions by flipping their $y$-bit.
\end{abstract}


\date{\today}

\maketitle

\section{Introduction}
\label{intro-sec}

$\SO(10)$,
introduced by \cite{Georgi:1975,Fritzsch:1975},
or equivalently its covering group $\Spin(10)$,
remains a popular candidate for grand unification.
The group $\Spin(10)$ is the grandest of the three grand unified groups
proposed in the 1970s,
containing as subgroups the other two grand unified groups
$\SU(5)$
\cite{Georgi:1974},
and the Pati-Salam group
$\SU(2)_\Rchiral \times \SU(2)_\Lchiral \times \SU(4)_c$
\cite{Pati:1974},
which is isomorphic to
$\Spin(4)_w \times \Spin(6)_c$.

There is a large and steadily growing literature on $\Spin(10)$
as a grand unified group, with or without supersymmetry.
Besides unifying each generation of fermions in a single spinor multiplet,
$\Spin(10)$
predicts a right-handed neutrino
\cite{Drewes:2013}
which, through the see-saw mechanism
\cite{Yanagida:1980,GellMann:1979},
allows the left-handed neutrino to acquire a small mass.
$\Spin(10)$ is consistent with the Super-Kamiokande limit
$\gtrsim 1.6 \times 10^{34} \unit{yr}$
on proton lifetime
\cite{Takenaka:2020}
provided that the scale of grand unification exceeds about
$4 \times 10^{15} \unit{GeV}$
\cite{King:2021b}.
$\Spin(10)$
can accommodate a neutrino sector with mixing angles consistent with experiment
\cite{Altarelli:2011}.
Out-of-equilibrium $CP$-violating decay of three generations
of right-handed neutrino in the early Universe can lead
to an asymmetry in baryon minus lepton number $B{-}L$,
a process called leptogenesis
(because the process generates $L$ but not $B$)
\cite{Fukugita:1986,Buchmuller:2004,Buchmuller:2005,Davidson:2008,Blanchet:2012,Fong:2012,Fong:2015,Fong:2022,Cline:2018,Mummidi:2021,Fu:2022},
which can then induce baryogenesis by sphaleron processes
\cite{Klinkhamer:1984,Kuzmin:1985,Buchmuller:2005},
which tend to erase $B{+}L$.

The principal uncertainty over $\Spin(10)$ is the
Higgs fields and symmetry breaking chain
that break it to the standard model
$\U(1)_Y \times \SU(2)_\Lchiral \times \SU(3)_c$
\cite{Harvey:1980,delAguila:1981,DiLuzio:2011,Dueck:2013,Ajaib:2013,Altarelli:2013a,Altarelli:2013b,Babu:2015,Jarkovska:2022,Furey:2022b,Krasnov:2022b}.
The various choices lead to the prediction of various
scalar or other particles,
which could potentially form the dark matter
\cite{Kadastik:2009,Kadastik:2010,Hambye:2010,Kadastik:2010,Parida:2017,Ferrari:2019,Cho:2022,Li:2022},
or produce experimentally detectable signatures
\cite{Badziak:2012,Aydemir:2020,Aydemir:2022,Antusch:2020,Aboubrahim:2021,Preda:2022,Gedeonova:2022}.
Yet other possibilities arise if the grand symmetry group is enlarged.
For example, axions, which could be the dark matter
\cite{Chadhaday:2021},
or not \cite{Ai:2021},
could be accommodated by expanding the group to the product
$\Spin(10) \times \U(1)_\textrm{PQ}$
\cite{Bajc:2006,Boucenna:2019,Agrawal:2022,Lazarides:2022}
where $\U(1)_\textrm{PQ}$ is a Peccei-Quinn symmetry.

The present paper follows a different path,
which seems to have been overlooked,
namely the possibility that the $\Spin(10)$ group itself
unifies nontrivially with the Lorentz group $\Spin(3,1)$
in $\Spin(11,1)$.
The proposed unification is
in a sense
obvious.
Each of the $2^5 = 32$ spinors in a fermion multiplet of $\Spin(10)$
is actually a 2-component chiral (massless) Weyl spinor,
so each fermion multiplet contains $2^6 = 64$ components.
Each 2-component Weyl spinor transforms under the Lorentz group $\Spin(3,1)$
of spacetime.
It is natural to ask whether $\Spin(10)$ and $\Spin(3,1)$ might combine
into a common spin group,
which given that the spinor representation must have 64 components,
and there must be a timelike dimension,
must be the group $\Spin(11,1)$ of rotations in 11+1 spacetime dimensions.
The reason it is possible to adjoin to $\Spin(10)$ just 2 extra dimensions,
and not the full 4 dimensions of spacetime,
is that $\Spin(10)$ and $\Spin(3,1)$ redundantly contain the degrees
of freedom associated with flipping chirality.
The redundancy is expressed mathematically by the coincidence
of the $\Spin(10)$ and $\Spin(3,1)$ chiral operators, equation~(\ref{II10}).
The viability of $\Spin(11,1)$ as a hypothetical unified group
is affirmed by the circumstance that the discrete symmetries
(of the spinor metric, which in the Dirac algebra is antisymmetric,
and of the conjugation operator, which in the Dirac algebra is symmetric)
of $\Spin(11,1)$ agree with those of $\Spin(3,1)$,
as follows from the well-known Cartan-Bott dimension-8 periodicity
of Clifford algebras
\cite{Cartan:1908,Bott:1970,Coquereaux:1982}.


Nesti \& Percacci \cite{Nesti:2010}
and
Krasnov \cite{Krasnov:2022a}
have previously proposed that $\Spin(10)$ and the Lorentz group $\Spin(3,1)$
are unified in $\Spin(11,3)$,
and that the 64 spinors of a generation comprise one of the two chiral
components of the $2^7 = 128$ spinors of $\Spin(11,3)$.
This possibility is discussed further in \S\ref{spin113-sec}.

Why might the unification of $\Spin(10)$ and $\Spin(3,1)$ in $\Spin(11,1)$
have been missed?
A possible reason is the Coleman-Mandula no-go theorem
\cite{Coleman:1967,Mandula:2015},
which states that any symmetry group that contains the Poincar\'e group
and that admits nontrivial analytic scattering amplitudes
must be a direct product of the Poincar\'e group and an internal group.
The Coleman-Mandula theorem generalizes to higher dimensions
\cite{Pelc:1997}.
$\Spin(11,1)$ satisfies the generalized Coleman-Mandula theorem
because then the spacetime group and the internal group are one and the same.

The Coleman-Mandula theorem refers to the Poincar\'e group, which is the global
group of rotations and translations in rigid Minkowski (flat) spacetime.
Indeed Minkowski space is the arena of textbook quantum field theory.
But the tangent space of general relativity is not that of the
Poincar\'e group or the associated Poincar\'e algebra.
Rather, the tangent algebra of general relativity is the Dirac algebra,
which is the geometric algebra (Clifford algebra) associated with $\Spin(3,1)$.
The Dirac algebra is needed to describe spinors in general relativity;
and general relativity can be expressed elegantly in the language of
multivector-valued differential forms \cite{Hamilton:2016mmc},
an approach pioneered by Cartan
\cite{Cartan:2001}.

After symmetry breaking,
the Coleman-Mandula theorem requires only that
the generators of {\em unbroken\/} symmetries commute with those of spacetime.
The unification proposed in the present paper achieves just that:
as shown in \S\ref{commutingsubalgebras-sec},
the Dirac algebra
and
the algebra of unbroken symmetries of $\Spin(10)$
are commuting subalgebras of the geometric (Clifford) algebra
of $\Spin(11,1)$.

The usual assumption
that $\Spin(10)$ combines with $\Spin(3,1)$ as a direct product
is equivalent to the assumption that
each of the 32 fermions of a generation in $\Spin(10)$ is a Weyl spinor.
The two components of a Weyl spinor
transform into each other under proper Lorentz transformations,
therefore necessarily have the same standard-model charges.
In $\Spin(10)$, standard-model charges are related to $\Spin(10)$ charges
by equations~(\ref{YIC}).
As elucidated in \S\ref{commutingsubalgebras-sec},
unifying $\Spin(10)$ and $\Spin(3,1)$ in $\Spin(11,1)$
requires abandoning the assumption that equations~(\ref{YIC}) hold
without modification.
Remarkably,
there exists an adjustment~(\ref{adjustsmbivectorsI6}) of the basis bivectors of
the $\Spin(10)$ algebra
that modifies standard-model charge operators in such a way that
the standard-model and Dirac algebras become commuting subalgebras
of the $\Spin(11,1)$ geometric algebra,
yielding a nontrivial unification of $\Spin(10)$ and $\Spin(3,1)$
in $\Spin(11,1)$.
The modification~(\ref{adjustsmbivectorsI6}) is not obvious,
and may be another reason why unification in $\Spin(11,1)$
has not been noticed previously.
The modification~(\ref{adjustsmbivectorsI6}) is crucial;
without it, the unification proposed in this paper would fail,
and this paper would not have been written.

The $\Spin(11,1)$ model is tightly constrained,
and as detailed in sections~\ref{spin5spin6-sec}--\ref{energyunification-sec},
there appears to be a unique minimal model.
The minimal Higgs sector
consists of the dimension~66 adjoint representation of $\Spin(11,1)$.
This is simpler than the standard $\Spin(10)$ model,
whose Higgs sector requires
\cite{King:2021b}
several distinct multiplets,
typically a bivector (45) to break grand symmetry,
a vector (10) to break electroweak symmetry,
and
a pentavector (126) to break $B{-}L$ symmetry to allow the right-handed
neutrino to acquire a Majorana mass.
The minimal symmetry breaking chain in $\Spin(11,1)$
proceeds by the Pati-Salam group $\Spin(4)_w \times \Spin(6)_c$,
as first proposed by \cite{Harvey:1980},
and advocated by \cite{Altarelli:2013b,Babu:2015},
albeit with a different Higgs sector,
\begin{align}
\label{symbreaking}
  &
  \Spin(11,1)
  \ \underset{??}{\parbox{2em}{\rightarrowfill}} \
  \Spin(10,1)
  \ \underset{10^{15} \unit{GeV}}{\parbox{3em}{\rightarrowfill}} \
  \Spin(4)_w \times \Spin(6)_c \times \Spin(3,1)
  \ \underset{10^{12} \unit{GeV}}{\parbox{3em}{\rightarrowfill}} \
\nonumber
\\
  &\
  \U(1)_Y \times \SU(2)_\Lchiral \times \SU(3)_c \times \Spin(3,1)
  \ \underset{160 \unit{GeV}}{\parbox{3em}{\rightarrowfill}} \
  \U(1)_Q \times \SU(3)_c \times \Spin(3,1)
  \ .
\end{align}
The top line of the chain~(\ref{symbreaking}) is the prediction,
while the bottom line is the standard model.
Note that, as discussed in sections~\ref{grandHiggs-sec} and~\ref{Spin101-sec},
the grand unified group is $\Spin(10,1)$ rather than $\Spin(11,1)$ itself.

The predicted energy of grand unification from $\Spin(10,1)$
is $10^{15} \unit{GeV}$, equation~(\ref{mubreak}),
which is less than the lower limit of
$4 \times 10^{15} \unit{GeV}$
obtained by
\cite{King:2021b}
from the Super-Kamiokande lower limit on proton lifetime \cite{Takenaka:2020},
so the $\Spin(11,1)$ model may already be ruled out by proton decay.
\cite{King:2021b}'s limit seems robust,
because the proton lifetime $\tau$ depends rather steeply
on the unification scale $M_X$, as $\tau \propto M_X^4$.
However,
the grand symmetry breaking scale of $10^{15} \unit{GeV}$ predicted here,
\S\ref{energyunification-sec},
is based on a simple model
(1-loop renormalization,
and an abrupt transition at the Pati-Salam symmetry breaking scale),
and moreover the $\Spin(11,1)$ Higgs sector is different from the
$\Spin(10)$ models considered by \cite{King:2021b}.
It would seem wise to incorporate $\Spin(11,1)$ into the pantheon
of beyond-standard-model models,
and to subject it to the same kind of scrutiny that $\Spin(10)$ has received.

The plan of this paper is as follows.
Section~\ref{spin10-sec}
recasts the charges of fermions in the standard model
in terms of a $\Spin(10)$ bitcode,
as first pointed out by \cite{Wilczek:1998},
and advocated by \cite{Hamilton:2023a},
and points out some patterns in the $\Spin(10)$ chart~(\ref{yzrgbtab})
of fermions that suggest that the Dirac and $\Spin(10)$ algebras might
fit nontrivially in a unified algebra.
Section~\ref{extradim-sec}
argues that if the unified algebra is a spin algebra,
then that algebra must have 11+1 spacetime dimensions.
Section~\ref{commutingsubalgebras-sec}
derives the main result of this paper,
that the Dirac and standard-model algebras are commuting subalgebras
of the $\Spin(11,1)$ geometric algebra,
provided that standard-model bivectors are
modified per~(\ref{adjustsmbivectorsI6}).
Section~\ref{spin5spin6-sec} considers
the possible unification path from the standard model toward $\Spin(11,1)$,
concluding that the only option for unification between electroweak
and grand unification is the Pati-Salam group $\Spin(4)_w \times \Spin(6)_c$.
Section~\ref{higgsew-sec}
addresses electroweak symmetry breaking,
\S\ref{grandsym-sec}
grand symmetry breaking,
and \S\ref{patisalam-sec} Pati-Salam symmetry breaking.
Section~\ref{energyunification-sec}
shows from the running of coupling parameters that
Pati-Salam unification occurs at $10^{12} \unit{GeV}$,
and grand unification at $10^{15} \unit{GeV}$.
Section~\ref{predflaw-sec} summarizes the predictions,
and highlights areas where the critical reader might look for flaws.
Section~\ref{conclusions-sec} summarizes the conclusions.


The notation in this paper follows that of \cite{Hamilton:2023a}.

\section{Spin(10)}
\label{spin10-sec}

\subsection{Spin(10) charges}

Introduced originally by Cartan \cite{Cartan:1913,Cartan:1938},
a spinor is the fundamental representation of the group of rotations
in $N$ spacetime dimensions.
As emphasized by \cite{Hamilton:2023a},
spinors have the intriguing property that their index is a bitcode,
with $[N/2]$ bits in $N$ spacetime dimensions.
Mathematically,
the dimension of the spinor representation is $2^{[N/2]}$
(strictly, in even $N$ dimensions
there are two isomorphic irreducible spinor representations,
of opposite chirality,
each of dimension $2^{[N/2]-1}$;
chirality counts whether the number of up-bits is odd or even).
The halving of dimensions is associated with the
fact that spinors have a natural complex structure.
In two or three dimensions, the number of bits is one, a Pauli spinor,
with $2^1 = 2$ complex components.
In four dimensions, the number of bits is two, a Dirac spinor,
with $2^2 = 4$ complex components.
A Dirac spinor in 3+1 spacetime dimensions has a spin bit and a boost bit,
\ref{diracalgebra-app}.
The Dirac spinor is called right-handed if the spin and boost bits are aligned,
left-handed if anti-aligned.
A massive Dirac spinor is a linear combination of right- and left-handed
components,
but only the left-handed component couples to the weak force
$\SU(2)_\Lchiral$.

As first pointed out by
\cite{Wilczek:1998},
and reviewed by \cite{Baez:2009dj},
$\Spin(10)$ describes a generation of fermions of the standard model
with a bitcode with $[10/2] = 5$ bits $y , z , r , g , b$
consisting of two weak bits $y$ and $z$, and three colour bits $r , g , b$.
\ref{namingbits-sec} comments on the naming of bits.
Each bit can be either up or down,
signifying a charge of $+\tfrac{1}{2}$ or $-\tfrac{1}{2}$.
The standard model has 5 conserved charges consisting of
hypercharge $Y$, weak isospin $I_\Lchiral$,
and three colours $R$, $G$, and $B$.
The relation between standard-model charges and $\Spin(10)$ charges is
\begin{subequations}
\label{YIC}
\begin{align}
\label{YICY}
  Y
  &=
  y + z - \tfrac{2}{3} ( r + g + b )
  \ ,
\\
\label{YICI}
  I_\Lchiral
  &=
  \tfrac{1}{2} ( z - y )
  \ ,
\\
\label{YICc}
  C
  &=
  c + \tfrac{1}{2}
  \quad
  ( C = R , G , B , \  c = r , g , b )
  \ .
\end{align}
\end{subequations}
The relation~(\ref{YIC}) between charges is possible
only thanks to certain coincidences
\cite{Baez:2009dj}
in the pattern of charges of fermions in the standard model.
Traditionally a quark has colour charge $C$ consisting of one unit of
either $R$, $G$, or $B$.
$\Spin(10)$ on the other hand says that an $r$ quark (for example) has
$rgb$ bits $\uparrow\downarrow\downarrow$,
meaning that its $r$ charge is $+\tfrac{1}{2}$
while its $g$ and $b$ charges are $-\tfrac{1}{2}$.
In the $\Spin(10)$ picture,
when an $r$ quark turns into a $g$ quark
through interaction with a $g{\bar r}$ gluon,
its $r$ charge flips from $+\tfrac{1}{2}$ to $-\tfrac{1}{2}$
while its $g$ charge flips from $-\tfrac{1}{2}$ to $+\tfrac{1}{2}$,
\begin{equation}
  \begin{array}{cccc}
  \uparrow\downarrow\downarrow &
  \parbox{4em}{\rightarrowfill} &
  \downarrow\uparrow\downarrow &
  .
  \\
  r~\textrm{quark} &
  g{\bar r}~\textrm{gluon} &
  g~\textrm{quark} &
  \end{array}
\end{equation}
In so doing, the quark loses one unit of $r$ charge,
and gains one unit of $g$ charge,
consistent with the traditional picture.

Standard-model transformations also preserve baryon number $B$
and lepton number $L$.
Mysteriously, although standard-model transformations preserve
$B$ and $L$, neither $\U(1)_B$ nor $\U(1)_L$, nor any combination thereof,
is a local symmetry of the standard model
(there is no force associated with these symmetries).
The combination $\U(1)_{B{-}L}$ is however a subgroup of $\Spin(10)$,
so $B{-}L$ is a conserved charge of $\Spin(10)$,
\begin{equation}
\label{BLcharge}
  B - L
  =
  - \tfrac{2}{3} ( r + g + b )
  \ .
\end{equation}
$\Spin(10)$ does not preserve baryon and lepton number individually.

After electroweak symmetry breaking,
the electroweak group $\U(1)_Y \times \SU(2)_\Lchiral$
breaks down to the electromagnetic group $\U(1)_Q$.
The electromagnetic charge $Q$ is
\begin{equation}
\label{QIc}
  Q
  =
  \tfrac{1}{2} Y + I_\Lchiral
  =
  z - \tfrac{1}{3} ( r + g + b )
  \ .
\end{equation}
Electroweak symmetry breaking is a loss of $y$-symmetry,
a loss of conservation of $y$-charge.

\spintenchartfig

\subsection{Spin(10) chart of spinors}

Figure~\ref{spin10chart} illustrates
one generation (the electron generation) of fermions of the standard model
arranged according to their $\Spin(10)$ $yzrgb$ charges.

The following $\Spin(10)$ chart
shows the electron generation of fermions of the standard model
arrayed in columns according to the number of up-bits
(compare Table~4 of \cite{Baez:2009dj}; see also \cite{Wilczek:1998}).
The left element of each entry (before the colon) signifies which bits
are up, from -- (no bits up, or
$\downarrow\downarrow\downarrow\downarrow\downarrow$)
in the leftmost (0) column,
to $yzrgb$ (all bits up, or
$\uparrow\uparrow\uparrow\uparrow\uparrow$)
in the rightmost (5) column;
the right element of each entry is the corresponding fermion,
which comprise (electron) neutrinos $\nu$, electrons $e$,
and up and down quarks $u$ and $d$,
each in right- and left-handed Dirac chiralities $\Rchiral$ and $\Lchiral$,
and each in (unbarred) particle and (barred) anti-particle species,
a total of $2^5 = 32$ fermions:
\begin{equation}
\label{yzrgbtab}
  \begin{array}{c@{\quad\quad}c@{\quad\quad}c@{\quad\quad}c@{\quad\quad}c@{\quad\quad}c}
  \hline
  \multicolumn{6}{c}{\mbox{Fermions and their $\Spin(10)$ bitcodes, arranged by the number of up-bits}}
  \\
  0 & 1 & 2 & 3 & 4 & 5
  \\
  \hline
  \noalign{\vskip .5ex}
  \mbox{--} : \
  \bar{\nu}_\Lchiral
  &
  y : \ 
  \bar{\nu}_\Rchiral
  &
  \phantom{d}\bar{c} : \ 
  \bar{u}_\Lchiral^{\bar c}
  &
  \phantom{d}y\bar{c} : \
  \bar{u}_\Rchiral^{\bar c}
  &
  zrgb : \ 
  \nu_\Lchiral
  &
  yzrgb : \ 
  \nu_\Rchiral
\\[.5ex]
  &
  z : \ 
  \bar{e}_\Rchiral
  &
  yz : \ 
  \bar{e}_\Lchiral
  &
  rgb : \ 
  e_\Rchiral
  &
  yrgb : \ 
  e_\Lchiral
  &
  \\[.5ex]
  &
  c : \ 
  d_\Rchiral^c
  &
  yc : \ 
  d_\Lchiral^c
  &
  \phantom{d}z\bar{c} : \ 
  \bar{d}_\Rchiral^{\bar{c}}
  &
  \phantom{d}yz\bar{c} : \ 
  \bar{d}_\Lchiral^{\bar{c}}
\\[.5ex]
  &
  &
  zc : \ 
  u_\Lchiral^c
  &
  yzc : \ 
  u_\Rchiral^c
  &
  &
\\[.5ex]
  \hline
  \end{array}
\end{equation}
Here $c$ denotes any of the three colours $r$, $g$, or $b$
(one colour bit up),
while ${\bar c}$ denotes any of the three anticolours $gb$, $br$, or $rg$
(two colour bits up, the bit flip of a one-colour-bit-up spinor).
Every compact spin group $\Spin(N)$ contains a subgroup $\SU([N/2])$
that preserves the number of up-bits \cite{Atiyah:1964}.
The columns of the chart~(\ref{yzrgbtab})
are $\SU(5)$ multiplets within $\Spin(10)$,
with dimensions respectively 1, 5, 10, 10, 5, 1.
The standard-model group is a subgroup of $\SU(5)$.
All standard-model interactions preserve the number of $\Spin(10)$ up-bits.

Each entry in the chart~(\ref{yzrgbtab})
is a 2-component massless Weyl fermion.
Lorentz transformations transform the 2 components
of a Weyl fermion among each other.
Flipping all 5 $yzrgb$ bits turns a fermion into its antifermion partner.


\subsection{Notable features of the Spin(10) chart}
\label{striking-sec}

The $\Spin(10)$ chart~(\ref{yzrgbtab}) shows some notable features.
The most striking feature is that $\Spin(10)$ chirality
coincides with Dirac chirality.
Chirality counts whether the number of up-bits is even or odd,
with right-handed defined as all bits up.
The odd and even columns of the $\Spin(10)$ chart~(\ref{yzrgbtab})
have respectively right-handed ($\Rchiral$) and left-handed ($\Lchiral$)
Dirac chirality.
Modulo a phase,
chirality is (the eigenvalue of) the pseudoscalar of the algebra,
the
product of all the $N$ vectors in the $N$-dimensional geometry.
The Dirac chiral operator is traditionally denoted $\bgamma_5$,
which equals $-\im$ times the Dirac pseudoscalar $I$.
The $\Spin(10)$ chiral operator can be denoted $\varkappa_{10}$,
which equals $-\im$ times the $\Spin(10)$ pseudoscalar $I_{10}$.
The coincidence
(signified $\smash{\overset{!}{=}}$)
between Dirac and $\Spin(10)$ chirality in the chart~(\ref{yzrgbtab}) is
\begin{align}
\label{II10}
  \gamma_5
  &\equiv
  -\im
  I
  \equiv
  -\im
  \bgamma_0 \bgamma_1
  \bgamma_2 \bgamma_3
\nonumber
\\
  \overset{!}{=}
  \varkappa_{10}
  &\equiv
  -\im
  I_{10}
  \equiv
  -\im
  \bgamma^+_y \bgamma^-_y
  \bgamma^+_z \bgamma^-_z
  \bgamma^+_r \bgamma^-_r
  \bgamma^+_g \bgamma^-_g
  \bgamma^+_b \bgamma^-_b
  \ .
\end{align}
The coincidence of Dirac and $\Spin(10)$ chiralities
suggests that the vectors $\bgamma_m$,  $(m = 0, 1, 2, 3)$ of spacetime
are related to the vectors $\bgamma^\pm_k$ $(k = y, z, r, g, b)$ of $\Spin(10)$,
in contrast to the usual assumption that the generators of spacetime
are unrelated to (commute with) those of grand unified symmetries.

A second notable feature of the chart~(\ref{yzrgbtab})
is that standard-model transformations are arrayed vertically,
whereas the four components of fermions of the same species,
such as electrons $e_\Rchiral$ and $e_\Lchiral$
and their positron partners $\bar{e}_\Lchiral$ and $\bar{e}_\Rchiral$,
are arrayed (mostly) horizontally.
In Dirac theory, a Dirac spinor such as an electron
has four complex components that
are distinguished by their properties under Lorentz transformations.
With standard-model transformations vertical
and spacetime transformations horizontal,
the chart~(\ref{yzrgbtab}) seems to invite the idea
that $\Spin(10)$ somehow contains both standard-model and
spacetime transformations under one tent.

A third notable feature
is that flipping the $y$-bit preserves the identity of the spinor
but flips its chirality;
for example the electron is flipped $e_\Rchiral \,\leftrightarrow\, e_\Lchiral$.
Electroweak symmetry breaking is a loss of $y$-symmetry
and a loss of conservation of $y$-charge.
In the standard model, electroweak symmetry breaking is accomplished
by a Higgs field that carries $y$-charge and gives mass to fermions
by flipping their $y$-bit.

A fourth tea leaf is that fermions (unbarred) and antifermions (barred)
in the chart~(\ref{yzrgbtab})
have colour chirality $\varkappa_{rgb}$ respectively positive and negative,
\begin{equation}
\label{krgb}
  \varkappa_{rgb}
  \equiv
  \im
  I_{rgb}
  \equiv
  \im
  \bgamma^+_r \bgamma^-_r \bgamma^+_g \bgamma^-_g \bgamma^+_b \bgamma^-_b
  \ .
\end{equation}
Colour chirality $\varkappa_{rgb}$ is positive or negative
as the number of colour up-bits is odd or even.



\subsection{Spin(10) gauge fields}
\label{Spin10gaugefields-sec}

Associated with a spin group $\Spin(N)$ is
a Clifford algebra \cite{Clifford:1878},
or geometric algebra \cite{Hestenes:1966,Hestenes:1987}.
In 3+1 spacetime dimensions,
the geometric algebra is the Dirac algebra
of Dirac $\gamma$-matrices and their products
(whence the notation $\bgamma_a$ for the vectors of a geometric algebra),
\ref{diracalgebra-app}.
The orthonormal vectors $\bgamma_a$, $a = 1 , ... , N$
of a geometric algebra in $N$ dimensions satisfy the product rule
\begin{equation}
  \bgamma_a \bgamma_b
  =
  \left\{
  \begin{array}{ll}
  \delta_{ab} & ( a = b ) \ ,
  \\
  - \bgamma_b \bgamma_a & ( a \neq b ) \ .
  \end{array}
  \right.
\end{equation}
If a dimension is timelike, as in the Dirac algebra,
then it is convenient to treat the timelike dimension
as the imaginary $\im$ times a spacelike dimension,
as in equation~(\ref{gammadiracpm}).
An antisymmetric product of vectors is commonly written with a wedge sign,
$\bgamma_a \wedgie \bgamma_b$,
although in the present paper the wedge sign is often omitted for brevity
when there is no ambiguity.
Physically,
a wedge product $\bgamma_{a_1} \wedgie ... \wedgie \bgamma_{a_p}$
of $p$ vectors,
a multivector of grade $p$,
or $p$-vector,
represents a $p$-dimensional element in the $N$-dimensional space.
The geometric algebra is the algebra of linear combinations of
multivectors of all possible grades $p = 0$ to $N$.

In the presence of spinors,
the orthonormal basis vectors $\bgamma_a$
of a geometric algebra group naturally into pairs
\cite{Hamilton:2023a},
conveniently denoted $\bgamma^+_k$ and $\bgamma^-_k$.
Spinors with $k$ bit up and down transform with opposite phases
$\ee^{\mp \im \theta/2}$
(or opposite boosts $\ee^{\pm \theta/2}$)
under rotations in the 2-dimensional $\bgamma^+_k \bgamma^-_k$~plane.
The 10 orthonormal basis vectors of the $\Spin(10)$ geometric algebra are
$\bgamma^\pm_k$,
$k = y, z, r, g, b$.
Chiral combinations of the orthonormal basis vectors are defined by
\begin{equation}
\label{gammakpm}
  \bgamma_k
  \equiv
  {\bgamma^+_k + \im \bgamma^-_k \over \sqrt{2}}
  \ , \quad
  \bgamma_{\bar k}
  \equiv
  {\bgamma^+_k - \im \bgamma^-_k \over \sqrt{2}}
  \ .
\end{equation}
The orthonormal vectors
$\bgamma^+_k$ and $\im \bgamma^-_k$
can be thought of as, modulo a normalization,
the real and imaginary parts of a complex vector
$\bgamma_k$ whose complex conjugate is $\bgamma_{\bar k}$.
The chiral vectors $\bgamma_k$ and $\bgamma_{\bar k}$
carry respectively $+$ and $-$ one unit of $k$ charge;
they transform with opposite phases
$\ee^{\mp \im \theta}$
(or opposite boosts $\ee^{\pm \theta}$)
under rotations in the 2-dimensional $\bgamma^+_k \bgamma^-_k$~plane,
and they raise and lower the charge of the object (spinor or multivector)
that they act on by one unit of $k$ charge.

The gauge fields associated with any gauge group form a multiplet
labelled by the generators of the group.
The generators of the $\Spin(2n)$ group with $n = 5$
are its $n ( 2n {-} 1 ) = 45$ orthonormal bivectors
(products of two orthonormal vectors),
comprising the $2 n ( n {-} 1 ) = 40$ bivectors
\begin{subequations}
\label{spin10bivectorsab}
\begin{align}
\label{spin10bivectorsabpp}
  \bgamma^+_k \wedgie \bgamma^+_l
  &=
  \tfrac{1}{2} ( \bgamma_k + \bgamma_{\bar k} )
  \wedgie
  ( \bgamma_l + \bgamma_{\bar l} )
  \ ,
\\
\label{spin10bivectorsabpm}
  \bgamma^+_k \wedgie \bgamma^-_l
  &=
  \tfrac{1}{2\im} ( \bgamma_k + \bgamma_{\bar k} )
  \wedgie
  ( \bgamma_l - \bgamma_{\bar l} )
  \ ,
\\
\label{spin10bivectorsabmp}
  \bgamma^-_k \wedgie \bgamma^+_l
  &=
  \tfrac{1}{2\im} ( \bgamma_k - \bgamma_{\bar k} )
  \wedgie
  ( \bgamma_l + \bgamma_{\bar l} )
  \ ,
\\
\label{spin10bivectorsabmm}
  \bgamma^-_k \wedgie \bgamma^-_l
  &=
  - \tfrac{1}{2} ( \bgamma_k - \bgamma_{\bar k} )
  \wedgie
  ( \bgamma_l - \bgamma_{\bar l} )
  \ ,
\end{align}
\end{subequations}
with distinct indices $k$ and $l$ running over
$y, z, r, g, b$,
together with the $n = 5$ diagonal bivectors
\begin{equation}
\label{spin10bivectorsa}
  \tfrac{1}{2} \,
  \bgamma^+_k \wedgie \bgamma^-_k
  =
  \tfrac{\im}{2} \,
  \bgamma_k \wedgie \bgamma_{\bar k}
  \ ,
\end{equation}
with indices $k$ running over
$y, z, r, g, b$.

The generators of a gauge group serve two roles.
On the one hand they generate the symmetries that rotate fields.
On the other hand,
the generators are themselves fields
that are rotated by the symmetries they generate.
To appreciate the distinction,
consider the diagonal chiral bivector
\begin{equation}
\label{diagbivector}
  \tfrac{1}{2} \,
  \bgamma_k \wedgie \bgamma_{\bar k}
  \ .
\end{equation}
On the one hand, the diagonal bivector~(\ref{diagbivector}) acts as an operator
whose eigenvalues equal the $k$ charge of the objects it acts on
(the normalization factor of $\tfrac{1}{2}$ serves to ensure
that is true).
As an operator, a generator acts on its argument by commutation
(equivalent to left multiplication
if the argument is a column spinor,
or right multiplication
if the argument is a row spinor).
On the other hand, the diagonal bivector~(\ref{diagbivector})
is itself an object, a field, whose $k$ charge is zero.
As a field, a generator is itself acted on by commutation.

Expressions for gauge fields of subgroups of $\Spin(10)$,
including for $\SU(5)$ and the standard-model,
are collected in \ref{gaugesm-sec}.

\section{Extra dimensions}
\label{extradim-sec}

The primary purpose of this paper is to explore whether the Dirac
and $\Spin(10)$ algebras unify nontrivially in a common geometric algebra.
This section argues that the minimal unified geometric algebra is
that associated with $\Spin(11,1)$ in 11+1 spacetime dimensions.

\subsection{Two extra dimensions, and one extra bit}
\label{eleventhdim-sec}

Each spinor multiplet of $\Spin(10)$ contains $2^{[10/2]} = 2^5 = 32$ fermions.
Each of the 32 fermions of the multiplet is itself a Weyl (massless, chiral)
spinor, containing 2 degrees of freedom,
for a total of $2^6 = 64$ degrees of freedom.
Each 2-component Weyl fermion transforms under the Lorentz group $\Spin(3,1)$
of spacetime.
If the spinors of $\Spin(10)$ and $\Spin(3,1)$ are to be unified
in a minimal fashion,
in a common spin group with no superfluous degrees of freedom,
then that group must have a spinor multiplet of dimension $2^6$.
Certainly, there must be one timelike dimension,
to accommodate the time dimension of the Dirac algebra.
Moreover,
the total number of spacetime dimensions must be even,
because the spinor multiplet must be chiral,
and only in even dimensions
is the chirality of a spinor invariant under rotations.

Before immediately concluding that the minimal spin group must therefore have
11+1 spacetime dimensions,
bear in mind that the Dirac algebra has $2^2 = 4$ complex components,
equivalent to $2^3 = 8$ real degrees of freedom.
The 32 2-component Weyl spinors of $\Spin(10)$ comprise 64 real, not complex,
degrees of freedom,
because, as is evident in the chart~(\ref{yzrgbtab}),
the 64 components include the degrees of freedom associated with both
fermions and antifermions, which are conjugates of each other.
Therefore it might be possible that the 64 spinors might
constitute $2^5 = 32$ complex components of spinors
in 9+1 spacetime dimensions.

The possibility of 9+1 spacetime dimensions
is ruled out by chirality.
In the standard model,
conjugation flips chirality:
the conjugate of a right-handed fermion is a left-handed antifermion.
In any geometric algebra,
the conjugation operator is, modulo a phase, the product of the spinor metric
$\varepsilon$ with the product of all time dimensions \cite{Hamilton:2023a}.
The spinor metric has the property that it flips all bits,
and in even $N$ dimensions
each vector (multivector of grade~1) of the algebra flips one bit.
Therefore, if there is one time dimension
(or more generally, an odd number of time dimensions),
conjugation flips chirality if and only if the number $[N/2]$ of bits is even.
This rules out 9+1 spacetime dimensions,
which has one time dimension and an odd number 5 of bits.

Therefore,
the minimal spin group must have 11+1 spacetime dimensions.
Adding two extra dimensions adjoins an additional, 6th, $t$-bit,
or time bit,
to the 5 $yzrgb$ bits of a $\Spin(10)$ spinor.
Like the other 5 bits, the $t$-charge of a $\Spin(11,1)$ spinor
can be either $+\tfrac{1}{2}$ ($t$-bit up) or $-\tfrac{1}{2}$ ($t$-bit down).

In Dirac theory,
massive spinors and antispinors are complex conjugates of each other,
and massive spinors at rest are eigenvectors of the time axis,
equations~(\ref{gammadirac}) and~(\ref{gammadiracpm}).
These two conditions require interpreting
the 12th dimension $\bgamma^-_t$,
not the 11th dimension $\bgamma^+_t$,
as providing a timelike dimension
$\bgamma_0$,
\begin{equation}
  \bgamma_0
  \equiv
  \im
  \bgamma^-_t
  \ .
\end{equation}

\subsection{The spinor metric and the conjugation operator}
\label{spin101metricandCp-sec}

Any spin algebra contains two operators,
the spinor metric $\varepsilon$
and the conjugation operator $\Cp$,
that are invariant under rotations
\cite{Hamilton:2023a}.
A consistent translation between Dirac and $\Spin(11,1)$ representations
must agree on the behaviour of these two operators.

The Dirac spinor metric $\varepsilon$, equations~(\ref{echiraldirac}),
and conjugation operator $\Cp$, equations~(\ref{Cpchiraldirac}),
are respectively antisymmetric ($\varepsilon^2 = -1$)
and symmetric ($\Cp \Cp^\ast = 1$).
Consistency requires that the $\Spin(11,1)$ spinor metric
and conjugation operator be similarly antisymmetric and symmetric.
Consultation of Tables~1 and~4 of \cite{Hamilton:2023a}
shows that in 11+1 dimensions
only the standard choice $\varepsilon$ of spinor metric
and associated conjugation operator $\Cp$
possess the desired antisymmetry and symmetry.
The standard spinor metric in 11+1 dimensions is
\begin{equation}
\label{spinormetricspin10p}
  \varepsilon
  =
  \bgamma^+_t \bgamma^+_y \bgamma^+_z \bgamma^+_r \bgamma^+_g \bgamma^+_b
  \ .
\end{equation}
Below it will be found that the representation of the spatial rotation
generator $\Ityzrgb \sigma_2$, equation~(\ref{Jsigma2}), coincides with
the representation of the spinor metric~(\ref{spinormetricspin10p}),
which is similar to the coincidence~(\ref{echiraldirac}) between $I \sigma_2$
and the spinor metric $\varepsilon$
in the chiral representation of the Dirac algebra.
Given the $\Spin(11,1)$ spinor metric~(\ref{spinormetricspin10p}),
and with the time axis $\bgamma_0 = \im \bgamma^-_t$,
the $\Spin(11,1)$ conjugation operator is
\begin{equation}
\label{Cpspin10p}
  \Cp
  =
  - \im \varepsilon \bgamma_0^\transpose
  =
  \im \varepsilon \bgamma_0
  =
  - \varepsilon \bgamma^-_t
  =
  \bgamma^+_t \bgamma^-_t \bgamma^+_y \bgamma^+_z \bgamma^+_r \bgamma^+_g \bgamma^+_b
  \ .
\end{equation}
Whereas the $\Spin(11,1)$ spinor metric~(\ref{spinormetricspin10p})
flips all bits,
the $\Spin(11,1)$ conjugation operator~(\ref{Cpspin10p})
flips all bits except $t$, that is, it flips $yzrgb$.
This is the same as the $\Spin(10)$ conjugation operator,
which flips the five $yzrgb$ bits.

That the $\Spin(11,1)$ and Dirac ($\Spin(3,1)$) geometric algebras
share the same symmetries is no coincidence:
it stems from the period-8 Cartan-Bott periodicity
\cite{Cartan:1908,Bott:1970,Coquereaux:1982}
of geometric algebras,
evident in Tables~1 and~4 of \cite{Hamilton:2023a}.


\subsection{Spin(11,3)?}
\label{spin113-sec}


Nesti \& Percacci \cite{Nesti:2010}
and
Krasnov \cite{Krasnov:2022a}
have previously proposed that $\Spin(10)$ and the Lorentz group $\Spin(3,1)$
are unified in $\Spin(11,3)$,
and that the 64 spinors of a generation comprise one of the two chiral
components of the $2^7 = 128$ spinors of $\Spin(11,3)$.
The 64 spinors comprise 32 spinors that are direct products of
right-handed $\Spin(10)$ spinors with right-handed Dirac spinors,
and their 32 antispinor conjugate partners that are direct products of
left-handed $\Spin(10)$ spinors with left-handed Dirac spinors.
It is precisely the coincidence of $\Spin(10)$ and Dirac chirality
that ensures the consistency of the construction.
The Coleman-Mandula theorem is satisfied because
the even $\Spin(10)$ geometric algebra
and
the Dirac algebra
are commuting subalgebras of the $\Spin(11,3)$ geometric algebra.

Percacci \cite{Percacci:1990wy}
(see \cite{Nesti:2007ka})
has previously proposed that the grand unified group $\SO(10)$
and the Lorentz group $\SO(3,1)$ are unified in $\SO(13,1)$.
That proposal runs into the difficulty
that the conjugation operator in 13+1 spacetime dimensions is antisymmetric,
whereas the conjugation operator in the Dirac algebra is symmetric
(see Table~4 of \cite{Hamilton:2023a}).

\section{The Dirac and standard-model algebras are commuting subalgebras of the Spin(11,1) geometric algebra}
\label{commutingsubalgebras-sec}

\subsection{Spin(11,1) chart of spinors}
\label{spin111chart-sec}

The challenge now is to reinterpret
the $\Spin(10)$ chart~(\ref{yzrgbtab}) of fermions
as living in $\Spin(11,1)$.
Recall that in the standard picture
each entry in the $\Spin(10)$ chart is assumed to be a 2-component Weyl spinor,
which is equivalent to the usual assumption
that spacetime and $\Spin(10)$ combine as a direct product.
That picture must be abandoned here,
because the hypothesis of this paper is precisely
that the Dirac and $\Spin(10)$ groups do not combine as a direct product.
In the $\Spin(11,1)$ picture,
the two components of each Weyl spinor must be distributed into
two separate entries.
Translated into $\Spin(11,1)$,
each entry in the $\Spin(10)$ chart~(\ref{yzrgbtab})
must still contain 2 components,
one with $t$-bit up, the other with $t$-bit down,
but those 2 components do not Lorentz-transform into each other.

At first sight it might seem that it would be impossible to distribute
the two components of a Weyl spinor in separate entries,
because the two components of a Weyl spinor necessarily have the same charge
(being related by a proper Lorentz transformation),
whereas different entries in the $\Spin(10)$ chart~(\ref{yzrgbtab})
carry different charges.
However, there is a trick that gets around that difficulty
(this is a key trick, without which this paper would not have been written).
The trick is suggested by the fourth tea leaf of \S\ref{striking-sec},
that the spinors of the $\Spin(10)$ chart~(\ref{yzrgbtab})
are fermions (unbarred) or antifermions (barred)
as the colour chirality $\varkappa_{rgb}$, equation~(\ref{krgb}),
is positive or negative.
The five $\Spin(10)$ charges of a spinor are eigenvalues of the five diagonal
bivectors~(\ref{spin10bivectorsa}).
If these diagonal bivectors are modified by multiplying them
by $\varkappa_{rgb}$, then their eigenvalues will measure the charge
of the fermion, not the antifermion, in all entries of the $\Spin(10)$ chart.
A key point that allows this adjustment to be made consistently
is that $\varkappa_{rgb}$ commutes with all standard-model bivectors.
Notably, $\varkappa_{rgb}$ does not commute
with $\SU(5)$ bivectors that transform between leptons and quarks;
but that is fine,
because $\SU(5)$ is not an unbroken symmetry of the standard model.
A consistent way to implement this modification,
that leaves the bivector algebra of the standard model
(but not of $\SU(5)$) unchanged,
is to multiply all imaginary bivectors $\bgamma^+_k \bgamma^-_l$
by $\varkappa_{rgb}$,
while leaving all real bivectors
$\bgamma^+_k \bgamma^+_l$ and $\bgamma^-_k \bgamma^-_l$
unchanged,
\begin{equation}
\label{adjustsmbivectorsI6}
  \bgamma^+_k
  \bgamma^-_l
  \rightarrow
  \bgamma^+_k
  \bgamma^-_l
  \varkappa_{rgb}
  \ , \quad
  k,l = t,y,z,r,g,b
  \ .
\end{equation}
Equivalently,
replace the imaginary $\im$ in all $\Spin(10)$ multivectors
by the colour pseudoscalar $- I_{rgb} = \im \varkappa_{rgb}$,
equation~(\ref{krgb}).
Although at this point the modification~(\ref{adjustsmbivectorsI6})
is needed only for bivectors in the $\Spin(10)$ algebra,
for which $k,l = y,z,r,g,b$
(excluding $t$),
it turns out that the electroweak and grand Higgs fields,
equations~(\ref{Higgsvac}) and (\ref{Higgsnuvac}),
have the correct symmetry-breaking behaviour
only if the modification~(\ref{adjustsmbivectorsI6})
is extended to all $\Spin(11,1)$ bivectors, that is,
for all of $k,l = t,y,z,r,g,b$.

The purpose of making the modification~(\ref{adjustsmbivectorsI6})
was to allow Lorentz transformations
to connect fermions across different entries of
the $\Spin(10)$ chart~(\ref{yzrgbtab}),
and that works, as will now be shown.
The modification~(\ref{adjustsmbivectorsI6})
serves to replace each antifermion in the chart with the corresponding fermion.
For example, the positron entries
$\bar{e}_\Rchiral$ and $\bar{e}_\Lchiral$
are replaced by electrons
$e_\Lchiral$ and $e_\Rchiral$.
What about antifermions?
Where have they gone?
The answer is that antifermions are obtained from fermions in the usual way
\cite{Hamilton:2023a},
by taking their complex conjugates and multiplying by the conjugation operator
$\Cp$, equation~(\ref{Cpspin10p}),
$\conj{\psi} \equiv \Cp \psi^\ast$.
Thus antifermions appear in a second copy of
the $\Spin(10)$ chart~(\ref{yzrgbtab}),
a conjugated version in which all fermions are replaced by antifermions.

It requires some work, \S\ref{spin111justification},
to establish the correct assignment of Dirac boost ($\Uparrow$ or $\Downarrow$)
and spin ($\uparrow$ or $\downarrow$) bits,
but the end result is the following $\Spin(11,1)$ chart of spinors,
arranged in columns by the number of $\Spin(10)$ up-bits
as in the earlier chart~(\ref{yzrgbtab}):
\begin{equation}
\label{tyzrgbtab}
  \begin{array}{c@{\quad}c@{\quad}c@{\quad}c@{\quad}c@{\quad}c}
  \hline
  0 & 1 & 2 & 3 & 4 & 5
  \\
  \hline
  \noalign{\vskip .5ex}
  \mbox{--} : \ 
  \substack{\displaystyle
  \bar{\nu}_{\Updown}
  \\\displaystyle
  \nu_{\Downdown}
  }
  &
  y : \ 
  \substack{\displaystyle
  \bar{\nu}_{\Downdown}
  \\\displaystyle
  \nu_{\Updown}
  }
  &
  \phantom{d}\bar{c} : \ 
  \substack{\displaystyle
  \bar{u}_{\Updown}^{\, \bar{c}}
  \\\displaystyle
  u_{\Downdown}^{\, c}
  }
  &
  \phantom{d}y\bar{c} : \
  \substack{\displaystyle
  \bar{u}_{\Downdown}^{\, \bar{c}}
  \\\displaystyle
  u_{\Updown}^{\, c}
  }
  &
  zrgb : \ 
  \substack{\displaystyle
  \nu_{\Downup}
  \\\displaystyle
  \bar{\nu}_{\Upup}
  }
  &
  yzrgb : \ 
  \substack{\displaystyle
  \nu_{\Upup}
  \\\displaystyle
  \bar{\nu}_{\Downup}
  }
  \\[2ex]
  &
  z : \ 
  \substack{\displaystyle
  \bar{e}_{\Downdown}
  \\\displaystyle
  e_{\Updown}
  }
  &
  yz : \ 
  \substack{\displaystyle
  \bar{e}_{\Updown}
  \\\displaystyle
  e_{\Downdown}
  }
  &
  rgb : \ 
  \substack{\displaystyle
  e_{\Upup}
  \\\displaystyle
  \bar{e}_{\Downup}
  }
  &
  yrgb : \ 
  \substack{\displaystyle
  e_{\Downup}
  \\\displaystyle
  \bar{e}_{\Upup}
  }
  &
  \\[2ex]
  &
  c : \ 
  \substack{\displaystyle
  d_{\Upup}^{\, c}
  \\\displaystyle
  \bar{d}_{\Downup}^{\, \bar{c}}
  }
  &
  yc : \ 
  \substack{\displaystyle
  d_{\Downup}^{\, c}
  \\\displaystyle
  \bar{d}_{\Upup}^{\, \bar{c}}
  }
  &
  \phantom{d}z\bar{c} : \ 
  \substack{\displaystyle
  \bar{d}_{\Downdown}^{\, \bar{c}}
  \\\displaystyle
  d_{\Updown}^{\, c}
  }
  &
  \phantom{d}yz\bar{c} : \ 
  \substack{\displaystyle
  \bar{d}_{\Updown}^{\, \bar{c}}
  \\\displaystyle
  d_{\Downdown}^{\, c}
  }
  \\[2ex]
  &
  &
  zc : \ 
  \substack{\displaystyle
  u_{\Downup}^{\, c}
  \\\displaystyle
  \bar{u}_{\Upup}^{\, \bar{c}}
  }
  &
  yzc : \ 
  \substack{\displaystyle
  u_{\Upup}^{\, c}
  \\\displaystyle
  \bar{u}_{\Downup}^{\, \bar{c}}
  }
  &
  &
  \\[1.7ex]
  \hline
  \end{array}
\end{equation}
Whereas in the original $\Spin(10)$ chart~(\ref{yzrgbtab})
each entry was a 2-component Weyl spinor,
in the $\Spin(11,1)$ chart~(\ref{tyzrgbtab})
the 2~components of each Weyl spinor appear in bit-flipped entries.
For example, the right-handed electron $e_\Rchiral$ of the original chart
is replaced by $e_{\Upup}$,
and its spatially rotated partner $e_{\Downdown}$ of the same chirality
appears in the all-bit-flipped entry.
Each entry still has two components,
but in the $\Spin(11,1)$ chart those two components differ by their $t$-bit:
the upper component has $t$-bit up,
the lower $t$-bit down.
The net number of degrees of freedom remains the same, $2^6 = 64$.
Flipping the $t$-bit transforms a fermion
into its antifermionic partner of opposite boost.
Flipping all bits except the $t$-bit transforms a fermion
into its antifermionic partner of opposite spin.
The Dirac boost bit in the $\Spin(11,1)$ chart~(\ref{tyzrgbtab}) is
$\Uparrow$ or $\Downarrow$ as $\varkappa_{tyz}$ is positive or negative,
that is, as the number of $tyz$ up-bits is odd or even.
The Dirac spin bit is $\uparrow$ or $\downarrow$
as $\varkappa_{rgb}$ is positive or negative,
that is, as the number of $rgb$ up-bits is odd or even.
Whereas in the original $\Spin(10)$ chart~(\ref{yzrgbtab})
a spinor was a fermion or antifermion
as $\varkappa_{rgb}$ was positive or negative,
in the $\Spin(11,1)$ chart~(\ref{tyzrgbtab}),
a spinor is a fermion or antifermion
as the time-colour chirality $\varkappa_{trgb}$ is positive or negative,
that is, as the number of $trgb$ up-bits is even or odd.

\spinelevenonechartfig

A Dirac spinor has 4 complex components, for a total of 8 degrees of freedom.
As previously in the $\Spin(10)$ chart~(\ref{yzrgbtab}),
each Dirac spinor in the $\Spin(11,1)$ chart~(\ref{tyzrgbtab})
has 8 components, so each component must represent one real degree of freedom.
In any geometric algebra with one time dimension,
conjugation flips all bits except the boost bit.
This is true in the Dirac algebra,
equation~(\ref{Cpchiraldirac}),
and it is also true in the $\Spin(11,1)$ algebra,
where the boost bit is the time bit $t$.

Figure~\ref{spin111chart} illustrates
one generation (the electron generation) of fermions of the standard model
arranged according to their $\Spin(11,1)$ $tyzrgb$ charges.

The correctness of the assignment of Dirac boost and spin bits
in the $\Spin(11,1)$ chart~(\ref{tyzrgbtab}),
and the consistency of the entire construction,
will now be established.

\subsection{Justification of the Spin(11,1) chart}
\label{spin111justification}

The assignment of Dirac boost and spin bits for each individual species
(electrons, for example)
in the $\Spin(11,1)$ chart~(\ref{tyzrgbtab})
is determined by two conditions,
that
conjugation flips spin $\uparrow \,\leftrightarrow\, \downarrow$,
while
flipping the $y$-bit flips boost $\Uparrow \,\leftrightarrow\, \Downarrow$.
The first condition holds because
the expression~(\ref{Cpchiraldirac}) for the Dirac conjugation operator $\Cp$
in the chiral representation shows that Dirac conjugation flips
chirality and spin, but not boost.
The second condition holds because,
after electroweak symmetry breaking,
flipping the $y$-bit flips spinors
between right- and left-handed Dirac chiralities of the same species,
for example $e_\Rchiral \,\leftrightarrow\, e_\Lchiral$.
Massive spinors are linear combinations of the two chiralities.
Since massive spinors have definite spin, either $\uparrow$ or $\downarrow$,
flipping the $y$-bit must flip the Dirac boost bit
while preserving the spin bit,
for example,
$e_{\Upup} \,\leftrightarrow\, e_{\Downup}$.

The two conditions
suffice to determine the translation between
Dirac and $\Spin(11,1)$ spinors of the same species,
but they do not fix the translation across different species.
The translation across different species
is determined by the condition that Lorentz transformations commute
with standard-model transformations,
in accordance with the Coleman-Mandula theorem.

The Dirac algebra contains two mutually commuting operators,
$\bgamma_0 \bgamma_3$ and $\bgamma_1 \bgamma_2$,
that respectively generate a boost and a spatial rotation
without flipping any bits.
The operator $\bgamma_0 \bgamma_3$ generates a boost
in the $\Uparrow$-$\Downarrow$ boost plane:
a boost by rapidity $\theta$
multiplies $\Uparrow$ and $\Downarrow$ spinors
by a real number $\ee^{\pm \theta/2}$.
The operator $\bgamma_1 \bgamma_2$ generates a spatial rotation
in the $\uparrow$-$\downarrow$ spin plane:
a rotation by angle $\theta$
multiplies $\uparrow$ and $\downarrow$ spinors
by a phase $\ee^{\mp \im \theta/2}$.

The $\Spin(11,1)$ geometric algebra contains three mutually commuting
operators that flip no bits,
and at the same time commute with all standard-model transformations,
namely the time pseudoscalar $I_{t}$,
the weak pseudoscalar $I_{yz}$,
and the colour pseudoscalar $I_{rgb}$
(the definition~(\ref{Irgb}) essentially repeats the earlier
definition~(\ref{krgb})),
\begin{subequations}
\label{Ityzs}
\begin{alignat}{5}
\label{It}
  I_t
  &\equiv
  -\im \bgamma^+_t \bgamma^-_t
  &&=\,
  &\varkappa_t
  &\equiv
  \bgamma_t \wedgie \bgamma_{\bar t}
  \ ,
\\
\label{Idu}
  I_{yz}
  &\equiv
  \bgamma^+_y \bgamma^-_y \bgamma^+_z \bgamma^-_z
  &&=\,
  &- \varkappa_{yz}
  &\equiv
  - \bgamma_y \wedgie \bgamma_{\bar y} \wedgie
  \bgamma_z \wedgie \bgamma_{\bar z}
  \ ,
\\
\label{Irgb}
  I_{rgb}
  &\equiv
  \bgamma^+_r \bgamma^-_r \bgamma^+_g \bgamma^-_g \bgamma^+_b \bgamma^-_b
  &&=\,
  &-\im \varkappa_{rgb}
  &\equiv
  -\im
  \bgamma_r \wedgie \bgamma_{\bar r} \wedgie
  \bgamma_g \wedgie \bgamma_{\bar g} \wedgie
  \bgamma_b \wedgie \bgamma_{\bar b}
  \ .
\end{alignat}
\end{subequations}
The weak pseudoscalar $I_{yz}$ changes sign when an odd number of $yz$ bits
are flipped,
and, as remarked above,
flipping an odd number of $yz$ bits flips the Dirac boost bit.
But $I_{yz}$ is a spacelike operator
(it generates a rotation by a phase),
so cannot by itself generate a boost.
On the other hand the time pseudoscalar $I_t$ does generate a Lorentz boost,
whose action is independent of the $yz$ bits.
Thus the combination $I_{tyz}$ generates a Lorentz boost
that acts oppositely on spinors of opposite weak chirality,
consistent with the behaviour of the Dirac boost operator $\bgamma_0 \bgamma_3$.
The $\Spin(11,1)$ boost operator that can be identified with
the Dirac boost operator $\bgamma_0 \bgamma_3$ is
\begin{equation}
\label{Ityz}
  I_{tyz}
  \equiv
  -\im \bgamma^+_t \bgamma^-_t \bgamma^+_y \bgamma^-_y \bgamma^+_z \bgamma^-_z
  =
  - \varkappa_{tyz}
  \equiv
  -
  \bgamma_t \wedgie \bgamma_{\bar t} \wedgie
  \bgamma_y \wedgie \bgamma_{\bar y} \wedgie
  \bgamma_z \wedgie \bgamma_{\bar z}
  \ .
\end{equation}

It was remarked just before the definition~(\ref{krgb})
of colour chirality
that $\Spin(10)$-conjugate spinors have opposite colour chirality
$\varkappa_{rgb}$.
As remarked above, in the Dirac algebra,
conjugation flips spin but not boost.
Therefore the colour pseudoscalar $I_{rgb}$ can be identified with
the Dirac spin operator $\bgamma_1 \bgamma_2$.

The product of the boost operator $I_{tyz}$
and the spin operator $I_{rgb}$ equals
the 12-dimensional pseudoscalar $\Ityzrgb$,
\begin{align}
\label{JItyzrgb}
  \Ityzrgb
  &\equiv
  I_{tyz} I_{rgb}
  =
  -\im
  \bgamma^+_t \bgamma^-_t
  \bgamma^+_y \bgamma^-_y \bgamma^+_z \bgamma^-_z
  \bgamma^+_r \bgamma^-_r \bgamma^+_g \bgamma^-_g \bgamma^+_b \bgamma^-_b
\nonumber
\\
  &=
  \im \varkappa_{12}
  \equiv
  \im
  \bgamma_t \wedgie \bgamma_{\bar t} \wedgie
  \bgamma_y \wedgie \bgamma_{\bar y} \wedgie
  \bgamma_z \wedgie \bgamma_{\bar z} \wedgie
  \bgamma_r \wedgie \bgamma_{\bar r} \wedgie
  \bgamma_g \wedgie \bgamma_{\bar g} \wedgie
  \bgamma_b \wedgie \bgamma_{\bar b}
  \ .
\end{align}
The 12-dimensional chiral operator $\varkappa_{12}$
corresponding to the Dirac chiral operator $\gamma_5 = -\im I$ is
\begin{equation}
\label{kappa12}
  \varkappa_{12}
  =
  -\im \Ityzrgb
  \ .
\end{equation}
It is the 12-dimensional chiral operator $\varkappa_{12}$
that should be identified with the Dirac chiral operator $\gamma_5$,
not the 10-dimensional chiral operator $\varkappa_{10}$
as suggested by the coincidence~(\ref{II10}).

The final ingredient to complete the translation between
Dirac and $\Spin(11,1)$ algebras is to identify an operator that
connects the two Weyl components of each spinor
in the $\Spin(11,1)$ chart~(\ref{tyzrgbtab})
to each other.
In the Dirac algebra,
the two Weyl components are connected by flipping both bits.
A Dirac operator that flips both bits is
$I \sigma_2 = - \bgamma_1 \bgamma_3$,
equation~(\ref{ichiral}),
which is also the generator of a spatial rotation.
The corresponding operator that flips all bits in the $\Spin(11,1)$ algebra is
\begin{equation}
\label{Jsigma2}
  \Ityzrgb \sigma_2
  \equiv
  \bgamma^+_t \bgamma^+_y \bgamma^+_z \bgamma^+_r \bgamma^+_g \bgamma^+_b
  \ ,
\end{equation}
where $\Ityzrgb$ is the pseudoscalar~(\ref{JItyzrgb}).
Equation~(\ref{Jsigma2}) can be regarded as defining $\sigma_2$.
Below, equation~(\ref{lorentzspin10ews2}),
$\sigma_2$ will be identified as a generator of a Lorentz boost.
The expression~(\ref{Jsigma2}) for $\Ityzrgb \sigma_2$ coincides with
that for the $\Spin(11,1)$ spinor metric $\varepsilon$,
equation~(\ref{spinormetricspin10p}),
but the two are not the same because $\Ityzrgb \sigma_2$ transforms as a multivector
whereas the spinor metric $\varepsilon$ transforms
as a (Lorentz-invariant) spinor tensor.
The coincidence of the expressions for $\Ityzrgb \sigma_2$ and $\varepsilon$
is similar to the coincidence~(\ref{echiraldirac}) between $I \sigma_2$
and the spinor metric $\varepsilon$
in the chiral representation of the Dirac algebra.

To qualify as a satisfactory operator that generates a spatial rotation,
the operator $\Ityzrgb \sigma_2$ defined by equation~(\ref{Jsigma2})
must satisfy two conditions.
First,
consistent with the expected anticommutation of generators of spatial rotations,
$\Ityzrgb \sigma_2$ must anticommute with $I_{rgb}$,
which was identified above as generating a spatial rotation.
Second,
in accordance with the Coleman-Mandula theorem,
$\Ityzrgb \sigma_2$ must commute with all standard-model bivectors
modified per~(\ref{adjustsmbivectorsI6}).
Both conditions hold, and that should not be too surprising.
Recall that the modification~(\ref{adjustsmbivectorsI6}) of $\Spin(10)$
bivectors was done precisely to enable the existence of an operator
that connects the two Weyl components of a spinor by a bit-flip;
and the two Weyl components of a spinor are related by a spatial rotation.

Note that whereas the generator $\Ityzrgb \sigma_2$, equation~(\ref{Jsigma2}),
of a spatial rotation flips all six $tyzrgb$ bits,
and preserves standard-model charges,
the $\Spin(11,1)$ conjugation operator $\Cp$, equation~(\ref{Cpspin10p}),
which transforms a fermion into its antifermionic partner,
flips all bits except $t$, and flips all standard-model charges,
consistent with the assignment of fermions
in the $\Spin(11,1)$ chart~(\ref{tyzrgbtab}).

\subsection{The Dirac algebra as a subalgebra of the Spin(11,1) geometric algebra}
\label{diracinspin10-sec}

The Dirac algebra can now be expressed in terms of the
$\Spin(11,1)$ geometric algebra.

The generators
$I_{rgb}$ and $\Ityzrgb \sigma_2$,
equations~(\ref{Irgb}) and~(\ref{Jsigma2}),
and their product
constitute a set of 3 anticommuting generators of spatial rotations that
commute with all standard-model bivectors
modified per~(\ref{adjustsmbivectorsI6}).
The pseudoscalar $\Ityzrgb$ is given by equation~(\ref{JItyzrgb}).
The full set of 6 Lorentz generators,
consisting of 3 spatial generators $\Ityzrgb \sigma_a$
and 3 boost generators $\sigma_a$, is
\begin{subequations}
\label{lorentzspin10ew}
\begin{align}
\label{lorentzspin10ewJs1}
  \Ityzrgb \sigma_1
  &=
  - \bgamma^+_t \bgamma^+_y \bgamma^+_z \bgamma^-_r \bgamma^-_g \bgamma^-_b
  \ ,
\\
\label{lorentzspin10ewJs2}
  \Ityzrgb \sigma_2
  &=
  \bgamma^+_t \bgamma^+_y \bgamma^+_z \bgamma^+_r \bgamma^+_g \bgamma^+_b
  \ ,
\\
\label{lorentzspin10ewJs3}
  I_{rgb}
  =
  \Ityzrgb \sigma_3
  &=
  \bgamma^+_r \bgamma^-_r \bgamma^+_g \bgamma^-_g \bgamma^+_b \bgamma^-_b
  \ ,
\\
\label{lorentzspin10ews1}
  \sigma_1
  &=
  \im \bgamma^-_t \bgamma^-_y \bgamma^-_z \bgamma^+_r \bgamma^+_g \bgamma^+_b
  \ ,
\\
\label{lorentzspin10ews2}
  \sigma_2
  &=
  \im \bgamma^-_t \bgamma^-_y \bgamma^-_z \bgamma^-_r \bgamma^-_g \bgamma^-_b
  \ ,
\\
\label{lorentzspin10ews3}
  I_{tyz}
  =
  \sigma_3
  &=
  -\im \bgamma^+_t \bgamma^-_t \bgamma^+_y \bgamma^-_y \bgamma^+_z \bgamma^-_z
  \ .
\end{align}
\end{subequations}
The 6 Lorentz generators all have grade 6.
They are not bivectors,
but they nevertheless generate Lorentz transformations.
The 8 basis elements of the complete Lie algebra of Lorentz transformations
comprise the 6 Lorentz generators~(\ref{lorentzspin10ew})
along with the unit element and the pseudoscalar $\Ityzrgb$
given by equation~(\ref{JItyzrgb}).
The commutation rules of the elements of the Lie algebra
are those of the Lorentz algebra.
With the modification~(\ref{adjustsmbivectorsI6}) to standard-model generators,
all the Lorentz generators commute with all standard-model generators.

Given a time vector $\bgamma_0$
and a set of generators $\sigma_a$ of Lorentz boosts,
spatial vectors $\bgamma_a$
can be deduced by Lorentz transforming $\bgamma_0$ appropriately.
Since the boost generators satisfy
$\sigma_a = \bgamma_0 \bgamma_a$,
spatial vectors satisfy
$\bgamma_a = - \bgamma_0 \sigma_a$.
With the time axis $\bgamma_0 = \im \bgamma^-_t$
and the expressions~(\ref{lorentzspin10ew}) for $\sigma_a$,
the full set of 4 orthonormal spacetime vectors $\bgamma_m$ is
\begin{subequations}
\label{vectorsspin10ew}
\begin{align}
\label{vectorsspin10ew0}
  \bgamma_0
  &=
  \im \bgamma^-_t
  \ ,
\\
\label{vectorsspin10ew1}
  \bgamma_1
  &=
  \bgamma^-_y \bgamma^-_z \bgamma^+_r \bgamma^+_g \bgamma^+_b
  \ ,
\\
\label{vectorsspin10ew2}
  \bgamma_2
  &=
  \bgamma^-_y \bgamma^-_z \bgamma^-_r \bgamma^-_g \bgamma^-_b
  \ ,
\\
\label{vectorsspin10ew3}
  \bgamma_3
  &=
  \bgamma^+_t \bgamma^+_y \bgamma^-_y \bgamma^+_z \bgamma^-_z
  \ .
\end{align}
\end{subequations}
The vectors~(\ref{vectorsspin10ew}) all have grade 1~mod~4.
The multiplication rules for
the vectors $\bgamma_m$ given by equations~(\ref{vectorsspin10ew})
agree with the usual multiplication rules for Dirac $\gamma$-matrices:
the vectors $\bgamma_m$ anticommute,
and their scalar products form the Minkowski metric.
All the spacetime vectors $\bgamma_m$ commute with all standard-model bivectors
modified per~(\ref{adjustsmbivectorsI6}).
The Dirac pseudoscalar $I$ coincides with
the $\Spin(11,1)$ pseudoscalar $\Ityzrgb$
defined by equation~(\ref{JItyzrgb}),
\begin{equation}
\label{IJ}
  I
  \equiv
  \bgamma_0 \bgamma_1 \bgamma_2 \bgamma_3
  =
  \Ityzrgb
  \ .
\end{equation}
Equivalently,
the Dirac chiral operator $\gamma_5 \equiv - \im I$
coincides with the $\Spin(11,1)$ chiral operator
$\varkappa_{12} \equiv - \im \Ityzrgb$.

Thus the Dirac and standard-model algebras are subalgebras
of the $\Spin(11,1)$ geometric algebra,
such that all Dirac generators commute with all standard-model generators
modified per~(\ref{adjustsmbivectorsI6}),
as was to be proved.

The time dimension~(\ref{vectorsspin10ew0}) is just a simple vector
in the $\Spin(11,1)$ algebra,
but the 3 spatial dimensions~(\ref{vectorsspin10ew1})--(\ref{vectorsspin10ew3})
are all 5-dimensional.
The spatial dimensions
share a common 2-dimensional factor $\bgamma^-_y \bgamma^-_z$.
Aside from that common factor,
each of the 3 spatial dimensions is itself 3-dimensional:
$\bgamma^+_r \bgamma^+_g \bgamma^+_b$,
$\bgamma^-_r \bgamma^-_g \bgamma^-_b$,
and
$\bgamma^+_t \bgamma^+_y \bgamma^+_z$.

\section{The path from the standard model to grand unification}
\label{spin5spin6-sec}

The previous section~\ref{commutingsubalgebras-sec}
established the main result of this paper,
that the Dirac and standard-model algebras are commuting subalgebras
of the $\Spin(11,1)$ geometric algebra.
It would be nice to declare victory at this point.
However, there is more hard work to do,
to determine whether there exists a viable symmetry breaking chain
from $\Spin(11,1)$ to the standard model.
As described in the Introduction, \S\ref{intro-sec},
there is a substantial literature on possible symmetry breaking chains
from $\Spin(10)$.

The bottom line of the next several
sections~\ref{higgsew-sec}--\ref{energyunification-sec},
is that there does appear to be a symmetry breaking chain~(\ref{symbreaking}),
and the Higgs sector that mediates it is encouragingly economical,
namely a 66-component $\Spin(11,1)$ bivector multiplet
modified per~(\ref{adjustsmbivectorsI6}).
In other words, the Higgs field is the scalar (spin~0)
counterpart of the vector (spin~1) gauge fields in $\Spin(11,1)$.
A key property of Higgs fields is that an unbroken massless gauge field
is broken and becomes massive by absorbing a Higgs scalar into its
longitudinal component.
With the Higgs sector matching the gauge sector,
for each gauge field in $\Spin(11,1)$
there is a matching Higgs field available to make it massive if needed.
It is possible that there is a more complicated route
that involves adjoining fields outside the realm of $\Spin(11,1)$,
but we have not pursued that possibility.
If there are other possibilities,
the model with a 66-component $\Spin(11,1)$ Higgs sector
is the minimal model.


Symmetry breaking from $\Spin(11,1)$ is constrained
by the condition that the spacetime and internal algebras commute
at every step of unification,
as required by the Coleman-Mandula theorem.
For example,
although the Dirac vectors $\bgamma_m$,
equations~(\ref{vectorsspin10ew}),
all commute with the all the bivector generators of the
$\U(1)_Y \times \SU(2)_\Lchiral \times \SU(3)_c$
standard model,
modified per~(\ref{adjustsmbivectorsI6}),
the Dirac vectors $\bgamma_m$
do {\em not\/} commute with general $\SU(5)$ transformations.
As long as spacetime is 4-dimensional and spacetime transformations
commute with internal rotations,
$\SU(5)$ cannot be an internal symmetry.

An exhaustive search establishes that the largest subgroup of $\Spin(11,1)$
whose bivector generators,
modified per~(\ref{adjustsmbivectorsI6}),
all commute with the spacetime vectors~(\ref{vectorsspin10ew})
is a product of weak and colour groups
$\Spin(5)_w \times \Spin(6)_c$,
in which the generators of $\Spin(5)_w$ are the 10 bivectors formed
from antisymmetric products of the five vectors
$\bgamma^+_t$,
$\bgamma^+_y$, $\bgamma^-_y$,
$\bgamma^+_z$, $\bgamma^-_z$,
while the generators of $\Spin(6)_c$ are the 15 bivectors formed
from antisymmetric products of the six vectors
$\bgamma^+_k$ and $\bgamma^-_l$
with $k$ and $l$ running over $r,g,b$.
More generally,
the largest subalgebra of the $\Spin(11,1)$ geometric algebra whose generators,
modified per~(\ref{adjustsmbivectorsI6}),
all commute with the spacetime vectors~(\ref{vectorsspin10ew})
is the $2^4 \times 2^5 = 2^9$ dimensional direct product of the even
geometric algebras associated with respectively $\Spin(5)_w$ and $\Spin(6)_c$.
Even more generally,
in order to commute with all spacetime vectors~(\ref{vectorsspin10ew}),
a tensor of (modified) multivectors
must have an even total number of $t^-$ indices,
and an even total number of $t^+,y^\pm,z^\pm$ indices,
and an even total number of $r^\pm,g^\pm,b^\pm$ indices.

This suggests that $\Spin(5)_w \times \Spin(6)_c$ could be on the path
up to grand unification while spacetime is still 4-dimensional.
As discussed in \S\ref{higgsew-sec},
the 4 modified bivectors involving the 11th spatial dimension $\bgamma^+_t$,
namely $\bgamma^+_t \bgamma^\pm_k$, $k = y,z$,
modified per~(\ref{adjustsmbivectorsI6}),
call attention to themselves because they match the
properties of the Lorentz-scalar electroweak Higgs multiplet.

Although the 4 (modified) bivectors
$\bgamma^+_t \bgamma^\pm_k$, $k = y,z$,
can generate the electroweak Higgs field, it turns out that they
cannot in addition generate a gauge field,
because they do not commute
with the grand Higgs field~(\ref{Higgsnuvac})
that causes grand symmetry breaking.
This leaves the Pati-Salam group
$\Spin(4)_w \times \Spin(6)_c$
as the largest possible gauge group on the path
to grand unification while spacetime is still 4-dimensional.

If the $\Spin(11,1)$ model of this paper is correct,
then surely intermediate unification to $\Spin(4)_w \times \Spin(6)_c$ must occur,
since otherwise,
in the absence of new physics such as supersymmetry \cite{Martin:1998},
the three coupling parameters of the standard model do not meet,
and grand unification does not occur.

A gauge field is a Lorentz vector
because it arises as a connection in a gauge-covariant spacetime derivative.
A Higgs field on the other hand must be a Lorentz scalar, since otherwise
it would break the Lorentz symmetry of the vacuum
(it would impose a preferred spatial direction and/or rest frame),
contrary to observation.
For a gauge field,
a scalar Lagrangian can be constructed from
the commutator of the gauge-covariant derivative.
For a Higgs field,
any generator that commutes with Lorentz symmetries~(\ref{lorentzspin10ew})
defines a Lorentz scalar that could play the role of a scalar field
in the Lagrangian.
Whereas gauge fields 
must transform according to the adjoint representation of the group,
Higgs generators could potentially transform in any representation of the group.
However, as remarked above,
the minimal Higgs sector in the $\Spin(11,1)$ model turns out to
transform as the adjoint representation, just like the gauge fields.

The general principles underlying symmetry breaking by the Higgs mechanism
\cite{Englert:1964,Higgs:1964}
are:
(1) the Higgs field before symmetry breaking is a scalar (spin~0) multiplet
of the unbroken symmetry;
(2) one component of the Higgs multiplet acquires a nonzero vacuum
expectation value, and that component is invariant (uncharged, a singlet)
under the symmetries that remain after symmetry breaking;
(3) components of the Higgs multiplet whose symmetry is broken
are absorbed into longitudinal components of the broken gauge (spin~1) fields
by the Goldstone mechanism \cite{Goldstone:1962},
giving those gauge fields mass;
and (4) unbroken components of the Higgs multiplet persist as scalar fields,
potentially available to mediate the next level of symmetry breaking.

There are three stages of symmetry breaking in the $\Spin(11,1)$ model,
namely grand symmetry breaking,
Pati-Salam ($\Spin(4)_w \times \Spin(6)_c$) symmetry breaking,
and electroweak symmetry breaking.
All the Higgs fields involved in symmetry breaking lie
in a common 66-component $\Spin(11,1)$ multiplet of bivectors
modified per~(\ref{adjustsmbivectorsI6}).
Denote this Higgs multiplet $\bE$,
after Englert-Brout
\cite{Englert:1964},
who proposed the Higgs mechanism at the same time as
(marginally before) Higgs \cite{Higgs:1964},
\begin{equation}
\label{spin111Emultiplet}
  \bE
  \equiv
  E^{k^\pm l^\pm} \bgamma^\pm_k \bgamma^\pm_l
  \ , \quad  k, l \mbox{~in~} t,y,z,r,g,b
  \ ,
\end{equation}
in which it is to be understood that all imaginary bivectors
($\bgamma^+_k \bgamma^-_l$ or $\bgamma^-_k \bgamma^+_l$)
are to be multiplied by $\varkappa_{rgb}$,
per the modification~(\ref{adjustsmbivectorsI6}).
The modification~(\ref{adjustsmbivectorsI6}) is necessary to ensure that,
after grand symmetry breaking,
all unbroken fields commute with all Lorentz bivectors~(\ref{lorentzspin10ew}).

There is a complication to the simple story of the previous paragraph.
The 4 (modified) bivectors
$\bgamma^+_t \bgamma^\pm_k$, $k = y,z$,
that belong to the $\Spin(5)_w$ but not $\Spin(4)_w$ algebra
appear to be special.
The grand Higgs field $\langle \bT \rangle$,
equation~(\ref{Higgsnuvac}),
fails to commute with these 4 bivectors,
yet the associated scalar fields are not absorbed into their
partner vector fields,
but rather persist as the 4-component electroweak Higgs multiplet.
The only way we have been able to make sense of that complication
is to postulate that the dimension $\bgamma^+_t$ is a scalar dimension
that does not generate any symmetry,
a possibility discussed in \S4.4 of \cite{Hamilton:2023a}.
In other words,
the gauge group before grand symmetry breaking is not $\Spin(11,1)$,
but rather $\Spin(10,1)$,
the group generated by the 11 bivectors formed from
the time vector $\bgamma^-_t$
and the 10 $\Spin(10)$ vectors $\bgamma^\pm_k$, $k = y,z,r,g,b$.
This conclusion is discussed further in \S\ref{Spin101-sec}.

\section{Electroweak symmetry breaking}
\label{higgsew-sec}

The 4 bivector generators $\bgamma^+_t \bgamma^\pm_k$ with $k = y,z$
call attention to themselves
because they transform spinors by one unit
of standard-model charge $y$ or $z$,
whereas the remaining $6 + 15 = 21$ of the $10 + 15 = 25$ bivector generators
of $\Spin(5)_w \times \Spin(6)_c$,
which generate the Pati-Salam group
$\Spin(4)_w \times \Spin(6)_c$ \cite{Pati:1974},
transform spinors by an even number of standard-model charges $yzrgb$.
The electroweak Higgs field responsible for breaking $y$-symmetry
carries one unit of $y$-charge.
The electroweak Higgs field gives masses to fermions
by flipping their $y$-bit.

The Weinberg theory of electroweak symmetry breaking \cite{Weinberg:1967}
requires the electroweak Higgs field to be part of a multiplet of 4 fields
that transform into each other under
$\U(1)_Y \times \SU(2)_\Lchiral$.
Indeed the 4 bivector generators $\bgamma^+_t \bgamma^\pm_k$ with $k = y,z$
provide such a set of fields.
The 4-component Higgs field $\bH$ may be defined by
\begin{equation}
\label{Higgsmultiplet}
  \bH
  \equiv
  H^{k^\pm} \bgamma^+_t \bgamma^\pm_k
  \ , \quad
  k = y, z
  \ ,
\end{equation}
where the imaginary bivectors
$\bgamma^+_t \bgamma^-_k$, $k = y,z$,
are to be understood as being multiplied by $\varkappa_{rgb}$
per the modification~(\ref{adjustsmbivectorsI6}).
Electroweak symmetry breaking occurs when the Higgs field
acquires a vacuum expectation value $\langle \bH \rangle$
proportional to $\bgamma^+_t \bgamma^-_y$,
\begin{equation}
\label{Higgsvac}
  \langle \bH \rangle
  =
  \langle H \rangle \bgamma^+_t \bgamma^-_y \varkappa_{rgb}
  \ ,
\end{equation}
in which the modification factor $\varkappa_{rgb}$ has been included
explicitly to avoid possible misunderstanding.
When combined with the time axis $-\im \bgamma_0 \equiv \bgamma^-_t$
in a fermion mass term
$\conj{\psi} \cdot \langle \bH \rangle \psi = - \im \psi^\dagger \bgamma_0 \langle \bH \rangle \psi$,
the vacuum Higgs field~(\ref{Higgsvac}) yields a Dirac mass term
\begin{equation}
\label{HiggsDiracmass}
  \bgamma^-_t\langle \bH \rangle
  =
  \langle H \rangle
  \bgamma^-_t \bgamma^+_t \bgamma^-_y \varkappa_{rgb}
  \ ,
\end{equation}
which indeed flips the $y$-bit as it should.
The vacuum Higgs field~(\ref{Higgsvac}) is proportional to
$\bgamma^-_y$
not $\bgamma^+_y$
because the factor $\bgamma^-_y$ in the mass term~(\ref{HiggsDiracmass})
preserves the spinor identity,
whereas $\bgamma^+_y$ flips between spinor and antispinor,
in much the same way that in the Dirac algebra
the time axis $\bgamma_0$ maps massive spinors and antispinors to themselves,
whereas the time axis' boost partner $\bgamma_3$
flips between massive spinor and massive antispinor,
equations~(\ref{gammadirac}).
Note that whereas the Higgs multiplet $\bH$, equation~(\ref{Higgsmultiplet}),
commutes with the time axis $\bgamma_0$,
the pseudoscalar multiplet $\Ityzrgb \bH$ does not,
so it is $\langle \bH \rangle$ and not $\Ityzrgb \langle \bH \rangle$
that generates a Dirac mass.

The electroweak Higgs field~(\ref{Higgsmultiplet}) possesses
$\Spin(4)_w$ rotational symmetry in the 4-dimensional space of $y,z$ vectors.
The manner in which the $\Spin(4)_w$ symmetry
is broken to the electroweak symmetry
$\U(1)_Y \times \SU(2)_\Lchiral$
is noted in \S\ref{PSmass-sec},
equation~(\ref{HiggsXcommutators}).


The remainder of this section~\ref{higgsew-sec}
is essentially a standard exposition of electroweak symmetry breaking,
adapted to the present notation;
see for example \cite[Ch.~20]{Peskin:1995}
for a pedagogical account.
To avoid overloading the notation,
all the gauge and Higgs generators
in the rest of this section~\ref{higgsew-sec}
are treated as bivectors,
whereas correctly they should be modified
per~(\ref{adjustsmbivectorsI6}),
by multiplying all imaginary bivectors by the colour chiral operator
$\varkappa_{rgb}$.
The operator $\varkappa_{rgb}$ commutes with all standard-model
bivectors, so its exclusion makes no difference to the algebra.

For a spinor field $\psi$, the gauge-covariant derivative
with respect to $\U(1)_Y \times \SU(2)_\Lchiral$ transformations is
\begin{equation}
\label{DBWpsi}
  \bDD_m \psi
  =
  \left(
  \partial_m
  +
  g_Y \bB_m
  +
  g_w \bW_m
  \right)
  \psi
  \ ,
\end{equation}
where $\bB_m$ and $\bW_m$
are the $\U(1)_Y$ and $\SU(2)_\Lchiral$ gauge fields
\begin{equation}
\label{BYWtau}
  \bB_m
  \equiv
  \im B_m Y
  \ , \quad
  \bW_m
  \equiv
  \im W_m^k \tau_k
  \ ,
\end{equation}
and $g_Y$ and $g_w$ are dimensionless coupling strengths
for those fields.
Here $\im Y$, equation~(\ref{uY1bivectora}),
is the generator of the hypercharge symmetry $\U(1)_Y$,
while $\im \tau_k$
with $\tau_k$ satisfying the Pauli algebra,
equations~(\ref{SU2bivectorspauli}),
are the 3 generators of $\SU(2)_\Lchiral$.
The weak Pauli matrix $\tau_3$ acting on a spinor
has eigenvalue equal to twice the weak isospin $2 I_\Lchiral = z - y$,
equation~(\ref{YICI}).
The electromagnetic charge generator $\im Q$, equation~(\ref{Qbivectora}),
is related to the hypercharge and weak generators $\im Y$ and $\im \tau_3$ by,
equation~(\ref{QIc}),
\begin{equation}
  Q
  =
  \tfrac{1}{2} ( Y + \tau_3 )
  \ .
\end{equation}
The sum $W_m^k \tau_k$ in the gauge field $\bW_m$, equation~(\ref{BYWtau}),
can be expressed with respect to either an orthonormal or a chiral basis,
\begin{equation}
\label{Wtauweak}
  W_m^k \tau_k
  =
  W_m^1 \tau_1 + W_m^2 \tau_2 + W_m^3 \tau_3
  =
  W_m^+ \tau_+ + W_m^- \tau_- + W_m^3 \tau_3
  \ ,
\end{equation}
where
$W_m^\pm \equiv ( W_m^1 \mp \im W_m^2 ) / \sqrt{2}$,
and the chiral Pauli operators $\tau_\pm$ are
\begin{equation}
\label{taupmweak}
  \tau_+
  \equiv
  {\tau_1 + \im \tau_2 \over \sqrt{2}}
  =
  {\bgamma_z \wedgie \bgamma_{\bar y} \over \sqrt{2}}
  \ , \quad
  \tau_-
  \equiv
  {\tau_1 - \im \tau_2 \over \sqrt{2}}
  =
  {\bgamma_y \wedgie \bgamma_{\bar z} \over \sqrt{2}}
  \ .
\end{equation}
The operator $\tau_+$
increases $z$-charge by 1
and decreases $y$-charge by 1,
and therefore carries $+1$ unit
of each of electric charge $Q$ and weak isospin $I_\Lchiral$.
Conversely,
$\tau_-$
decreases $z$-charge by 1
and increases $y$-charge by 1,
and therefore carries $-1$ unit
of each of electric charge $Q$ and weak isospin $I_\Lchiral$.
The operators $Y$ and $\tau_3$ leave $y$- and $z$-charge unchanged,
so carry zero electric charge $Q$ and weak isospin $I_\Lchiral$.

Introduce the weak mixing, or Weinberg, angle $\theta_w$ defined by
\begin{equation}
\label{weakmixingangle}
  \sin\theta_w
  \equiv
  {g_Y \over g}
  \ , \quad
  \cos\theta_w
  \equiv
  {g_w \over g}
  \ , \quad
  g
  \equiv
  \sqrt{g_Y^2 + g_w^2}
  \ .
\end{equation}
Define the electromagnetic and weak fields $A_m$ and $Z_m$
to be the orthogonal linear combinations of
$B_m$ and $W_m^3$,
\begin{equation}
\label{AZBW}
  \left(
  \begin{array}{c}
  A_m \\ Z_m
  \end{array}
  \right)
  \equiv
  \left(
  \begin{array}{cc}
  \cos\theta_w & \sin\theta_w \\
  -\sin\theta_w & \cos\theta_w
  \end{array}
  \right)
  \left(
  \begin{array}{c}
  B_m \\ W_m^3
  \end{array}
  \right)
  \ .
\end{equation}
In terms of the rotated electromagnetic and weak fields $A_m$ and $Z_m$,
the electroweak gauge connection is
\begin{equation}
  g_Y \bB_m
  +
  g_w \bW_m
  =
  \im \Bigl(
  2 e A_m Q
  +
  2 g Z_m ( I_\Lchiral - \sin^2\!\theta_w Q )
  +
  g_w ( W_m^+ \tau_+ + W_m^- \tau_- )
  \Bigr)
  \ ,
\end{equation}
where the electromagnetic coupling $e$ is
\begin{equation}
  e
  =
  {g_Y g_w \over g}
  =
  g_Y \cos\theta_w
  =
  g_w \sin\theta_w
  =
  g \cos\theta_w \sin\theta_w
  \ .
\end{equation}
The electromagnetic coupling $e$ is related to the fine-structure constant
$\alpha$ by $e^2 = 4\pi\alpha$.

The gauge-covariant derivative of the 4-component Higgs field $\bH$
with respect to $\U(1)_Y \times \SU(2)_\Lchiral$ transformations is
\begin{equation}
\label{DBWHiggs}
  \DD_m \bH
  =
  \partial_m \bH
  +
  g_Y [ \bB_m , \bH ]
  +
  g_w [ \bW_m , \bH ]
  \ .
\end{equation}
Whereas in the covariant derivative~(\ref{DBWpsi}) of a spinor $\psi$,
the fields $\bB_m$ and $\bW_m$ act directly on the spinor,
in the covariant derivative~(\ref{DBWHiggs}) of the Higgs field $\bH$,
the fields act as a commutator,
because whereas a spinor $\psi$ transforms as
$\psi \rightarrow R \psi$
under a rotor $R$ (an element of the group $\U(1)_Y \times \SU(2)_\Lchiral$),
a multivector such as the Higgs field $\bH$ transforms as
$\bH \rightarrow R \bH \reverse{R}$,
with $\reverse{R} = R^{-1}$ the reverse of $R$.
The covariant derivative of the expectation value~(\ref{Higgsvac})
of the Higgs field is
\begin{equation}
\label{DHiggs}
  \DD_m \langle \bH \rangle
  =
  \langle H \rangle
  \left(
  g_Y B_m [ \im Y , \bgamma^+_t \bgamma^-_y ]
  +
  g_w W_m^k [ \im \tau_k , \bgamma^+_t \bgamma^-_y ]
  \right)
  \ .
\end{equation}
The relevant commutators
of the generators $\im Y$ of $\U(1)_Y$,
equation~(\ref{uY1bivectora}),
and $\im \tau_k$ of $\SU(2)_\Lchiral$,
equations~(\ref{SU2bivectorspauli}),
with the electroweak Higgs field $\bgamma^+_t \bgamma^-_y$ are
\begin{subequations}
\label{YtauHiggscommutators}
\begin{alignat}{3}
\label{YHiggscommutator}
  [ \im Y , \bgamma^+_t \bgamma^-_y ]
  &=
  \phantom{-}
  \bgamma^+_y \bgamma^-_y
  \bgamma^+_t \bgamma^-_y
  &&=
  \phantom{-}
  \bgamma^+_t \bgamma^+_y
  \ ,
\\
  [ \im \tau_1 , \bgamma^+_t \bgamma^-_y ]
  &=
  -
  \bgamma^-_y \bgamma^+_z
  \bgamma^+_t \bgamma^-_y
  &&=
  \phantom{-}
  \bgamma^+_t \bgamma^+_z
  \ ,
\\
  [ \im \tau_2 , \bgamma^+_t \bgamma^-_y ]
  &=
  -
  \bgamma^-_y \bgamma^-_z
  \bgamma^+_t \bgamma^-_y
  &&=
  \phantom{-}
  \bgamma^+_t \bgamma^-_z
  \ ,
\\
\label{tau3Higgscommutator}
  [ \im \tau_3 , \bgamma^+_t \bgamma^-_y ]
  &=
  -
  \bgamma^+_y \bgamma^-_y
  \bgamma^+_t \bgamma^-_y
  &&=
  -
  \bgamma^+_t \bgamma^+_y
  \ .
\end{alignat}
\end{subequations}
With the commutators~(\ref{YtauHiggscommutators}),
the covariant derivative~(\ref{DHiggs})
of the expectation value of the Higgs field becomes
\begin{align}
\label{DHiggsu}
  \DD_m \langle \bH \rangle
  &=
  \langle H \rangle
  \Bigl(
  ( g_Y B_m - g_w W_m^3 )
  \bgamma^+_t \bgamma^+_y
  +
  g_w (
  W_m^1 \bgamma^+_t \bgamma^+_z
  +
  W_m^2 \bgamma^+_t \bgamma^-_z
  )
  \Bigr)
\nonumber
\\
  &=
  \langle H \rangle
  \Bigl(
  - \,
  g
  Z_m
  \bgamma^+_t \bgamma^+_y
  +
  g_w (
  W_m^+ \bgamma^+_t \bgamma_z + W_m^- \bgamma^+_t \bgamma_{\bar z}
  )
  \Bigr)
  \ .
\end{align}
Notice that the expression~(\ref{DHiggsu})
for the covariant derivative $\DD_m \langle \bH \rangle$
does not depend on the electromagnetic field $A_m$,
which stems from the fact that $\langle \bH \rangle$
commutes with the electric charge operator $Q$.
The kinetic term in the scalar Higgs Lagrangian is
$-\tfrac{1}{2} ( \DD^m \reverse{\bH} ) \cdot ( \DD_m \bH )$,
where $\reverse{\bH}$ is the reverse of $\bH$.
When the Higgs field acquires a nonzero expectation value
$\langle \bH \rangle$,
it contributes to the Lagrangian a kinetic term proportional to
(abbreviating $Z^m Z_m = ( Z_m )^2$ and so forth)
\begin{equation}
\label{DHiggsvac}
  \left( \DD^m \langle \reverse{\bH} \rangle \right)
  \cdot
  \left( \DD_m \langle \bH \rangle \right)
  =
  \langle H \rangle^2
  \Bigl(
  g^2 ( Z_m )^2
  +
  g_w^2 \left( ( W_m^+ )^2 + ( W_m^- )^2 \right)
  \Bigr)
  \ .
\end{equation}
The contribution~(\ref{DHiggsvac}) to the Lagrangian has the form
of mass-squared terms for the $Z_m$ and $W^\pm_m$ electroweak fields.
The Higgs field thus generates masses
$m_Z$ and $m_W$
for the $Z_m$ and $W^\pm_m$ fields,
\begin{equation}
\label{mZW}
  m_Z
  \equiv
  \langle H \rangle
  g
  \ , \quad
  m_W
  \equiv
  \langle H \rangle
  g_w
  \ .
\end{equation}
The electromagnetic field $A_m$ remains massless.
In accordance with the remarks after equations~(\ref{taupmweak}),
the electromagnetic field $A_m$ and weak field $Z_m$
both carry zero electric charge and weak isospin,
while the weak fields $W_m^\pm$ carry respectively $\pm 1$ unit of each of
electric charge $Q$ and weak isospin $I_\Lchiral$.

\section{Grand symmetry breaking}
\label{grandsym-sec}

\subsection{Grand Higgs field}
\label{grandHiggs-sec}

As set forth in \S\ref{spin5spin6-sec},
the largest subgroup of $\Spin(11,1)$ that commutes with all the
spacetime vectors~(\ref{vectorsspin10ew})
and can therefore be on the path to grand unification
while spacetime is still 4-dimensional is
$\Spin(5)_w \times \Spin(6)_c$.
The transition to this group,
or possibly to some maximal subgroup of it
(which proves to be the Pati-Salam group
$\Spin(4)_w \times \Spin(6)_c$),
marks grand symmetry breaking.
In the $\Spin(11,1)$ model,
grand unification involves a reduction of the number of spacetime dimensions
to the 3+1 observed today.

According to the general rules governing symmetry breaking by a Higgs field,
the grand Higgs field
must acquire a nonzero expectation value that
(1) is a Lorentz scalar,
commuting with all Lorentz bivectors~(\ref{lorentzspin10ew}),
(2) commutes with all generators of the unbroken group,
and (3) fails to commute with other generators.
There is just one multivector of the $\Spin(11,1)$ algebra
that fits the aspiration to be the grand Higgs field,
namely the time bivector
$\bgamma^+_t \bgamma^-_t$
multiplied by the colour chiral operator $\varkappa_{rgb}$,
\begin{equation}
\label{Higgsnuvac}
  \langle \bT \rangle
  =
  - \im
  \langle T \rangle
  \bgamma^+_t \bgamma^-_t
  \varkappa_{rgb}
  \ ,
\end{equation}
the imaginary $\im$ coming from the time vector being timelike,
$\bgamma_0 = \im \bgamma^-_t$.
The grand Higgs field~(\ref{Higgsnuvac})
(1) commutes with all Lorentz bivectors~(\ref{lorentzspin10ew})
and is therefore a Lorentz scalar,
(2) commutes with the 21 Pati-Salam $\Spin(4)_w \times \Spin(6)_c$ bivectors
modified per~(\ref{adjustsmbivectorsI6}),
and (3) fails to commute with the 24 modifed bivectors of $\Spin(10)$
that are not Pati-Salam bivectors.
Although $\langle \bT \rangle$ is a Lorentz scalar,
commuting with all Lorentz bivectors~(\ref{lorentzspin10ew}),
it anticommutes with all Dirac vectors~(\ref{vectorsspin10ew}),
but that is fine because, unlike a gauge field,
a Higgs field need only be a Lorentz scalar, not a Dirac scalar.
In particular, $\langle \bT \rangle$ fails to commute with the time axis
$\bgamma_0 = \im \bgamma^-_t$,
and therefore breaks $t$-symmetry,
which has the happy side effect of generating a Majorana mass
for the right-handed neutrino, \S\ref{grandMajorana-sec}.

Besides the modified bivectors of $\Spin(10)$,
the grand Higgs field~(\ref{Higgsnuvac}) commutes with
the 12 modified bivectors $\bgamma^\pm_t \bgamma^\pm_k$ with $k = r,g,b$,
and it fails to commute with
the 8 modified bivectors $\bgamma^\pm_t \bgamma^\pm_k$ with $k = y,z$.

The grand Higgs field $\langle \bT \rangle$,
equation~(\ref{Higgsnuvac}),
is a bivector modified per~(\ref{adjustsmbivectorsI6}),
just like the electroweak Higgs field~(\ref{Higgsvac}),
and just like the gauge bivectors of the standard model
discussed in \S\ref{spin111chart-sec}.
According to the general rules,
the grand Higgs field $\langle \bT \rangle$
must be a component of a multiplet of the unbroken $\Spin(11,1)$ symmetry,
and the gauge (vector) and Higgs (scalar) fields
corresponding to the symmetries that are broken by the grand Higgs field
must transform consistently,
so that the scalars and vectors can combine into massive gauge bosons.
These requirements point to the conclusion
that the Higgs field is a 66-component $\Spin(11,1)$ field $\bE$
of bivectors modified per~(\ref{adjustsmbivectorsI6}),
transforming in the adjoint representation,
as anticipated in equation~(\ref{spin111Emultiplet}).

Here is a synopsis of what happens to the various fields as a result
of grand symmetry breaking
(all bivectors should be understood as modified per~(\ref{adjustsmbivectorsI6})).
\begin{itemize}
\item
The $6+15 = 21$ bivectors
of $\Spin(4)_w \times \Spin(6)_c$,
namely $\bgamma^\pm_k \bgamma^\pm_l$ with $k$ and $l$ both in $y,z$,
or $k$ and $l$ both in $r,g,b$,
commute with the grand Higgs field~(\ref{Higgsnuvac})
and with the Dirac vectors~(\ref{vectorsspin10ew}),
and therefore yield unbroken scalar and gauge fields.
$\Spin(4)_w \times \Spin(6)_c$
preserves time, weak, and colour chiralities
$\varkappa_t$, $\varkappa_{yz}$, and $\varkappa_{rgb}$,
equations~(\ref{Ityzs}),
so $\Spin(4)_w$ and $\Spin(6)_c$ gauge bosons
allow transformations between fermions
of the same boost ($\varkappa_{tyz}$) and spin ($\varkappa_{rgb}$),
\begin{subequations}
\label{nueudUpup}
\begin{alignat}{5}
\label{nueudUpupw}
  \SU(2)_\Rchiral
  &:
  \quad
  &
  \nu_{\Upup}
  &
  \,\leftrightarrow\,
  e_{\Upup}
  \ , \quad
  &
  u^c_{\Upup}
  &
  \,\leftrightarrow\,
  d^c_{\Upup}
  \ ,
\\
\label{nueudUpupc}
  \Spin(6)_c / \SU(3)_c
  &:
  \quad
  &
  \nu_{\Upup}
  &
  \,\leftrightarrow\,
  u^c_{\Upup}
  \ , \quad
  &
  e_{\Upup}
  &
  \,\leftrightarrow\,
  d^c_{\Upup}
  \ .
\end{alignat}
\end{subequations}
The $\SU(2)_\Rchiral$ transitions~(\ref{nueudUpupw})
are right-handed versions of analogous $\SU(2)_\Lchiral$ left-handed transitions
in the standard model.
The $\Spin(6)_c / \SU(3)_c$ transitions~(\ref{nueudUpupc})
and their left-handed equivalents
transform between leptons and quarks,
for which reason the associated gauge bosons are called leptoquarks.
\item
The $4 \times 6 = 24$ bivectors
of $\Spin(10) / ( \Spin(4)_w \times \Spin(6)_c )$,
namely $\bgamma^\pm_k \bgamma^\pm_l$ with $k$ in $y,z$, and $l$ in $r,g,b$,
fail to commute with the grand Higgs field~(\ref{Higgsnuvac}),
and they also fail to commute
with some Lorentz bivectors~(\ref{lorentzspin10ew}).
The transitions
preserve time and $\Spin(10)$ chiralities
$\varkappa_t$ and $\varkappa_{yzrgb}$,
but break the individual weak and colour chiralities:
\begin{equation}
  \begin{matrix}
  \bar{\nu}_{\Downdown}
  \\
  \bar{e}_{\Downdown}
  \end{matrix}
  \,\leftrightarrow\,
  \begin{matrix}
  u^c_{\Upup}
  \\
  d^c_{\Upup}
  \end{matrix}
  \,\leftrightarrow\,
  \begin{matrix}
  \bar{u}^{\bar{c}}_{\Downdown}
  \\
  \bar{d}^{\bar{c}}_{\Downdown}
  \end{matrix}
  \,\leftrightarrow\,
  \begin{matrix}
  \nu_{\Upup}
  \\
  e_{\Upup}
  \end{matrix}
  \ .
\end{equation}
The 24 scalar bivectors combine with the 24 vector bivectors
by the Goldstone mechanism \cite{Goldstone:1962}
to yield 24 massive charged gauge bosons,
which include 12 bivectors in $\SU(5)$,
with charges $k{\bar l}$ or ${\bar k}l$,
and 12 bivectors not in $\SU(5)$,
with charges $k l$ or ${\bar k}{\bar l}$.
Observations and experiment show no sign of Lorentz violation
\cite{Mattingly:2005}.
To avoid a violation of Lorentz invariance,
the gauge bosons must presumably acquire a mass
of the order of the grand unified scale.
\item
The 4 bivectors
$\bgamma^+_t \bgamma^\pm_k$, $k = y,z$,
fail to commute with the grand Higgs field~(\ref{Higgsnuvac}),
but commute with the Dirac vectors~(\ref{vectorsspin10ew}).
If the grand gauge group were $\Spin(11,1)$,
then the 4 scalar bivectors would combine with the 4 vector bivectors
by the Goldstone mechanism
to yield 4 massive gauge bosons at grand unification.
But in practice the 4 bivectors
$\bgamma^+_t \bgamma^\pm_k$, $k = y,z$
apparently persist to produce the electroweak Higgs
multiplet~(\ref{Higgsmultiplet}) that mediates electroweak symmetry breaking.
The only way we have been able to make sense of this complication
is to postulate that the dimension $\bgamma^+_t$ is a scalar dimension
that does not generate any symmetry,
a possibility discussed in \S4.4 of \cite{Hamilton:2023a}.
If there is no gauge symmetry,
then the Goldstone mechanism \cite{Goldstone:1962} does not operate,
and the scalar field can persist.
\item
The 6 bivectors
$\bgamma^+_t \bgamma^\pm_k$, $k = r,g,b$,
commute with the grand Higgs field~(\ref{Higgsnuvac}),
but fail to commute with several Lorentz bivectors~(\ref{lorentzspin10ew}).
If $\bgamma^+_t$ is a scalar dimension,
then the bivectors do not generate a gauge symmetry,
but there is still a 6-component ``scalar'' multiplet.
To avoid a violation of Lorentz invariance,
the multiplet must presumably become massive,
of the order of the grand unified scale,
as the dimensionality of spacetime drops from 10+1 to 3+1.
\item
The 4 bivectors
$\bgamma^-_t \bgamma^\pm_k$, $k = y,z$,
fail to commute with the grand Higgs field~(\ref{Higgsnuvac}),
but commute with the Lorentz bivectors~(\ref{lorentzspin10ew})
(though not with the Dirac vectors~(\ref{vectorsspin10ew})).
The 4 scalar bivectors combine with the 4 vector bivectors
to yield 4 massive gauge bosons at grand symmetry breaking.
\item
The 6 bivectors
$\bgamma^-_t \bgamma^\pm_k$, $k = r,g,b$,
commute with the grand Higgs field~(\ref{Higgsnuvac}),
but fail to commute with several 
Lorentz bivectors~(\ref{lorentzspin10ew}).
To avoid a violation of Lorentz invariance,
the 6 scalar bivectors must combine with the 6 vector bivectors
to yield 6 massive gauge bosons at grand symmetry breaking.
\end{itemize}

\subsection{Cosmological inflation}

A leading idea of the standard model of cosmology is that inflation
in the early universe was driven by the energy associated with
grand unification.
There is a vast literature on the subject,
e.g.~\cite{Martin:2013,Baumann:2014,Kumar:2018,Horn:2020}.

The grand Higgs field $\langle \bT \rangle$, equation~(\ref{Higgsnuvac}),
is available to drive inflation.
The mechanism of grand symmetry breaking by a single Higgs field
is consistent with the simplest conventional models of inflation.
It is perhaps surprising that the simplest model works,
since the restructuring of spacetime
could potentially be a complicated process
\cite{Baumann:2014}.

\subsection{Gauge boson masses at grand symmetry breaking}

Subsection~\ref{grandHiggs-sec} argued that
grand symmetry breaking breaks 34 of the 55 bivectors of $\Spin(10,1)$,
generating 34 massive gauge bosons.
Of the 34 broken bivectors,
28 are broken by virtue of failing to commute with
the grand Higgs field $\langle \bT \rangle$, equation~(\ref{Higgsnuvac}),
while the other 6 are broken by virtue of failing to commute with
Lorentz bivectors~(\ref{lorentzspin10ew}).
Exactly how the 6 are broken as the dimension of spacetime drops from 10+1
to 3+1 lies outside the realm of standard Higgs theory,
and beyond the scope of the present paper to model.

The $\Spin(10,1)$ covariant derivative of the expectation value
$\langle \bT \rangle$, equation~(\ref{Higgsnuvac}),
of the grand Higgs field is
\begin{equation}
\label{DN}
  D_m \langle \bT \rangle
  =
  g
  \sum_A
  [ \bC_m , \langle \bT \rangle ]
  =
  -\im
  \langle T \rangle g
  \sum_A
  C_m^A [ \bgamma_A , \bgamma^+_t \bgamma^-_t \varkappa_{rgb} ]
  \ ,
\end{equation}
where $g$ is a grand coupling parameter,
and
$\bC_m = C^A_m \bgamma_A$
are the gauge fields associated with the 55 $\Spin(10,1)$ bivectors $\bgamma_A$
modified per~(\ref{adjustsmbivectorsI6}).
The nonvanishing commutators of the $\Spin(10,1)$ bivectors with
the grand Higgs generator
$\bgamma^+_t \bgamma^-_t \varkappa_{rgb}$
are the $24 + 4 = 28$ commutators
\begin{subequations}
\label{commT}
\begin{alignat}{3}
  [ \bgamma^\pm_k \bgamma^\pm_l , \bgamma^+_t \bgamma^-_t \varkappa_{rgb} ]
  &=
  2 \bgamma^\pm_k \bgamma^\pm_l \bgamma^+_t \bgamma^-_t \varkappa_{rgb}
  &&
  \ \  ( k \mbox{~in~} y,z , \ l \mbox{~in~} r,g,b )
  \ ,
\\
  [ \bgamma^-_t \bgamma^\pm_l , \bgamma^+_t \bgamma^-_t \varkappa_{rgb} ]
  &=
  2 \bgamma^-_t \bgamma^\pm_l \bgamma^+_t \bgamma^-_t \varkappa_{rgb}
  &&
  \ \  ( l \mbox{~in~} y,z )
  \ .
\end{alignat}
\end{subequations}
The bivector factors $\bgamma^\pm_k \bgamma^\pm_l$
should strictly be modified per~(\ref{adjustsmbivectorsI6}),
but this modification can be omitted here because $\bgamma^+_t \bgamma^-_t$
commutes with $\varkappa_{rgb}$, so the commutation rules~(\ref{commT})
are unaffected by the possible extra factor of $\varkappa_{rgb}$
on $\bgamma^\pm_k \bgamma^\pm_l$.
When the grand Higgs field acquires a nonzero expectation value
$\langle \bT \rangle$,
it contributes to the Lagrangian a kinetic term proportional to
\begin{equation}
\label{DHiggsnuvac}
  \bigl( \DD^m \langle \reverse{\bT} \rangle \bigr)
  \cdot
  \bigl( \DD_m \langle \bT \rangle \bigr)
  =
  4
  \langle T \rangle^2
  g^2
  \sum_A
  ( C_m^A )^2
  \ ,
\end{equation}
in which the sum over $A$ is over noncommuting gauge fields $\bC_m$.
The contribution~(\ref{DHiggsnuvac})
takes the form of a mass-squared term for each of the noncommuting fields,
so the grand Higgs field $\langle \bT \rangle$ generates a mass
\begin{equation}
\label{mT}
  m_T
  =
  2 \langle T \rangle g
\end{equation}
for each of the 28 noncommuting gauge fields.

It was noted in \S\ref{grandHiggs-sec}
that the 24 bivectors
$\bgamma^\pm_k \bgamma^\pm_l$ with $k$ in $y,z$, and $l$ in $r,g,b$,
fail to commute not only with the grand Higgs field $\langle \bT \rangle$,
but also with the Lorentz bivectors~(\ref{lorentzspin10ew}),
so the 24 bivectors do not form Lorentz-covariant multiplets,
and the covariant derivative $\DD_m$ of these bivectors in equation~(\ref{DN})
cannot simply be a standard 3+1 dimensional Lorentz-covariant derivative.
However,
the dynamics of the various fields must presumably emerge
from a suitable scalar Lagrangian,
so the expression~(\ref{DHiggsnuvac})
must be a valid spacetime-scalar equation,
and the resulting gauge boson mass $m_T$ given by equation~(\ref{mT})
must be a scalar.
Again, it is beyond the scope of this paper to dig more deeply into
just what happens as the dimensionality of spacetime drops from 10+1 to 3+1.

\subsection{The 11th dimension as a scalar}
\label{Spin101-sec}

It was argued in \S\ref{grandHiggs-sec} that the 4 bivectors
$\bgamma^+_t \bgamma^\pm_k$, $k = y,z$
cannot generate a gauge symmetry,
because if they did,
then that gauge symmetry would be broken
by the grand Higgs field $\langle \bT \rangle$ at grand symmetry breaking,
and the associated scalar multiplet would be absorbed by the Goldstone mechanism
into 4 massive gauge bosons,
whereas the scalar multiplet
apparently persists to produce the electroweak Higgs
multiplet~(\ref{Higgsmultiplet}) that mediates electroweak symmetry breaking.

The only solution that we can see to this problem
is to postulate that the dimension $\bgamma^+_t$ is a scalar dimension
that does not generate any symmetry,
a possibility discussed in \S4.4 of \cite{Hamilton:2023a}.
The dimension $\bgamma^+_t$ stands out
as the only spacelike vector of $\Spin(11,1)$
that is not a factor in any unbroken gauge symmetry of the standard model.
As first shown by \cite{Brauer:1935},
and expounded by \cite{Hamilton:2023a},
the algebra of outer products of spinors is isomorphic
to the geometric (Clifford) algebra,
in any number of spacetime dimensions.
The natural complex structure of spinors means that
geometric algebras live naturally in even dimensions.
In odd dimensions,
there are two ways to realize a geometric algebra as an outer product of
spinors.
One is to project the odd-dimensional algebra into one dimension lower;
the second is to embed the odd-dimensional algebra into one dimension higher,
and to treat the extra dimension as a scalar.
The first option, projecting the odd algebra into one lower dimension,
is the usual one.
For example, the pseudoscalar of the geometric algebra in 3 dimensions,
the Pauli algebra,
is identified with $\im$ times the unit matrix,
$\sigma_1 \sigma_2 \sigma_3 = \im 1$.
The disadvantage of the first option is that the geometric algebra does not
contain within itself a time-reversal or parity-reversal operator.
By contrast, in the second option,
the extra axis, here $\bgamma^+_t$,
which as usual anticommutes with all other axes $\bgamma^\pm_k$,
serves the role of a time-reversal operator,
because it flips the 10 other spatial axes (leaving spatial parity unchanged),
and also flips the 1 time axis, reversing time.
A time-reversal operator is important for a consistent quantum field theory.

\subsection{The grand Higgs field generates a Majorana mass for the right-handed neutrino}
\label{grandMajorana-sec}

The grand Higgs field $\langle \bT \rangle$, equation~(\ref{Higgsnuvac}),
has the felicitous property that it is able to generate a Majorana mass
for the right-handed neutrino.
The Majorana mass term is
\begin{equation}
\label{mnu}
  \conj{\nu} \cdot \langle \bT \rangle \nu
  =
  - \im \nu^\dagger \bgamma_0 \langle \bT \rangle \nu
  =
  \nu^\dagger \bgamma^-_t \langle \bT \rangle \nu
  \ ,
\end{equation}
which flips the right-handed neutrino's $t$-bit.
The right-handed neutrino,
the all-bit-up entry in the $\Spin(11,1)$ chart~(\ref{tyzrgbtab}),
has two components,
a right-handed neutrino $\nu_{\Upup}$ with $t$-bit up,
and a left-handed antineutrino partner $\bar{\nu}_{\Downup}$ with $t$-bit down.
The two components have opposite boost, but the same spin,
so flipping the $t$-bit generates a mass for the right-handed neutrino.
Only the right-handed neutrino can acquire mass this way,
because only the right-handed neutrino has zero couplings to the gauge groups
$\U(1)_Y \times \SU(2)_\Lchiral \times \SU(3)_c$
of the standard model.
For other fermions,
conservation of standard-model charges prohibits flipping the $t$-bit.

Although the grand Higgs field $\langle \bT \rangle$ aquires its nonzero
expectation value at grand symmetry breaking,
it is only later, at Pati-Salam symmetry breaking, \S\ref{patisalam-sec},
that the grand Higgs field
can generate a Majorana mass for the right-handed neutrino.
Before Pati-Salam symmetry breaking,
$\Spin(4)_w \times \Spin(6)_c$ transitions~(\ref{nueudUpup})
unite right-handed neutrinos with other right-handed leptons and quarks.
$\Spin(4)_w \times \Spin(6)_c$ transitions preserve chirality,
and likewise Lorentz transformations preserve chirality,
so before Pati-Salam symmetry breaking
all fermions are massless chiral fermions,
including the right-handed neutrino,
which is part of the 16-component right-handed lepton-quark multiplet
$\{ \nu_\Rchiral , e_\Rchiral , u^c_\Rchiral , d^c_\Rchiral \}$.

After Pati-Salam symmetry breaking into the standard model,
the right-handed $\Spin(4)_w \times \Spin(6)_c$ transitions~(\ref{nueudUpup})
no longer occur.
The right-handed neutrino becomes isolated into a singlet of the standard model,
and standard-model symmetries do not prohibit
the Majorana mass term~(\ref{mnu}) from flipping the right-handed neutrino's
$t$-bit.

The nonvanishing charges of the right-handed neutrino $\nu_\Upup$
and of its left-handed antineutrino partner $\bar{\nu}_\Downup$
are $\rho_3 = 1$ and $-1$ respectively,
and $B{-}L = -1$ and $+1$ respectively.
Thus the Majorana mass term~(\ref{mnu}) violates
baryon-minus-lepton $B{-}L$ symmetry.
According to the standard picture of leptogenesis
\cite{Buchmuller:2004},
out-of-equilibrium $CP$-violating decay of an initially thermal distribution of
massive right-handed Majorana neutrinos can generate a nonzero $B{-}L$ number,
which sphaleron processes then transform into a nonzero baryon number.

The argument has led to a conundrum.
The grand Higgs field $\langle \bT \rangle$,
equation~(\ref{Higgsnuvac}),
commutes with $\rho_3$ and $B{-}L$,
yet the mass term
$\conj{\nu} \cdot \langle \bT \rangle \nu$,
equation~(\ref{mnu}),
violates conservation of $\rho_3$ and $B{-}L$.
This would seem to violate the Noetherian \cite{Noether:1918}
principle relating symmetries to conservation laws.
The solution to the conundrum seems to be that
$\langle \bT \rangle$
breaks the $t$-symmetry that links fermions and antifermions,
by virtue of $\langle \bT \rangle$ not commuting with the time axis
$\bgamma_0 \equiv -\im \bgamma^-_t$
($\langle \bT \rangle$ is nevertheless a Lorentz scalar,
because it commutes with the Lorentz bivectors~(\ref{lorentzspin10ew})).
Breaking $t$-symmetry apparently allows
$\langle \bT \rangle$
to break $\rho_3$ and $B{-}L$ symmetries in spite of commuting with them.
This issue is revisited in \S\ref{scalarps-sec},
and called out as a potential flaw in \S\ref{flaws-sec}.

In the usual $\Spin(10)$ model,
the standard Higgs used to break $B{-}L$
so as to generate a Majorana mass is
\cite{Harvey:1980,delAguila:1981,King:2021b}
a pentavector, belonging to the dimension
$\binomial{10}{5} = 126 + \overline{126}$
representation of $\Spin(10)$.
Specifically, the pentavector
\begin{equation}
\label{BLpentavector}
  \bgamma_y \bgamma_z \bgamma_r \bgamma_g \bgamma_b
  +
  \bgamma_{\bar y} \bgamma_{\bar z} \bgamma_{\bar r} \bgamma_{\bar g} \bgamma_{\bar b}
\end{equation}
has the property that it flips between the all-bit-up right-handed neutrino
and its all-bit-down left-handed antineutrino partner
in the $\Spin(10)$ chart~(\ref{yzrgbtab}),
$\uparrow\uparrow\uparrow\uparrow\uparrow
\,\leftrightarrow\,
\downarrow\downarrow\downarrow\downarrow\downarrow$,
while yielding zero acting on any other spinor.
The pentavector~(\ref{BLpentavector})
commutes with $\SU(3)_c$ but not with $B{-}L$,
so explicitly breaks $B{-}L$ symmetry.

The Higgs pentavector~(\ref{BLpentavector}) that generates a Majorana
mass in $\Spin(10)$
does not work for the $\Spin(11,1)$ model
because the modification~(\ref{adjustsmbivectorsI6})
that allows $\Spin(10)$ and $\Spin(3,1)$ to unify into $\Spin(11,1)$
puts the left-handed antineutrino partner $\bar{\nu}_\Downup$
of the right-handed neutrino $\nu_\Upup$
into the same $\SU(5)$ column of
the $\Spin(11,1)$ chart~(\ref{tyzrgbtab}),
instead of into the opposite $\SU(5)$ columns of
the $\Spin(10)$ chart~(\ref{yzrgbtab}).
In $\Spin(11,1)$,
the multivector $\bgamma^-_t \langle \bT \rangle$ that transforms between
$\nu_\Upup$ and $\bar{\nu}_\Downup$
has $\Spin(10)$ grade 0, not~5.

\section{Pati-Salam symmetry breaking}
\label{patisalam-sec}

\subsection{The Higgs field that breaks Pati-Salam symmetry}
\label{higgsspin46-sec}

According to the general rules,
the Higgs field that breaks the Pati-Salam $\Spin(4)_w \times \Spin(6)_c$
symmetry must be part of a multiplet of Lorentz-scalar fields
that rotate into each other under $\Spin(4)_w \times \Spin(6)_c$.
Now the grand Higgs field $\langle \bT \rangle$, equation~(\ref{Higgsnuvac}),
that mediates grand symmetry breaking is part of a 66-component scalar
multiplet $\bE$, equation~(\ref{spin111Emultiplet}),
of $\Spin(11,1)$ bivectors modified per~(\ref{adjustsmbivectorsI6}).
After grand symmetry breaking to the Pati-Salam group,
the scalar multiplet $\bE$ contains an unbroken 21-component Pati-Salam
multiplet generated by the $6 + 15 = 21$ modified bivectors
of $\Spin(4)_w \times \Spin(6)_c$.
This scalar Pati-Salam multiplet could potentially contain a component $\bP$
that, when it acquires an expectation value $\langle \bP \rangle$,
breaks the Pati-Salam symmetry.
Indeed it does.

The Higgs field $\bP$ that breaks the Pati-Salam
symmetry must leave the standard-model group unbroken,
so must commute with standard-model bivectors
modified per~(\ref{adjustsmbivectorsI6}).
The weak subgroup $\SU(2)_\Lchiral$
permutes $yz$ charges while preserving the total charge $y{+}z$,
while the colour subgroup $\SU(3)_c$
permutes $rgb$ charges while preserving the total charge $r{+}g{+}b$.
Therefore the Higgs field $\bP$ must be a linear combination
of generators that are invariant under permutations of each of $yz$ and $rgb$.
There are two such basis bivectors in $\Spin(4)_w \times \Spin(6)_c$,
namely the third Pauli generator
$\im \rho_3$
that generates the $\U(1)_\Rchiral$ subgroup
of the right-handed weak group $\SU(2)_\Rchiral$,
equation~(\ref{SUR2bivectorspauli3});
and the baryon minus lepton operator $\im ( B{-}L )$
that generates the $\U(1)_c$ subgroup of the colour group $\Spin(6)_c$,
equation~(\ref{Sbivector}).
The corresponding conserved charges are
(the expression for $B{-}L$ repeats equation~(\ref{BLcharge}))
\begin{equation}
\label{RScharge}
  \rho_3
  =
  y + z
  \ , \quad
  B - L
  =
  - \tfrac{2}{3} ( r + g + b )
  \ .
\end{equation}
If the Higgs field $\bP$ lies inside the
$\Spin(4)_w \times \Spin(6)_c$ submultiplet of $\bE$,
then $\bP$ must be a linear combination of the bivectors
$\im \rho_3$ and $\im (B{-}L)$
modified per~(\ref{adjustsmbivectorsI6}),
\begin{equation}
\label{P}
  \bP
  =
  \im
  \left(
  P^3 \rho_3
  +
  P_m^{B{-}L} (B{-}L)
  \right)
  \kappa_{rgb}
  \ .
\end{equation}
The factor $\kappa_{rgb}$ is from
the modification~(\ref{adjustsmbivectorsI6}).

When the Higgs doublet $\bP$, equation~(\ref{P}),
acquires a vacuum expectation value $\langle \bP \rangle$,
it must leave unbroken the hypercharge group $\U(1)_Y$ of the standard model,
which is generated by the bivector $\im Y$,
equation~(\ref{uY1bivectora}).
The hypercharge $Y$
is a linear combination of the charges $\rho_3$ and $B{-}L$,
equation~(\ref{YICY}),
\begin{equation}
\label{YRS}
  Y
  =
  \rho_3 + B - L
  \ .
\end{equation}
When the Higgs doublet $\bP$ acquires its vacuum expectation value,
it condenses the two phases $R$ and $B{-}L$ into one,
\begin{equation}
\label{URBLY}
  \U(1)_R \times \U(1)_{B{-}L}
  \rightarrow
  \U(1)_Y
  \ .
\end{equation}
This is similar to the way that electroweak symmetry breaking involves
condensing the two phases of hypercharge and weak isospin into one,
$\U(1)_Y \times \U(1)_\Lchiral \rightarrow \U(1)_Q$.
The symmetry breaking~(\ref{URBLY}) breaks the Pati-Salam group
$\Spin(4)_w \times \Spin(6)_c$
to the standard-model group
$\U(1)_Y \times \SU(2)_\Lchiral \times \SU(3)_c$.

The linear combination of
$\im \rho_3$ and $\im B{-}L$
to which $\bP$ breaks can be deduced from the condition that the
unbroken symmetry is $\U(1)_Y$.
The
$\Spin(4)_w \times \Spin(6)_c$
gauge connection is
\begin{equation}
\label{connection46}
  g_r \bR_m
  +
  g_w \bW_m
  +
  g_c \bC_m
  \ ,
\end{equation}
where
$\bR_m$,
$\bW_m$,
and $\bC_m$ are respectively
gauge fields of
$\SU(2)_\Rchiral$,
$\SU(2)_\Lchiral$,
and $\Spin(6)_c$,
and
$g_r$,
$g_w$,
and $g_c$ are the respective
dimensionless coupling parameters.
The part of the $\Spin(4)_w \times \Spin(6)_c$
connection~(\ref{connection46})
associated with the $\U(1)_R \times \U(1)_{B{-}L}$ symmetry is
\begin{equation}
\label{connectionRS}
  \im
  \left(
  g_r R_m^3 \rho_3
  +
  g_c C_m^{B{-}L} (B{-}L)
  \right)
  \varkappa_{rgb}
  \ ,
\end{equation}
where $R_m^3$ and $C_m^{B{-}L}$ are the corresponding connection coefficients.
Analogously to the weak mixing angle $\theta_w$,
equations~(\ref{weakmixingangle}),
define the $\SU(2)_\Rchiral \times \Spin(6)_c$ mixing angle $\theta_r$ by
\begin{equation}
\label{spin46mixingangle}
  \sin\theta_r
  \equiv
  {g_r \over g}
  \ , \quad
  \cos\theta_r
  \equiv
  {g_c \over g}
  \ , \quad
  g \equiv \sqrt{g_r^2 + g_c^2}
  \ .
\end{equation}
And analogously to the rotated fields $A_m$ and $Z_m$ of electroweak theory,
equation~(\ref{AZBW}),
define gauge fields $B_m$ and $P_m$ to be orthogonal linear combinations
of $R_m^3$ and $C_m^{B{-}L}$,
\begin{equation}
\label{BFWC}
  \left(
  \begin{array}{c}
  B_m \\ P_m
  \end{array}
  \right)
  \equiv
  \left(
  \begin{array}{cc}
  \cos\theta_r & \sin\theta_r \\
  -\sin\theta_r & \cos\theta_r
  \end{array}
  \right)
  \left(
  \begin{array}{c}
  R_m^3 \\ C_m^{B{-}L}
  \end{array}
  \right)
  \ .
\end{equation}
In terms of the rotated fields $B_m$ and $P_m$,
the $\U(1)_R \times \U(1)_{B{-}L}$ connection~(\ref{connectionRS}) is
(a commuting factor of $\varkappa_{rgb}$
from the modification~(\ref{adjustsmbivectorsI6})
has been dropped for brevity
from both sides of equation~(\ref{connectionYE}))
\begin{equation}
\label{connectionYE}
  \im \left(
  g_r R_m^3 \rho_3
  +
  g_c C_m^{B{-}L} (B{-}L)
  \right)
  =
  \im \left(
  g_Y B_m Y
  -
  g
  P_m
  ( \sin^2\!\theta_r \rho_3 - \cos^2\!\theta_r (B{-}L) )
  \right)
  \ ,
\end{equation}
where the hypercharge coupling parameter $g_Y$ is
\begin{equation}
\label{gY}
  g_Y
  \equiv
  g_r \cos\theta_r
  =
  g_c \sin\theta_r
  =
  g
  \cos\theta_r \sin\theta_r
  \ .
\end{equation}
The term proportional to $B_m Y$ in the connection~(\ref{connectionYE})
has the correct form for the unbroken $\U(1)_Y$ hypercharge connection.
The remaining term on the right hand side of equation~(\ref{connectionYE})
represents the broken direction,
the direction along which the Higgs doublet $\bP$
acquires an expectation value $\langle \bP \rangle$,
\begin{equation}
\label{spin46Higgs}
  \langle \bP \rangle
  =
  \im \langle P \rangle
  ( \sin^2\!\theta_r \rho_3 - \cos^2\!\theta_r (B{-}L) ) \varkappa_{rgb}
  \ ,
\end{equation}
where the factor $\varkappa_{rgb}$
from the modification~(\ref{adjustsmbivectorsI6})
has been restored.

\subsection{Masses of Pati-Salam gauge bosons}
\label{PSmass-sec}

For simplicity, factors of $\varkappa_{rgb}$
from the modification~(\ref{adjustsmbivectorsI6})
are omitted throughout this subsection~(\ref{PSmass-sec}).
The factor $\varkappa_{rgb}$ commutes with all the generators,
so has no effect on the predictions~(\ref{mRC}) of masses
of Pati-Salam gauge bosons.

The $\Spin(4)_w \times \Spin(6)_c$ covariant derivative
of the expectation value $\langle \bP \rangle$
of the Higgs doublet $\bP$, equation~(\ref{spin46Higgs}), is
\begin{align}
\label{Dspin46Higgsvac}
  \DD_m \langle \bP \rangle
  &=
  \langle P \rangle
  \left(
  g_r \sin^2\!\theta_r [ \bR_m , \im \rho_3 ]
  -
  g_c \cos^2\!\theta_r [ \bC_m , \im (B{-}L) ]
  \right)
\nonumber
\\
  &=
  g \langle P \rangle
  \left(
  \sin^3\!\theta_r [ \bR_m , \im \rho_3 ]
  -
  \cos^3\!\theta_r [ \bC_m , \im (B{-}L) ]
  \right)
  \ ,
\end{align}
where
$\bR_m$ and $\bC_m$ are the gauge fields
of respectively
$\SU(2)_\Rchiral$ and $\Spin(6)_c$,
\begin{equation}
  \bR_m
  =
  \im R_m^k \rho_k
  \ , \quad
  \bC_m
  =
  C_m^{[k^\pm l^\pm]} \bgamma^\pm_k \bgamma^\pm_l
  \ \  ( k,l \mbox{~in~} r,g,b )
  \ .
\end{equation}
The nonvanishing commutators of the Pati-Salam
$\Spin(4)_w \times \Spin(6)_c$ gauge fields with the component generators
$\rho_3$ and $B{-}L$
of the expectation value $\langle \bP \rangle$ of the Higgs doublet are
\begin{subequations}
\label{spin46Xcommutators}
\begin{alignat}{3}
\label{weakRbivectorXppcommutators}
  [ \rho_2 , \rho_3 ]
  &=
  2 \im \rho_1
  \ ,
\\
\label{weakRbivectorXpmcommutators}
  [ \rho_1 , \rho_3 ]
  &=
  - 2 \im \rho_2
  \ ,
\\
\label{leptoquarkbivectorXppcommutators}
  [ \tfrac{1}{2} ( 1 + \varkappa_{kl} ) \bgamma^+_k \bgamma^+_l , B{-}L ]
  &=
  - \tfrac{2}{3}
  ( 1 + \varkappa_{kl} ) \bgamma^+_k \bgamma^-_l
  &&
  \ \ 
  ( k,l \mbox{~in~} r,g,b )
  \ ,
\\
\label{leptoquarkbivectorXpmcommutators}
  [ \tfrac{1}{2} ( 1 + \varkappa_{kl} ) \bgamma^+_k \bgamma^-_l , B{-}L ]
  &=
  \tfrac{2}{3}
  ( 1 + \varkappa_{kl} ) \bgamma^+_k \bgamma^+_l
  &&
  \ \ 
  ( k,l \mbox{~in~} r,g,b )
  \ .
\end{alignat}
\end{subequations}
The right-handed Pauli matrices $\rho_k$ are given by
equations~(\ref{SUR2bivectorspauli}).
The leptoquark bivectors of $\Spin(6)_c$ in
equations~(\ref{leptoquarkbivectorXppcommutators})
and~(\ref{leptoquarkbivectorXpmcommutators})
are given in expanded form by
equations~(\ref{notsu5bivectorsab}).

The scalar product of each commutator of equations~(\ref{spin46Xcommutators})
with its reverse equals
2 for the top two lines,
and $2 \times ( \tfrac{2}{3} )^2$ for the bottom two lines.
The square of the covariant derivative~(\ref{Dspin46Higgsvac}) is then
\begin{equation}
\label{DDXvacZ}
  ( \DD^m \langle \reverse{\bP} \rangle ) \cdot ( \DD_m \langle \bP \rangle )
  =
  g^2 \langle P \rangle^2
  \Bigl(
  2
  \sin^6\!\theta_r
  ( R_m^{k} )^2
  +
  2
  ( \tfrac{2}{3} )^2
  \cos^6\!\theta_r
  ( C_m^{[k^\pm l^\pm]} )^2
  \Bigr)
  \ ,
\end{equation}
where the fields on the right hand side are the subset of the weak and colour
fields 
that fail to commute with the Higgs field $\langle \bP \rangle$.
The two right-handed weak fields $R^k_m$, $k = 1,2$,
carry two units of $y,z$ charge and zero colour charge $r,g,b$;
they transform right-handed leptons and quarks into their
right-handed weak partners,
transformations~(\ref{nueudUpupw})
and their antiparticle equivalents.
The six leptoquark fields $C^{[k^\pm l^\pm]}$
carry zero $y$ or $z$ charge, and two units of $r,g,b$ charge;
they are called leptoquarks because they transform
between leptons and quarks,
transformations~(\ref{nueudUpupc})
and their antiparticle equivalents.
Equation~(\ref{DDXvacZ}) shows that the Higgs field $\langle \bP \rangle$
generates masses for the noncommuting fields,
\begin{equation}
\label{mRC}
  m_R
  =
  \sqrt{2} \,
  g
  \langle P \rangle
  \sin^3\!\theta_r
  \ , \quad
  m_C
  =
  \sqrt{2} \, \tfrac{2}{3}
  g
  \langle P \rangle
  \cos^3\!\theta_r
  \ .
\end{equation}
The mass $m_R$ is the mass of each of the two right-handed weak gauge bosons
after $\Spin(4)_w \times \Spin(6)_c$ symmetry breaking;
the mass is different from
the mass $m_W$ of the two left-handed charged weak gauge bosons.
The mass $m_C$ is the mass of each of the six leptoquark gauge bosons
after $\Spin(4)_w \times \Spin(6)_c$ symmetry breaking.
A prediction of the model is that the masses
$m_R$ and $m_C$
are related by, from equations~(\ref{mRC}),
\begin{equation}
\label{mRCratio}
  {m_R \over m_C}
  =
  \tfrac{3}{2}
  \tan^3\!\theta_r
  =
  1.1
  \ ,
\end{equation}
with $\tan\theta_r = 0.90$ from equation~(\ref{thetar}).

Besides not commuting with right-handed weak and leptoquark fields,
the $\Spin(4)_w \times \Spin(6)_c$ Higgs field $\langle \bP \rangle$
also fails to commute with
the electroweak Higgs multiplet $\bH$~(\ref{Higgsmultiplet}),
\begin{equation}
\label{HiggsXcommutators}
  [ \bgamma^+_t \bgamma^\pm_k , \rho_3 ]
  =
  \pm \bgamma^+_t \bgamma^\mp_k
  \quad
  k = y , z
  \ .
\end{equation}
But, as discussed in \S\ref{Spin101-sec},
the bivectors $\bgamma^+_t \bgamma^\mp_k$, $k = y,z$,
do not generate any gauge symmetry,
so $\DD_m \langle \bP \rangle$, equation~(\ref{Dspin46Higgsvac}),
goes not generate any mass term for the electroweak Higgs multiplet.
$\Spin(4)_w \times \Spin(6)_c$ symmetry breaking
breaks the original $\Spin(4)_w$ symmetry of
the electroweak Higgs multiplet
down to the electroweak symmetry $\U(1)_Y \times \SU(2)_\Lchiral$,
but the number of Higgs components, 4, remains unchanged.

\subsection{Scalar fields after Pati-Salam symmetry breaking}
\label{scalarps-sec}

Of the 21 Pati-Salam $\Spin(4)_w \times \Spin(6)_c$ gauge and scalar fields,
the $1{+}3{+}8 = 12$ standard-model
$\U(1)_Y \times \SU(2)_\Lchiral \times \SU(3)_c$
fields remain unbroken,
the $2+6=8$ right-handed and leptoquark fields are broken and become massive,
and the 1 Pati-Salam Higgs scalar $\langle \bP \rangle$ acquires
an expectation value.
What about the $\U(1)_P$ gauge field?


In the Coleman-Weinberg \cite{Coleman:1973,Chataignier:2018} mechanism,
a $\U(1)_P$ gauge field becomes massive
by coupling to a complex scalar field carrying $P$ charge,
which here is a linear combination~(\ref{P}) of $\rho_3$ and $B{-}L$ charge.
Now the field that apparently breaks $\rho_3$ symmetry and $B{-}L$ symmetry
while preserving $Y$ symmetry
is the grand Higgs field $\langle \bT \rangle$,
equation~(\ref{Higgsnuvac}).
As discussed in~\S\ref{grandMajorana-sec},
the grand Higgs field $\langle \bT \rangle$
carries no explicit $\rho_3$ and $B{-}L$ charge;
but it does break $t$-symmetry,
which allows it to generate a Majorana mass term
that flips between the right-handed neutrino with $\rho_3 = 1$ and $B{-}L = -1$,
and its left-handed antineutrino partner with $\rho_3 = -1$ and $B{-}L = +1$.
This is a novel situation that to our knowledge
has not been considered previously in the literature.

Our guess is that the behaviour of 
the $\langle \bT \rangle$ and $\langle \bP \rangle$ fields
during Pati-Salam symmetry breaking are closely related.
As argued in \S\ref{grandMajorana-sec},
even though the grand Higgs field $\langle \bT \rangle$
acquires its expectation value at grand symmetry breaking,
it can generate a Majorana mass for the right-handed neutrino
only later, at Pati-Salam symmetry breaking,
when the Pati-Salam Higgs field $\langle \bP \rangle$
acquires its expectation value
and breaks the symmetries~(\ref{nueudUpup})
that connect right-handed neutrinos to other right-handed fermions.
In turn,
the Majorana mass term~(\ref{mnu}) violates conservation of $\rho_3$
and $B{-}L$ while preserving $Y$,
potentially allowing $\U(1)_P$ symmetry to be broken.
We emphasize that this is just our best guess;
the scenario warrants further investigation.




After $\Spin(4)_w \times \Spin(6)_c$ symmetry breaking,
the components of the original scalar multiplet $\bE$,
equation~(\ref{spin111Emultiplet}),
that survive as potentially light scalar fields comprise
the 4 components of the electroweak Higgs multiplet~(\ref{Dspin46Higgsvac}),
and a standard-model $\U(1)_Y \times \SU(2)_\Lchiral \times \SU(3)_c$ component
of dimension $1{+}3{+}8 = 12$.

\subsection{Scalar fields after electroweak symmetry breaking}
\label{scalarew-sec}

The fate of the 4-component electroweak Higgs multiplet~(\ref{Dspin46Higgsvac})
at electroweak symmetry breaking is known:
three of the four degrees of freedom of the Higgs multiplet
generate masses for the $W^\pm$ and $Z$ electroweak gauge bosons,
while the fourth yields a single electroweak Higgs scalar,
of mass $125 \unit{GeV}$,
that was detected at the Large Hadron Collider (LHC) in 2012
\cite{Aad:2012tfa,Chatrchyan:2012}.

The 12-component standard-model $\U(1)_Y \times \SU(2)_\Lchiral \times \SU(3)_c$
remnant of the original $\Spin(11,1)$ $\bE$ scalar field,
equation~(\ref{spin111Emultiplet}),
persists after electroweak symmetry breaking.
It is the 4 components of the electroweak Higgs multiplet that are
rearranged by electroweak symmetry breaking,
not the corresponding components of the $\bE$ field.
This is just as well:
the world would be a very different place if the electroweak Higgs field
were not there to couple right- and left-handed fermions
and thereby make them massive.

Electroweak symmetry breaking breaks the symmetry of the
12-component standard-model remnant of $\bE$,
but the number of scalar components, 12, remains the same.
These 12 scalar fields
are scalar counterparts of standard-model gauge bosons,
with identical standard-model charges.
The masses of the scalar fields are unknown,
because they depend on the unknown potential energies of the scalar fields.
Presumably,
to have escaped detection,
the masses of most of these fields must be greater than the electroweak scale.

Of the 12 scalar fields, two,
namely the scalar counterparts of the photon or $Z$-boson,
carry zero standard-model charge,
and could potentially constitute the cosmological dark matter.
A crucial property of a dark matter particle is that it must be stable
against annihilation into lighter particles.
One possibility is that a scalar particle might be stabilized by a parity
symmetry \cite{Kadastik:2009,Kadastik:2010},
but no such symmetry is apparent in the present model.
An alternative is that the scalar particle is so light,
$m \sim 10^{-22} \unit{eV}$,
that its decay rate is tiny,
a hypothesis called Fuzzy Dark Matter
\cite{Hui:2017}.
As far as we can see,
in the plain $\Spin(11,1)$ model,
an ultralight scalar is the only possible candidate for dark matter.
Event Horizon Telescope observations of the M87 black hole
disfavor ultralight scalar dark matter with mass in the range
$\sim 3{-}5 \times 10^{-21} \unit{eV}$
\cite{Davoudias:2019}.



\subsection{Fermion masses}
\label{fermionmass-sec}

In the standard model,
after electroweak symmetry breaking,
fermions acquire nonzero rest masses by interacting with the
electroweak Higgs field.
The electroweak Higgs field $\langle \bH \rangle$, equation~(\ref{Higgsvac}),
carries weak charge $y$,
and flips fermions between right- and left-handed chiralities.
In the present picture,
the Higgs field $\langle \bH \rangle$ flips fermions between
Dirac boosts $\Uparrow \,\leftrightarrow\, \Downarrow$ while preserving spin,
in accordance with the $\Spin(11,1)$ chart~(\ref{tyzrgbtab}) of spinors.
It is the flipping between boosts that allows a fermion
to be a superposition of opposite boosts, and therefore to have rest mass.

Unlike the electroweak Higgs field,
the Pati-Salam Higgs field $\langle \bP \rangle$,
equation~(\ref{spin46Higgs}),
carries zero standard model charge,
and does not flip between chiralities.
The $\langle \bP \rangle$ Higgs field does not flip boosts,
and cannot give fermions rest masses.

The conclusion that fermions can acquire their mass
only after electroweak symmetry breaking
at $160 \unit{GeV}$ \cite{DOnofrio:2016}
is intriguing.
In the $\Spin(11,1)$ model,
each generation of fermions belongs to a spinor multiplet of $\Spin(11,1)$.
The masses of the three known generations of fermions (excluding neutrinos)
show a striking progression of masses,
with all $e$-generation fermions being less massive
than all $\mu$-generation masses, which are in turn all less massive
than $\tau$-generation masses.
The most massive fermion known is the $\tau$-generation top,
with mass $173 \unit{GeV}$
\cite{Tanabashi:2018}
just above the $160 \unit{GeV}$ electroweak scale.
If the progression of masses continues, then there is no fourth generation.
The known fermions are all there is.
This is consistent with the evidence that
there are only 3 neutrino types with masses less than
half the mass of the $Z$ neutral weak gauge boson,
$\tfrac{1}{2} m_Z \approx 45 \unit{GeV}$
\cite{Tanabashi:2018},
\begin{equation}
\label{Nnu}
  N_\nu =
  2.991 \pm 0.007 
  \ .
\end{equation}
Evidence from the cosmic microwave background
indicates that there are only 3 neutrino types
with masses less than about the electron mass
(the observations set limits on the number of neutrino types
post electron-positron annihilation)
\cite{Aghanim:2018eyx},
\begin{equation}
\label{Neffnu}
  N_\nu = 3.0 \pm 0.5
  \ .
\end{equation}

The question of why the rest masses of fermions have the pattern observed
remains one of the deepest mysteries of the standard model
\cite{Quigg:2007dt}.

\subsection{Neutrino masses}
\label{neutrinomass-sec}

Neutrinos cannot acquire their mass in the same way as the other fundamental
fermions, because only left-handed neutrinos (and right-handed antineutrinos)
are observed in Nature.
The leading standard solution to the puzzle of neutrino masses
is the see-saw mechanism
proposed by \cite{Yanagida:1980,GellMann:1979},
which requires that
there exists a right-handed neutrino with a large Majorana mass,
as is true in the present model,
\S\ref{grandMajorana-sec}.

The see-saw mechanism posits that,
after electroweak symmetry breaking,
neutrinos have Dirac mass $m_D$ like other fermions, but in addition,
alone among fermions,
a right-handed neutrino $\nu_\Rchiral$,
having no conserved standard-model charge,
has a Majorana mass $m_M$
that couples it to its
left-handed antineutrino partner $\conj{\nu}_\Lchiral$.
The neutrino mass matrix
has eigenvalues $m_\pm$ with
\begin{equation}
\label{mseesaw}
  m_\pm
  =
  \pm 
  \tfrac{1}{2}
  m_M
  +
  \sqrt{\left( \tfrac{1}{2} m_M \right)^2 \! + m_D^2}
  \ ,
\end{equation}
satisfying
$m_+ m_- = m_D^2$.
If the Majorana mass is much larger than the Dirac mass,
$m_M \gg m_D$,
then the masses of the massive and light neutrinos satisfy
the see-saw condition
\begin{equation}
\label{mseesawapprox}
  m_M \approx m_+
  \gg
  m_- \approx {m_D^2 \over m_M}
  \ .
\end{equation}
The massive eigenstate is mostly the right-handed neutrino
with a small admixture of its left-handed antineutrino partner.
The light eigenstate is mostly the left-handed neutrino
with a small admixture of its right-handed antineutrino partner.
The see-saw mechanism offers an explanation for why
observed left-handed, weakly interacting neutrinos have nonzero mass
but are nevertheless light compared to other fermions.

If the mass of the left-handed tauon neutrino
is estimated at $\sim 0.1 \unit{eV}$
\cite{Tortola:2012te},
and the Dirac mass of tauon leptons and quarks
is estimated at $\sim 10 \unit{GeV}$
\cite{Tanabashi:2018},
then the see-saw mechanism would predict
the mass of the right-handed tauon neutrino to be $\sim 10^{12} \unit{GeV}$.
This is consistent with the $10^{12} \unit{GeV}$
energy scale of Pati-Salam symmetry breaking
found in \S\ref{energyunification-sec}, equation~(\ref{mubreak}).

Massive right-handed neutrinos cannot be the cosmological dark matter
because they are unstable,
the small admixture of left-handedness allowing them to decay
by left-handed weak interactions
\cite{Drewes:2013,Bhupal:2016}.
Right-handed neutrinos are sometimes considered as candidates for dark matter,
but that requires their mixing with left-handed neutrinos to be tiny
(so-called sterile neutrinos), which is not true here.

\section{Energy scales of unification}
\label{energyunification-sec}

This section~\ref{energyunification-sec} shows
from the running of the coupling parameters,
Figure~\ref{coupling},
that the predicted energy scale of $\Spin(4)_w \times \Spin(6)_c$ unification is
$10^{12} \unit{GeV}$,
while that of grand unification is
$10^{15} \unit{GeV}$,
equations~(\ref{mubreak}).

The relations~(\ref{gY}) predict that
the hypercharge, right-handed weak, and colour coupling parameters
are related by
\begin{equation}
\label{ggYgrgc}
  {g g_Y \over g_r g_c}
  =
  1
  \ .
\end{equation}
In renormalization theory the coupling parameters
vary with the energy at which they are probed.
The condition~(\ref{ggYgrgc}) can be interpreted as determining
the energy scale of $\Spin(4)_w \times \Spin(6)_c$ symmetry breaking.

\couplingfig

$\Spin(4)_w$ is isomorphic to
the product $\SU(2)_\Rchiral \times \SU(2)_\Lchiral$
of right- and left-handed weak groups,
and there could potentially be two distinct coupling parameters
associated with the two groups.
This section starts out treating the two coupling parameters as distinct,
but then focuses on the special case of equal right- and left-handed
coupling parameters, corresponding to exact chiral symmetry.
Equal coupling parameters would be expected if $\Spin(4)_w$
is the broken remnant of a higher spin group,
as in the scenario considered in this paper.

According to renormalization theory,
to leading (one-loop) order,
the coupling parameter $g$ associated with a gauge group $G$ varies
with the log of the cutoff energy $\mu$ as
(e.g.\ \cite[eq.~(39)]{Peskin:1997},
\cite[eq.~(2.6)]{Kazakov:2000ra},
\cite[eq.~(5.15)]{Schienbein:2019},
or
\cite[eq.~(3.3)]{King:2021b})
\begin{equation}
\label{grunning}
  {\dd g^{-2} \over \dd \ln \mu}
  =
  {1 \over 8\pi^2}
  \Bigl(
  \tfrac{11}{3} S_2 ( G , \mbox{adjoint} )
  -
  \tfrac{2}{3}
  S_2 ( G , \mbox{spinor} )
  n_{\fback}
  -
  \tfrac{1}{6}
  S_2 ( G , \mbox{scalar} )
  n_s
  \Bigr)
  \ ,
\end{equation}
where $S_2 ( G , \mbox{rep} )$
is the Dynkin index
of the representation of the group,
and $n_{\fback}$ and $n_s$ are respectively the number of
fermion and real scalar multiplets that couple to the gauge group $G$.
For $\U(1)_Y$, the formula is
\cite[eq.~(40)]{Peskin:1997}
\begin{equation}
\label{gYrunning}
  {\dd g_Y^{-2} \over \dd \ln \mu}
  =
  {1 \over 8\pi^2}
  \Bigl(
  - \,
  \tfrac{2}{3}
  \sum_f
  \left( \tfrac{1}{2} Y \right)^2
  -
  \tfrac{1}{6}
  \sum_s
  \left( \tfrac{1}{2} Y \right)^2
  \Bigr)
  \ .
\end{equation}
The coupling parameters of the standard model,
evaluated at the energy $\mu = m_Z = 91.1884 \unit{GeV}$ of the Z boson,
are, from \cite{Tanabashi:2018},
\begin{equation}
  g_Y^2
  =
  {e^2 \over \cos^2\!\theta_w}
  =
  0.1278
  \ , \quad
  g_w^2
  =
  {e^2 \over \sin^2\!\theta_w}
  =
  0.4242
  \ , \quad
  g_c^2
  =
  4\pi \alpha_s
  =
  1.492
  \ ,
\end{equation}
where
the weak mixing angle satisfies
$\sin^2\!\theta_w = 0.2315$,
the electromagnetic coupling is
$e^2 = 4\pi \alpha$,
and the fine structure and strong coupling constants
$\alpha$ and $\alpha_s$ are
\begin{equation}
  \alpha(\mu = m_Z) = 1/127.955
  \ , \quad
  \alpha_s(\mu = m_Z) = 0.1187
  \ .
\end{equation}


The Dynkin index $S_2$ of a multiplet in the
representations relevant here is
\begin{subequations}
\label{dynkin2SM}
\begin{alignat}{3}
\label{dynkin2SMSU}
  S_2 ( \SU(N) ,\, \mbox{adjoint} )
  &=
  N
  \ , \quad
  &
  S_2 ( \SU(N) ,\, \mbox{spinor} )
  &=
  \tfrac{1}{2}
  \ ,
\\
\label{dynkin2SMSpin}
  S_2 ( \Spin(N) ,\, \mbox{grade $p$} )
  &=
  \biggl(
  \begin{matrix}
  N{-}2 \\ p{-}1
  \end{matrix}
  \biggr)
  \ , \quad
  &S_2 ( \Spin(N) ,\, \mbox{spinor} )
  &=
  2^{[(N{-}1)/2] - 3}
  \ .
\end{alignat}
\end{subequations}
The adjoint representation is the special representation where
the fields upon which the group acts are the group generators themselves.
For $\Spin(N)$, the adjoint representation is the bivector representation,
the multivector representation of grade $p = 2$.
$\Spin(N)$ has a unique spinor representation,
but $\SU(N)$ has spinor representations of spinor grades $p \leq N/2$
(for example the $\SU(5)$ representations given by the columns
of the $\Spin(10)$ chart~(\ref{yzrgbtab})).
The spinor representation of $\SU(N)$ given by
equation~(\ref{dynkin2SMSU})
is that for spinor grade $p = 1$,
which is the only nontrivial spinor grade
for the groups $\SU(2)$ and $\SU(3)_c$ relevant here.

The numbers of fermion, electroweak scalar, and adjoint scalar multiplets
in each of the standard-model and Pati-Salam regimes are as follows.
\begin{itemize}
\item
{\em Fermions in the standard-model regime.}
In the standard model,
the left-handed weak group $\SU(2)_\Lchiral$ acts on 4 fermion multiplets,
namely the $( \nu_\Lchiral , e_\Lchiral )$ left-handed lepton multiplet,
and the three $( u_\Lchiral^c , d_\Lchiral^c )$ left-handed quark multiplets
of colours $c = r, g, b$.
The colour group $\SU(3)_c$
acts on 4 fermion multiplets,
namely the right- and left-handed up and down quark multiplets
$u_\Rchiral$, $u_\Lchiral$, $d_\Rchiral$, and $d_\Lchiral$.
Each fermion multiplet comes in 3 generations,
so the number of fermion multiplets in equation~(\ref{grunning})
is $n_{\fback} = 4 \times 3 = 12$ for each of $\SU(2)_\Lchiral$ and $\SU(3)_c$.
\item
{\em Fermions in the Pati-Salam regime.}
In the Pati-Salam regime,
the weak right-handed group $\SU(2)_\Rchiral$ is similar to
the left-handed group $\SU(2)_\Lchiral$,
but acting on right-handed in place of left-handed multiplets.
The number of $\SU(2)_\Rchiral$ fermion multiplets
is the same as for $\SU(2)_\Lchiral$,
$n_{\fback} = 4 \times 3 = 12$.
The colour group $\Spin(6)_c$ acts on enlarged fermion multiplets
that contain leptons as well as quarks,
$( \nu_\Lchiral , u_\Lchiral )$, $( \nu_\Rchiral , u_\Rchiral )$,
$( e_\Lchiral , d_\Lchiral )$, and $( e_\Rchiral , d_\Rchiral )$,
but the number of fermion multiplets remains the same,
$n_{\fback} = 4 \times 3 = 12$.
\item
{\em Electroweak scalars in the standard-model regime.}
In the standard model,
the electroweak multiplet transforms
as a pair of 2-component $\SU(2)_\Lchiral$ spinors
with opposite hypercharges $Y = \pm 1$,
so $n_s = 2$ for each of $\U(1)_Y$ and $\SU(2)_\Lchiral$,
while $n_s = 0$ for $\SU(3)_c$.
This is the same as in the ``minimal'' standard model
of the electroweak Higgs field.
\item
{\em Electroweak scalars in the Pati-Salam regime.}
In the Pati-Salam regime,
the electroweak Higgs multiplet~(\ref{Higgsmultiplet}) transforms
as a 4-component vector under $\Spin(4)_w$.
But $\Spin(4)_w$ is isomorphic to $\SU(2)_\Rchiral \times \SU(2)_\Lchiral$,
and the electroweak multiplet effectively transforms
as a pair of 2-component spinors under each of
$\SU(2)_\Rchiral$ and $\SU(2)_\Lchiral$,
so $n_s = 2$ for each of $\SU(2)_\Rchiral$ and $\SU(2)_\Lchiral$.
The scalar Dynkin index is the spinor index $S_2 = \tfrac{1}{2}$.
\item
{\em Adjoint Higgs scalars in the standard-model regime.}
In the standard model regime,
the Higgs multiplet $\bE$ that breaks
the $\Spin(4)_w \times \Spin(6)_c$ symmetry,
equation~(\ref{spin111Emultiplet}),
transforms as a bivector under each of
$\SU(2)_\Lchiral$ and $\Spin(6)_c$.
The bivectors carry zero $Y$ charge.
Thus the $\bE$ multiplet contributes
$n_s = 1$ for each of $\SU(2)_\Lchiral$ and $\Spin(6)_c$,
with adjoint Dynkin index $S_2 = 2$ for $\SU(2)_\Lchiral$,
and adjoint index $S_2 = 3$ for $\Spin(6)_c$.
\item
{\em Adjoint Higgs scalars in the Pati-Salam regime.}
In the Pati-Salam regime,
the Higgs multiplet $\bE$ that breaks
the $\Spin(4)_w \times \Spin(6)_c$ symmetry
transforms as a bivector under each of
$\SU(2)_\Rchiral$, $\SU(2)_\Lchiral$, and $\Spin(6)_c$.
Thus the $\bE$ multiplet contributes
$n_s = 1$ for each of $\SU(2)_\Rchiral$, $\SU(2)_\Lchiral$, and $\Spin(6)_c$,
with adjoint Dynkin index $S_2 = 2$ for $\SU(2)_\Rchiral$ and $\SU(2)_\Lchiral$,
and adjoint index $S_2 = 4$ for $\Spin(6)_c$.
\end{itemize}

The above counting of scalars includes those from the adjoint Higgs scalar
multiplet $\bE$, equation~(\ref{spin111Emultiplet}).
Correctly, only scalars whose masses are less than the running energy scale
$\mu$ should be included.
In the literature,
in calculating the running of coupling parameters
it is common to ignore Higgs scalars
other than the known electroweak Higgs scalar,
on the grounds that any additional Higgs scalars may well be massive.
The calculation shown in Figure~\ref{coupling}
assumes contrariwise that the Higgs scalars coming from $\bE$
are less than the running energy scale $\mu$,
because this yields the largest energy of grand unification,
$10^{15} \unit{GeV}$, equations~(\ref{mubreak}),
more consistent with the constraint estimated by \cite{King:2021b}.


To summarize,
in the standard model,
the factor in parentheses on the right hand side of equation~(\ref{grunning})
for the running of coupling parameters is
\begin{subequations}
\label{brunningsm}
\begin{alignat}{7}
\label{bYrunningsm}
  \U(1)_Y : \
  &
  &\ -\ &
  \tfrac{2}{3} \times \tfrac{5}{6} \times 12
  &\ -\ &
  \tfrac{1}{6} \times \tfrac{1}{2} \times 2
  &\ =\ &
  {-}\tfrac{41}{6}
  \ ,
  \\
  \SU(2)_\Lchiral : \
  &
  \tfrac{11}{3} \times 2
  &\ -\ &
  \tfrac{2}{3} \times \tfrac{1}{2} \times 12
  &\ -\ &
  \tfrac{1}{6} \times ( \tfrac{1}{2} \times 2 + 2 \times 1 )
  &\ =\ &
  \tfrac{17}{6}
  \ ,
  \\
  \SU(3)_c : \
  &
  \tfrac{11}{3} \times 3
  &\ -\ &
  \tfrac{2}{3} \times \tfrac{1}{2} \times 12
  &\ -\ &
  \tfrac{1}{6} \times 3 \times 1
  &\ =\ &
  \tfrac{13}{2}
  \ .
\end{alignat}
\end{subequations}
The sum over fermions of squared hypercharge in equation~(\ref{gYrunning}) is
$\sum_f \left( \tfrac{1}{2} Y \right)^2 = \tfrac{10}{3}$
per generation, or an average of $\tfrac{5}{6}$
per $\SU(2)_\Lchiral$ or $\SU(3)_c$ fermion multiplet,
which accounts for
the factor $\tfrac{5}{6}$ in the $\U(1)_Y$ expression~(\ref{bYrunningsm})
(a common practice, not followed here,
is to multiply the hypercharge coupling parameter
$g_Y^2$ by $\tfrac{5}{3}$
so that the fermionic factor $\tfrac{5}{6}$ in~(\ref{bYrunningsm})
becomes $\tfrac{1}{2}$,
the same as the fermionic factors in $\SU(2)_\Lchiral$ and $\SU(3)_c$;
the scalar factor in~(\ref{bYrunningsm}) would then change
from $\tfrac{1}{6}$ to $\tfrac{1}{10}$,
and the overall value would change
from $-\tfrac{41}{6}$ to $-\tfrac{41}{10}$).
In the Pati-Salam regime
before $\Spin(4)_w \times \Spin(6)_c$ symmetry breaking,
the factor in the running of coupling parameters is
\begin{subequations}
\label{brunning46}
\begin{alignat}{7}
  \SU(2)_\Rchiral , \  \SU(2)_\Lchiral : \
  &
  \tfrac{11}{3} \times 2
  &\ -\ &
  \tfrac{2}{3} \times \tfrac{1}{2} \times 12
  &\ -\ &
  \tfrac{1}{6} \times ( \tfrac{1}{2} \times 2 + 2 \times 1 )
  &\ =\ &
  \tfrac{17}{6}
  \ ,
\\
  \Spin(6)_c : \ 
  &
  \tfrac{11}{3} \times 4
  &\ -\ &
  \tfrac{2}{3} \times \tfrac{1}{2} \times 12
  &\ -\ &
  \tfrac{1}{6} \times 4 \times 1
  &\ =\ &
  10
  \ .
\end{alignat}
\end{subequations}
The factors for $\SU(2)_\Rchiral$ and $\SU(2)_\Lchiral$
are the factors for each of the groups individually
(and are, naturally, the same);
the factors do not add.

If
the right-handed weak coupling parameter $g_r$ is treated as unknown,
then the condition~(\ref{ggYgrgc}) leads to no prediction
for the energy scale of $\Spin(4)_w \times \Spin(6)_c$ unification.
However,
in a scenario where $\Spin(4)_w$ is the broken remnant of
a higher-dimensional spin group, as considered in this paper,
the right- and left-handed coupling parameters $g_r$ and $g_w$ of $\Spin(4)_w$
are same,
\begin{equation}
\label{grw}
  g_r = g_w
  \ ,
\end{equation}
in which case the condition~(\ref{ggYgrgc}) leads to a definite prediction.

The left panel of Figure~\ref{coupling}
shows the combination
$g g_Y / ( g_w g_c )$,
equation~(\ref{ggYgrgc}),
which is predicted to be 1 at $\Spin(4)_w \times \Spin(6)_c$ symmetry breaking.
The right panel of
Figure~\ref{coupling} shows the running of the hypercharge, weak, and colour
coupling parameters $g_Y$, $g_w$, and $g_c$
as a function of the renormalization cutoff energy $\mu$.
The energy scale
of $\Spin(4)_w \times \Spin(6)_c$ symmetry breaking,
where $g g_Y / ( g_w g_c ) = 1$,
and the energy scale of grand unification,
where $g_w = g_c$,
are predicted to be
\begin{equation}
\label{mubreak}
  \mu_{\tfrac{g g_Y}{g_w g_c} = 1}
  =
  1.4 \times 10^{12} \unit{GeV}
  \ , \quad
  \mu_{g_w = g_c}
  =
   1.0 \times 10^{15} \unit{GeV}
  \ .
\end{equation}
The energy scales~(\ref{mubreak})
assume that the adjoint Higgs scalars are all light compared to the scale $\mu$.
If instead the adjoint Higgs scalars are all heavy compared to $\mu$,
then
\begin{equation}
\label{mubreaklo}
  \mu_{\tfrac{g g_Y}{g_w g_c} = 1}
  =
  4.3 \times 10^{11} \unit{GeV}
  \ , \quad
  \mu_{g_w = g_c}
  =
  3.4 \times 10^{14} \unit{GeV}
  \ .
\end{equation}

The predicted energy $\approx 10^{12} \unit{GeV}$ of Pati-Salam unification
is comparable to (slightly greater than) that of the
most energetic cosmic rays observed \cite{Abbasi:2014,AlvesBatista:2019}.

The predicted energy $\approx 10^{15} \unit{GeV}$ of grand unification
falls below the lower limit
\begin{equation}
  \mu_{\Spin(10)}
  \gtrsim 4 \times 10^{15} \unit{GeV}
\end{equation}
inferred for $\Spin(10)$ models by \cite{King:2021b}
from the Super-Kamiokande lower limit of $\gtrsim 1.6 \times 10^{34} \unit{yr}$
on the proton lifetime \cite{Takenaka:2020}.
Thus the $\Spin(11,1)$ model may already be ruled out.
However,
it should be borne in mind, first,
that the prediction of $10^{15} \unit{GeV}$
is based on the simplifying assumptions of 1-loop renormalization
and an abrupt transition at the Pati-Salam energy,
and, second, that the $\Spin(11,1)$ symmetry breaking chain
differs from the $\Spin(10)$ chains considered by \cite{King:2021b}.
An improved computation would be desirable.

The upper limit on the energy scale of cosmological inflation
inferred from the upper limit to $B$-mode polarization power
in the cosmic microwave background measured by the Planck satellite is
\cite[eq.~(26)]{Ade:2015lrj}
\begin{equation}
\label{muinfl}
  \mu_\textrm{inflation}
  \lesssim
  2 \times 10^{16} \unit{GeV}
  \ .
\end{equation}
The predicted energy scale~(\ref{mubreak}) of grand symmetry breaking
is within the upper limit~(\ref{muinfl}),
although it is not clear that the grand symmetry breaking
and inflationary scales can be compared directly.

The $\Spin(4)_w \times \Spin(6)_c$ mixing angle $\theta_r$
defined by equations~(\ref{spin46mixingangle})
is predicted to be
\begin{equation}
\label{thetar}
  \theta_r
  =
  0.73
  =
  42^\circ
  \ , \quad
  \tan\theta_r
  =
  0.90
  \ ,
\end{equation}
insensitive to whether adjoint scalars are taken to be light or massive
in the running of coupling parameters.

\section{Summary of predictions, and potential flaws}
\label{predflaw-sec}

Subsection~\ref{predictions-sec} summarizes the predictions of the proposed
$\Spin(11,1)$ model.
Subsection~\ref{flaws-sec} highlights possible flaws that merit scrutiny.

\subsection{Predictions}
\label{predictions-sec}

\begin{enumerate}
\item
$\Spin(11,1)$,
like $\Spin(10)$,
predicts a right-handed neutrino,
with consequences that have been well explored in the literature.
\item
The tight fit of the Dirac and standard-model algebras
in the $\Spin(11,1)$ geometric algebra
admits a unique minimal symmetry breaking chain~(\ref{symbreaking}),
proceeding via the Pati-Salam group $\Spin(4)_w \times \Spin(6)_c$.
\item
$\SU(5)$ cannot be on the path to grand unification,
because its generators fail to commute
with the spacetime generators~(\ref{vectorsspin10ew})
in $\Spin(11,1)$.
\item
All the Higgs fields involved in symmetry breaking,
including grand, Pati-Salam, and electroweak symmetry breaking,
lie in a common 66-component $\Spin(11,1)$ scalar multiplet $\bE$ of bivectors,
equation~(\ref{spin111Emultiplet}).
The Higgs fields are the scalar (spin~0) counterparts
of the vector (spin~1) gauge fields of $\Spin(11,1)$.
\item
The grand Higgs field $\langle \bT \rangle$, equation~(\ref{Higgsnuvac}),
is available to drive cosmological inflation.
\item
The generators of the electroweak Higgs field $\bH$,
equation~(\ref{Higgsmultiplet}),
fail to commute with the grand Higgs field $\langle \bT \rangle$.
To allow the electroweak Higgs field to persist without being
absorbed at grand symmetry breaking,
we argue, \S\ref{Spin101-sec},
that the 11th spatial dimension $\bgamma^+_t$ behaves as a scalar dimension,
as discussed in \S4.4 of \cite{Hamilton:2023a}.
Consequently the grand unified group is $\Spin(10,1)$ rather than $\Spin(11,1)$.
\item
Grand symmetry breaking is predicted to occur at $10^{15} \unit{GeV}$,
and Pati-Salam symmetry breaking at $10^{12} \unit{GeV}$,
equations~(\ref{mubreak}).
\item
The grand Higgs field $\langle \bT \rangle$ breaks $t$-symmetry,
which generates a Majorana mass term~(\ref{mnu}) for the right-handed neutrino,
although the right-handed neutrino acquires that Majorana mass
only at Pati-Salam symmetry breaking.
Later, after electroweak symmetry breaking,
the left-handed neutrino can acquire a small mass
by the standard see-saw mechanism
\cite{GellMann:1979}.
\item
The Pati-Salam group
$\Spin(4)_w \times \Spin(6)_c$
predicts 8 vector gauge boson fields beyond those of the standard model,
comprising
2 right-handed weak gauge bosons $R$,
and 6 leptoquark gauge bosons $C$.
At Pati-Salam symmetry breaking,
the weak and leptoquark bosons acquire masses in the ratio
$m_R / m_C = 1.1$,
equation~(\ref{mRCratio}).
\item
The $\langle \bP \rangle$ Higgs field, equation~(\ref{spin46Higgs}),
that breaks Pati-Salam symmetry carries zero standard model charge,
and in particular does not flip the Dirac boost bit,
so, unlike the electroweak Higgs field, cannot give masses to fermions,
\S\ref{fermionmass-sec}.
Fermions other than the right-handed neutrino
remain massless until electroweak symmetry breaking.
\item
As remarked at the end of \S\ref{scalarew-sec},
in the $\Spin(11,1)$ model
the only evident candidate for cosmological dark matter is an ultralight scalar
\cite{Hui:2017}
with vanishing standard model charge,
the scalar counterpart of either the photon or the $Z$-boson.
\item
\newcounter{predcount}
\setcounter{predcount}{\theenumi}
In the model proposed in this paper,
supersymmetry \cite{Martin:1998} is not needed to achieve grand unification.
\end{enumerate}

%
%

\subsection{Places to pick holes}
\label{flaws-sec}

Here we bring attention to areas that the critical reader might
examine for flaws.
\begin{quote}
``We are very much aware that we are exploring
unconventional ideas and that there may be some
basic flaw in our whole approach which we have
been too stupid to see.'' --- Sidney Coleman \& Erick Weinberg
\cite{Coleman:1973}.
\end{quote}

\begin{enumerate}
\setcounter{enumi}{\thepredcount}
\item
The adjustment~(\ref{adjustsmbivectorsI6})
of $\Spin(10)$ bivectors is essential
to permit unification of $\Spin(10)$ and
the Lorentz group $\Spin(3,1)$ in $\Spin(11,1)$.
Without this adjustment, the $\Spin(11,1)$ idea fails.
\item
To allow the electroweak Higgs bivectors to survive grand symmetry breaking,
it appears necessary to postulate, \S\ref{Spin101-sec},
that the 11th dimension $\bgamma^+_t$,
the spatial partner of the time vector $\bgamma^-_t$,
is a scalar that does not generate any gauge symmetry,
in which case the symmetry at grand symmetry breaking is $\Spin(10,1)$
rather than the parent symmetry $\Spin(11,1)$.
We have not speculated how $\Spin(11,1)$ might break to $\Spin(10,1)$.
\item
The 12 (modified) bivectors
$\bgamma^\pm_t \bgamma^\pm_k$, $k = r,g,b$,
are unbroken by the grand Higgs field $\langle \bT \rangle$,
but fail to commute with several Lorentz bivectors~(\ref{lorentzspin10ew}),
so cannot persist after grand symmetry breaking,
\S\ref{grandHiggs-sec}.
We suggest that these bivectors are broken and become massive
as the dimensionality of spacetime drops from 10+1 to 3+1,
but we have not offered a detailed model of this process.
\item
The grand Higgs field $\langle \bT \rangle$ breaks $t$-symmetry,
thereby generating a Majorana mass term
$\conj{\nu} \cdot \langle \bT \rangle \nu$, equation~(\ref{mnu}),
for the right-handed neutrino.
The result is that $B{-}L$ is violated
even though $\langle \bT \rangle$ commutes with $B{-}L$.
\item
The right-handed neutrino can acquire its Majorana mass only
after Pati-Salam symmetry breaking,
when the neutrino is isolated into a standard-model singlet.
But the Pati-Salam Higgs field $\langle \bP \rangle$
can acquire its expectation value only after $B{-}L$ is violated
by the right-handed neutrino acquiring its Majorana mass.
The two processes must be tightly coupled, \S\ref{scalarps-sec},
but we have not offered a detailed model.
\item
The predicted energy scale of grand unification is $10^{15} \unit{GeV}$,
equations~(\ref{mubreak}),
which is less than the lower limit of $4 \times 10^{15} \unit{GeV}$
inferred for $\Spin(10)$ models by \cite{King:2021b}
from the Super-Kamiokande lower limit on the proton lifetime
\cite{Takenaka:2020}.
A more careful calculation needs to be done that follows the
symmetry breaking chain of the $\Spin(11,1)$ model
rather than the $\Spin(10)$ models of \cite{King:2021b}.
\end{enumerate}

\section{Conclusions}
\label{conclusions-sec}

$\Spin(10)$ is well-known as a promising candidate for
a grand unified group that contains the standard-model group
$\U(1)_Y \times \SU(2)_\Lchiral \times \SU(3)_c$.
But the full elegance of $\Spin(10)$ emerges only when its elements
are recast in terms of spinors \cite{Wilczek:1998,Baez:2009dj,Hamilton:2023a}.
Spinors in $\Spin(N)$ are indexed by a bitcode with $[N/2]$ bits,
each bit representing a conserved charge,
and each bit taking the values $+\tfrac{1}{2}$ ($\uparrow$)
or $-\tfrac{1}{2}$ ($\downarrow$).
The 5 charges of the standard model
--- hypercharge, weak isospin, and three colours ---
are harmoniously re-expressed in terms of a 5-bit bitcode $yzrgb$,
with 2 weak bits $y$ and $z$, and 3 colour bits $r$, $g$, $b$,
equations~(\ref{YIC}).

This paper is motivated by some striking (to the authors) features,
\S\ref{striking-sec},
of the $\Spin(10)$ chart~(\ref{yzrgbtab}) of standard model fermions,
which seem to suggest that
the $\Spin(10)$ algebra
and
the Dirac algebra,
which is the geometric algebra of the Lorentz group $\Spin(3,1)$,
might unify nontrivially in a common spin algebra.
The standard treatment of $\Spin(10)$ as a grand unified group
takes each of the $2^5 = 32$ fermions of a generation to be a Weyl fermion,
a 2-component chiral (massless) fermion transforming under $\Spin(3,1)$,
for a total of $2^6$ degrees of freedom.
If the $\Spin(10)$ and $\Spin(3,1)$ algebras unify nontrivially,
then the unified algebra must have $2^6$ spinor degrees of freedom.
Given that there must be 1 time dimension,
the spinor algebra must be that of $\Spin(11,1)$.
This requires adding a 6th bit, the $t$-bit,
or time bit, to the 5 bits $yzrgb$ of $\Spin(10)$.

The usual assumption,
motivated by the Coleman-Mandula no-go theorem
\cite{Coleman:1967,Mandula:2015,Pelc:1997},
is that the generators of a grand unified group such as $\Spin(10)$
commute with those of spacetime.
But that assumption is unnecessarily strong:
the Coleman-Mandula theorem requires only that the generators of
unbroken internal symmetries commute with those of spacetime.
In the $\Spin(11,1)$ model,
before grand symmetry breaking
the internal group and the spacetime group are one and the same,
so the Coleman-Mandula theorem is satisfied trivially.
After grand symmetry breaking,
the $\Spin(11,1)$ algebra contains
the internal (initally Pati-Salam, then standard-model)
and Dirac algebras as commuting subalgebras,
in accordance with the Coleman-Mandula theorem.

The main result of this paper is the expressions~(\ref{vectorsspin10ew})
for the vectors of the Dirac algebra
in terms of multivectors of $\Spin(11,1)$.
The Dirac vectors~(\ref{vectorsspin10ew}) satisfy the Dirac algebra,
and they all commute with all standard model bivectors
modified per~(\ref{adjustsmbivectorsI6}),
in accordance with the Coleman-Mandula theorem.
The modification~(\ref{adjustsmbivectorsI6}),
which is the key trick of this paper,
is to multiply all imaginary standard model bivectors
by the colour chiral operator $\varkappa_{rgb}$,
which leaves the standard-model algebra unchanged,
but allows the Dirac algebra to be embedded in the $\Spin(11,1)$ algebra
in a nontrivial way.

The remainder of the paper,
\S\ref{spin5spin6-sec}--\S\ref{energyunification-sec},
identifies the symmetry breaking chain and associated Higgs sector
that breaks $\Spin(11,1)$ to the standard model.
The tight fit of the Dirac and standard-model algebras
in the $\Spin(11,1)$ algebra leaves little freedom of choice.
The minimal symmetry breaking chain~(\ref{symbreaking}) is unique,
proceeding via the Pati-Salam group.
The minimal Higgs sector is similarly unique,
consisting of a 66-component $\Spin(11,1)$ scalar multiplet
transforming in the adjoint represention.
The same Higgs multiplet mediates all symmetry breakings,
including grand, Pati-Salam, and electroweak.
The running of coupling parameters predicts that
grand symmetry breaking occurs at $10^{15} \unit{GeV}$,
and Pati-Salam symmetry breaking at $10^{12} \unit{GeV}$,
equations~(\ref{mubreak}).

The grand Higgs field $\langle \bT \rangle$,
equation~(\ref{Higgsnuvac}),
breaks $t$ symmetry,
is available to drive cosmological inflation at the grand unified scale,
and generates a large Majorana mass for the right-handed neutrino
by flipping its $t$-bit.
The electroweak Higgs field $\langle \bH \rangle$,
equation~(\ref{Higgsvac}),
breaks $y$ symmetry,
and generates fermion masses by flipping their $y$-bit.

Although the unified algebra is that of $\Spin(11,1)$,
the gauge field before grand symmetry breaking cannot be the full group
$\Spin(11,1)$,
because if it were,
then the subset of $\Spin(11,1)$ bivectors
comprising the 4 electroweak Higgs bivectors
$\bgamma^+_t \bgamma^\pm_k$, $k = y,z$,
equation~(\ref{Higgsmultiplet}),
would be absorbed into generating masses for the corresponding bivector
gauge fields,
whereas in practice those 4 electroweak Higgs bivectors persist,
remaining available to mediate electroweak symmetry breaking.
The solution proposed in \S\ref{Spin101-sec}
is that the spatial dimension $\bgamma^+_t$,
the partner of the time dimension $\bgamma^-_t$,
behaves as a scalar dimension, not participating in any gauge symmetry,
as discussed in \S4.4 of \cite{Hamilton:2023a}.
The grand unified group is then $\Spin(10,1)$, not $\Spin(11,1)$.

Section~\ref{predictions-sec} summarizes the various predictions
of the $\Spin(11,1)$ model,
while section~\ref{flaws-sec} highlights its possible flaws.

It is beyond the scope of this paper to address whether the proposed
$\Spin(11,1)$ model might be accommodated in string theory.
It can scarcely escape notice that $\Spin(10,1)$
has the same number 11 of spacetime dimensions as maximal supergravity
\cite{Freedman:2012,vanNieuwenhuizen:2004rh,Ferrara:2017hed,Deser:2018},
the low-energy limit of M theory \cite{Schwarz:1999,Becker:2007},
which is the conjectured extension of string theory
to include higher-dimensional objects, branes.
Extensions of string theory to 12 spacetime dimensions, F-theory,
have also been proposed \cite{Vafa:1996,Heckman:2010,Callaghan:2012,Weigand:2018}.
String-theory-inspired models usually assume that
the parent spacetime is, at least locally,
a product space, consisting of 4 large dimensions
multiplied by a space of compactified or hidden dimensions.
By contrast,
in the $\Spin(11,1)$ geometric algebra,
although the time dimension is a vector,
each of the 3 spatial Dirac dimensions
is a pentavector, a 5-dimensional multivector,
equations~(\ref{vectorsspin10ew}).
The spatial dimensions share a common 2-dimensional factor,
and beyond that are each 3-dimensional.
Is this arrangement viable in string theory?

\section*{Data statement}

No new data were created or analysed in this study.

\ack
We thank Pierre Ramond, John Baez, John Huerta, and Kirill Krasnov
for helpful conversations and correspondence.
This research was supported in part by FQXI mini-grant FQXI-MGB-1626.

\appendix

\section{Dirac algebra}
\label{diracalgebra-app}

The Dirac algebra is the geometric (Clifford) algebra associated with the group
$\Spin(3,1)$ of Lorentz transformations in 3+1 spacetime dimensions.
In the chiral representation,
the 4 chiral basis spinors
$\bepsilon_a \equiv \{ \bepsilon_{\Upup} , \, \bepsilon_{\Downdown} , \, \bepsilon_{\Downup} , \, \bepsilon_{\Updown} \}$
of the Dirac algebra are the column spinors\footnote{
The ordering of the Dirac bits is opposite from that in the construction
in \S3.1 of \cite{Hamilton:2023a},
and the sign of one of the spinors is flipped,
\begin{equation}
\label{diracvscanonical}
  \{ \bepsilon_{\Upup} , \, \bepsilon_{\Downdown} , \, \bepsilon_{\Downup} , \, \bepsilon_{\Updown} \}
  =
  \{ \bepsilon_{\upup} , \, \bepsilon_{\downdown} , \, \bepsilon_{\updown} , \, -\bepsilon_{\downup} \}
  \ .
\end{equation}
The reordering of bits and the sign flip ensures
the chiral and Dirac representations of the orthonormal Dirac $\gamma$-matrices
take a standard form,
equations~(\ref{gammachiral}) and~(\ref{gammadirac})}
\begin{equation}
\label{chiralbasisspinors}
  \bepsilon_{\Upup}
  =
  \left(
  \begin{array}{c}
    1 \\ 0 \\ 0 \\ 0
  \end{array}
  \right)
  \ , \quad
  \bepsilon_{\Downdown}
  =
  \left(
  \begin{array}{c}
    0 \\ 1 \\ 0 \\ 0
  \end{array}
  \right)
  \ , \quad
  \bepsilon_{\Downup}
  =
  \left(
  \begin{array}{c}
    0 \\ 0 \\ 1 \\ 0
  \end{array}
  \right)
  \ , \quad
  \bepsilon_{\Updown}
  =
  \left(
  \begin{array}{c}
    0 \\ 0 \\ 0 \\ 1
  \end{array}
  \right)
  \ .
\end{equation}
The indices
$\{ \Upup , \Downdown , \Downup , \Updown \}$
signify the transformation properties of the basis spinors:
$\Uparrow$ and $\Downarrow$
signify boost weight $+\frac{1}{2}$ and $-\frac{1}{2}$,
while
$\uparrow$ and $\downarrow$
signify spin weight $+\frac{1}{2}$ and $-\frac{1}{2}$.
The first two spinors,
$\bepsilon_{\Upup}$ and $\bepsilon_{\Downdown}$,
have right-handed Dirac chirality (even number of up-bits),
while the last two,
$\bepsilon_{\Downup}$ and $\bepsilon_{\Updown}$,
have left-handed Dirac chirality (odd number of up-bits).
Chiral spinors are natively massless.

Dirac basis spinors
$\bepsilon_a \equiv \{ \bepsilon_{\upTup} , \, \bepsilon_{\upTdown} , \, \bepsilon_{\downTup} , \, \bepsilon_{\downTdown} \}$
are massive linear combinations of chiral spinors,
\begin{equation}
\label{diracbasisspinors}
  \underset{\textrm{Dirac}}{\bepsilon_{a}}
  =
  X_{ab}
  \underset{\textrm{chiral}}{\bepsilon_{b}}
  \ ,
\end{equation}
where $X$ is the
symmetric ($X = X^\transpose$),
unitary ($X^{-1} = X^\dagger$)
matrix
\begin{equation}
\label{Xchiral}
  X
  \equiv
  {1 \over \sqrt{2}}
  \left(
  \begin{array}{cccc}
  1 & 0 & -\im & 0 \\
  0 & 1 & 0 & -\im \\
  -\im & 0 & 1 & 0 \\
  0 & -\im & 0 & 1
  \end{array}
  \right)
  \ .
\end{equation}
The Dirac basis spinors
$\bepsilon_{\upTup}$ and $\bepsilon_{\upTdown}$
represent massive spinors in their rest frames,
while
$\bepsilon_{\downTup}$ and $\bepsilon_{\downTdown}$
represent massive antispinors in their rest frames.

Orthonormal vectors
$\{ \bgamma_0 , \, \bgamma_1 , \, \bgamma_2 , \, \bgamma_3 \}$
in the chiral representation are the four $4 \times 4$ unitary matrices
\begin{equation}
\label{gammachiral}
  \bgamma_0 =
  \left(
  \begin{array}{cc}
     0 & 1 \\
     -1 & 0
  \end{array}
  \right)
  \ , \quad
  \bgamma_a =
  \left(
  \begin{array}{cc}
     0 & \sigma_a \\
     \sigma_a & 0
  \end{array}
  \right)
  \ ,
\end{equation}
where $\sigma_a$ are Pauli matrices, and 1 is the $2 \times 2$ unit matrix.
The advantage of the representation~(\ref{gammachiral})
compared to some others commonly found in the literature is that
the chiral basis vectors in the chiral representation are purely real,
equations~(\ref{gammachiralNP}).
The Dirac representation of orthonormal or chiral basis vectors
may be obtained from the chiral representation by the transformation
\begin{equation}
  \underset{\textrm{Dirac}}{\bgamma_m}
  =
  X
  \underset{\textrm{chiral}}{\bgamma_m}
  X^{-1}
  \ ,
\end{equation}
with $X$ from equation~(\ref{Xchiral}).
The orthonormal vectors in the Dirac representation are
\begin{equation}
\label{gammadirac}
  \bgamma_0 =
  \im
  \left(
  \begin{array}{cc}
     1 & 0 \\
     0 & -1
  \end{array}
  \right)
  \ , \quad
  \bgamma_a =
  \left(
  \begin{array}{cc}
     0 & \sigma_a \\
     \sigma_a & 0
  \end{array}
  \right)
  \ .
\end{equation}
In the notation $\bgamma^\pm_k$ of paired orthonormal vectors
used elsewhere in this paper, equations~(\ref{gammakpm}),
the orthonormal vectors $\bgamma_a$ are
\begin{equation}
\label{gammadiracpm}
  \{ \bgamma_1 , \, \bgamma_2 , \, \bgamma_3 , \, \bgamma_0 \}
  =
  \{ \bgamma_1^+ , \, \bgamma_1^- , \, \bgamma_2^+ , \, \im \bgamma_2^- \}
  \ .
\end{equation}
The bivectors $\bsigma_a$ and $I \bsigma_a$
($\bsigma_a$ are $4 \times 4$ matrices satisfying the same algebra
as Pauli matrices)
and the pseudoscalar $I$
in the chiral representation
are
\begin{equation}
\label{ichiral}
  \setlength\arraycolsep{2.5pt}
  \bgamma_0 \bgamma_a
  =
  \bsigma_a
  \equiv
  \left(
  \begin{array}{cc}
     \sigma_a & 0 \\
     0 & - \sigma_a
  \end{array}
  \right)
  \ , \quad
  \tfrac{1}{2}
  \varepsilon_{abc} \bgamma_b \bgamma_c
  =
  I \bsigma_a
  =
  \im
  \left(
  \begin{array}{cc}
     \sigma_a & 0 \\
     0 & \sigma_a
  \end{array}
  \right)
  \ , \quad
  I
  =
  \im
  \left(
  \begin{array}{cc}
     1 & 0 \\
     0 & -1
  \end{array}
  \right)
  \ .
\end{equation}
Chiral basis vectors in the chiral representation,
commonly called Newman-Penrose basis vectors,
are the real matrices
\begin{equation}
\label{gammachiralNP}
  \setlength\arraycolsep{2.5pt}
  \bgamma_v =
  \left(
  \begin{array}{cc}
     0 & \sigma_v \\
     - \sigma_u & 0
  \end{array}
  \right)
  \ , \quad
  \bgamma_u =
  \left(
  \begin{array}{cc}
     0 & \sigma_u \\
     - \sigma_v & 0
  \end{array}
  \right)
  \ , \quad
  \bgamma_+ =
  \left(
  \begin{array}{cc}
     0 & \sigma_+ \\
     \sigma_+ & 0
  \end{array}
  \right)
  \ , \quad
  \bgamma_- =
  \left(
  \begin{array}{cc}
     0 & \sigma_- \\
     \sigma_- & 0
  \end{array}
  \right)
  \ ,
\end{equation}
where $\sigma_m$ are the Newman-Penrose Pauli matrices
\begin{subequations}
\label{PaulimatricesNP}
\begin{align}
  \sigma_v
  \equiv
  {1 \over \sqrt{2}} \left( 1 + \sigma_3 \right)
  =
  \sqrt{2}
  \left(
  \begin{array}{cc}
     1 & 0 \\
     0 & 0
  \end{array}
  \right)
  \ &, \quad
  \sigma_u
  \equiv
  {1 \over \sqrt{2}} \left( 1 - \sigma_3 \right)
  =
  \sqrt{2}
  \left(
  \begin{array}{cc}
     0 & 0 \\
     0 & 1
  \end{array}
  \right)
  \ ,
\\
  \sigma_+
  \equiv
  {1 \over \sqrt{2}} \left( \sigma_1 + \im \sigma_2 \right)
  =
  \sqrt{2}
  \left(
  \begin{array}{cc}
     0 & 1 \\
     0 & 0
  \end{array}
  \right)
  \ &, \quad
  \sigma_-
  \equiv
  {1 \over \sqrt{2}} \left( \sigma_1 - \im \sigma_2 \right)
  =
  \sqrt{2}
  \left(
  \begin{array}{cc}
     0 & 0 \\
     1 & 0
  \end{array}
  \right)
  \ .
\end{align}
\end{subequations}
The labelling $v$, $u$, $+$, $-$ for the chiral basis vectors
is conventional in the Newman-Penrose community.
In the notation $\bgamma_k$ and $\bgamma_{\bar k}$
of paired chiral vectors
used elsewhere in this paper, equations~(\ref{gammakpm}),
the chiral (Newman-Penrose) basis vectors $\bgamma_a$ are
\begin{equation}
  \{ \bgamma_+ , \, \bgamma_- , \, \bgamma_v , \, \bgamma_u \}
  =
  \{ \bgamma_1 , \, \bgamma_{\bar 1} , \, \bgamma_2 , \, -\bgamma_{\bar 2} \}
  \ .
\end{equation}

The spinor metric $\varepsilon$
in the chiral
representation is \cite{Hamilton:2023a}
\begin{equation}
\label{echiraldirac}
  \varepsilon
  =
  I \bsigma_2
  =
  \left(
  \begin{array}{cccc}
  0 & 1 & 0 & 0 \\
  -1 & 0 & 0 & 0 \\
  0 & 0 & 0 & 1 \\
  0 & 0 & -1 & 0
  \end{array}
  \right)
  \ .
\end{equation}
The expression for $\varepsilon$
shows that the spinor metric flips boost and spin,
leaving chirality unchanged.
The conjugation operator is related to the spinor metric by
$\Cp = -\im \varepsilon \bgamma_0^\transpose$.
The conjugation operator in the chiral
representation is
\begin{equation}
\label{Cpchiraldirac}
  \Cp
  =
  \im I \bgamma_2
  =
  \left(
  \begin{array}{cccc}
  0 & 0 & 0 & \im \\
  0 & 0 & -\im & 0 \\
  0 & -\im & 0 & 0 \\
  \im & 0 & 0 & 0
  \end{array}
  \right)
  \ .
\end{equation}
The expression for $\Cp$
shows that the conjugation operator flips chirality and spin,
leaving boost unchanged.
The expressions for the spinor metric and conjugation operator
in terms of multivectors are to be understood as
holding in the particular representation.
The spinor metric and conjugation operators do not transform as multivectors
under Lorentz transformations;
rather, they are Lorentz invariant, unchanged under Lorentz transformations.

\section{Naming of bits}
\label{namingbits-sec}

This paper adopts the notation $t,y,z,r,g,b$
for the 6 bits of $\Spin(11,1)$ spinors,
consisting of the time bit $t$, two weak bits $y,z$,
and three colour bits $r,g,b$.
\cite{Wilczek:1998}
used $r,w,b$ (red, white, blue) for the colour bits,
and (arbitrarily) $p, g$ (purple, green) for the weak bits.
\cite{Baez:2009dj}
adopted $r,g,b$ (red, green, blue) for the colour bits,
and $d, u$ (down, up) for the weak bits,
on the grounds that down and up quarks $d^c_\Lchiral$ and $u^c_\Lchiral$
have up bits respectively $dc$ and $uc$
(with $c$ one of $r,g,b$),
per the $\Spin(10)$ chart~(\ref{yzrgbtab}).

The present paper follows \cite{Baez:2009dj} for the colour bits,
since it is consistent with the elegant mental picture that
equal amounts of the three colours $r,g,b$ are colourless,
as are all free fermionic bound states at low energy
(leptons, baryons, mesons).
However, \cite{Baez:2009dj}'s adoption of $d,u$ for the weak bits is potentially
confusing, first because each bit $d$ or $u$ can itself be either down or up,
and second because $d$ and $u$ might be misinterpreted as measuring
the number of $d$ and $u$ quarks,
potentially confusing weak isospin with isospin.

The present paper chooses $y$, $z$ for the weak bits
in part because it reduces the potential for misinterpretation,
and in part because $y$ and $z$ are infrared bands
to be used by the Vera Rubin Observatory (formerly the LSST)
\cite{Ivezic:2019},
for which astronomical first light is expected in 2024.
The sequence $yzrgb$ is in (inverse) order of wavelength,
\begin{equation}
  \{y,z,r,g,b\}
  \sim
  \{1000, 900, 600, 500, 400\} \unit{nm}
  \ .
\end{equation}
Note that hypercharge $Y$ carries one unit of $y$-charge,
in addition to other charges,
equation~(\ref{YICY}).

\section{Gauge fields of subgroups of $\Spin(10)$}
\label{gaugesm-sec}

The standard-model gauge group
$\U(1)_Y \times \SU(2)_\Lchiral \times \SU(3)_c$
is a subgroup of the $\SU(5)$ subgroup of $\Spin(10)$.
The gauge fields (generators) of $\SU(5)$ comprise
the subset of gauge fields of $\Spin(10)$
that leave the number of up-bits of a spinor unchanged.
The gauge bivectors of $\SU(n)$ with $n = 5$ constitute
$( n {+} 1 ) ( n {-} 1 ) = 24$ bivectors
comprising the $n ( n - 1 ) = 20$ off-diagonal bivectors
\begin{subequations}
\label{su5bivectorsab}
\begin{alignat}{3}
\label{su5bivectorsabpp}
  \tfrac{1}{2} ( 1 - \varkappa_{kl} )
  \bgamma^+_k \bgamma^+_l
  &=
  \tfrac{1}{2}
  ( \bgamma^+_k \bgamma^+_l
  +
  \bgamma^-_k \bgamma^-_l )
  &
  &=
  \tfrac{1}{2}
  ( \bgamma_k \wedgie \bgamma_{\bar l} + \bgamma_{\bar k} \wedgie \bgamma_l )
  \ ,
\\
\label{su5bivectorsabpm}
  \tfrac{1}{2} ( 1 - \varkappa_{kl} )
  \bgamma^+_k \bgamma^-_l
  &=
  \tfrac{1}{2}
  ( \bgamma^+_k \bgamma^-_l
  -
  \bgamma^-_k \bgamma^+_l )
  &
  &=
  \tfrac{\im}{2}
  ( \bgamma_k \wedgie \bgamma_{\bar l} - \bgamma_{\bar k} \wedgie \bgamma_l )
  \ ,
\end{alignat}
\end{subequations}
and the $n{-}1=4$ diagonal bivectors
\begin{equation}
\label{su5bivectorsa}
  \tfrac{1}{2} \,
  \bgamma^+_k \bgamma^-_k
  =
  \tfrac{\im}{2} \,
  \bgamma_k \wedgie \bgamma_{\bar k}
  \quad
  \mbox{modulo}
  \quad
  \tfrac{1}{2}
  \sum_k \bgamma^+_k \bgamma^-_k
  =
  \tfrac{\im}{2}
  \sum_k \bgamma_k \wedgie \bgamma_{\bar k}
  \ ,
\end{equation}
with indices $k$ and $l$ running over
$y$, $z$, $r$, $g$, $b$.
The quantity
$\varkappa_{kl} \equiv \bgamma_k \wedgie \bgamma_{\bar k} \wedgie \bgamma_l \wedgie \bgamma_{\bar l} = - \bgamma^+_k \bgamma^-_k \bgamma^+_l \bgamma^-_l$
in equations~(\ref{su5bivectorsab})
is the $kl$ chiral operator.
The factor $\tfrac{1}{2} ( 1 - \varkappa_{kl} )$
is a projection operator,
whose square is itself,
which serves to project its argument into the space where
the sum of $k$ and $l$ charges is zero.
The S in $\SU(5)$ restricts to $\U(5)$ matrices of unit determinant,
effectively removing the bivector
$\tfrac{1}{2} \sum_k \bgamma^+_k \bgamma^-_k$
that rotates all spinors in an $\SU(5)$ multiplet by a common phase.

The bivectors of $\Spin(10)$ that are not in $\SU(5)$
are the 20 off-diagonal bivectors
\begin{subequations}
\label{notsu5bivectorsab}
\begin{alignat}{3}
\label{notsu5bivectorsabpp}
  \tfrac{1}{2} ( 1 + \varkappa_{kl} )
  \bgamma^+_k \bgamma^+_l
  &=
  \tfrac{1}{2}
  ( \bgamma^+_k \bgamma^+_l
  -
  \bgamma^-_k \bgamma^-_l )
  &
  &=
  \tfrac{1}{2}
  ( \bgamma_k \wedgie \bgamma_l + \bgamma_{\bar k} \wedgie \bgamma_{\bar l} )
  \ ,
\\
\label{notsu5bivectorsabpm}
  \tfrac{1}{2} ( 1 + \varkappa_{kl} )
  \bgamma^+_k \bgamma^-_l
  &=
  \tfrac{1}{2}
  ( \bgamma^+_k \bgamma^-_l
  +
  \bgamma^-_k \bgamma^+_l )
  &
  &=
  - \tfrac{\im}{2}
  ( \bgamma_k \wedgie \bgamma_l - \bgamma_{\bar k} \wedgie \bgamma_{\bar l} )
  \ ,
\end{alignat}
\end{subequations}
and the 1 diagonal bivector
\begin{equation}
\label{Xbivectora}
  \im X
  \equiv
  \tfrac{1}{2}
  \sum_k \bgamma^+_k \bgamma^-_k
  =
  \tfrac{\im}{2}
  \sum_k \bgamma_k \wedgie \bgamma_{\bar k}
  \ ,
\end{equation}
with indices $k$ running over
$y, z, r, g, b$.
The bivector $X$ measures total $yzrgb$ charge $y+z+r+g+b$,
and $\im X$ is the generator of the $\U(1)$ factor
that would complete $\SU(5)$ to $\U(5)$.

The 6 bivectors involving the weak charges $y$ and $z$ generate the weak group
$\Spin(4)_w = \SU(2)_\Rchiral \times \SU(2)_\Lchiral$,
a product of right- and left-handed components.
The subgroup of $\Spin(4)_w$ that preserves the number of $yz$ up-bits
is the standard-model left-handed component $\SU(2)_\Lchiral$.
The bivectors of the left-handed weak group $\SU(2)_\Lchiral$
comprise the $2{+}(2{-}1) = 3$ bivectors~(\ref{su5bivectorsab})
and~(\ref{su5bivectorsa})
with $k$ and $l$ running over $y$ and $z$.
The bivectors are equivalent to $\im \tau_k$
with $\tau_k$ satisfying the Pauli algebra,
\begin{subequations}
\label{SU2bivectorspauli}
\begin{alignat}{5}
\label{SU2bivectorspauli1}
  \im \tau_1
  &\equiv
  \phantom{-} \tfrac{1}{2} ( 1 - \varkappa_{yz} ) \bgamma^+_y \bgamma^-_z
  &&=
  \phantom{-} \tfrac{1}{2}
  ( \bgamma^+_y \bgamma^-_z - \bgamma^-_y \bgamma^+_z )
  &&=
  \phantom{-} \tfrac{\im}{2}
  ( \bgamma_y \wedgie \bgamma_{\bar z} + \bgamma_z \wedgie \bgamma_{\bar y} )
  \ ,
\\
\label{SU2bivectorspauli2}
  \im \tau_2
  &\equiv
  - \tfrac{1}{2} ( 1 - \varkappa_{yz} ) \bgamma^+_y \bgamma^+_z
  &&=
  - \tfrac{1}{2}
  ( \bgamma^+_y \bgamma^+_z + \bgamma^-_y \bgamma^-_z )
  &&=
  - \tfrac{1}{2}
  ( \bgamma_y \wedgie \bgamma_{\bar z} - \bgamma_z \wedgie \bgamma_{\bar y} )
  \ ,
\\
\label{SU2bivectorspauli3}
  \im \tau_3
  &\equiv
  - \tfrac{1}{2} ( 1 - \varkappa_{yz} ) \bgamma^+_y \bgamma^-_y
  &&=
  - \tfrac{1}{2}
  ( \bgamma^+_y \bgamma^-_y - \bgamma^+_z \bgamma^-_z )
  &&=
  \phantom{-} \tfrac{\im}{2}
  ( \bgamma_z \wedgie \bgamma_{\bar z} - \bgamma_y \wedgie \bgamma_{\bar y} )
  \ ,
\end{alignat}
\end{subequations}
where
$\varkappa_{yz} \equiv \bgamma_y \wedgie \bgamma_{\bar y} \wedgie \bgamma_z \wedgie \bgamma_{\bar z}$
is the weak chiral operator.
The left-handed projection operator $\tfrac{1}{2} ( 1 - \varkappa_{yz} )$
equals 1 acting on left-handed weak chiral states,
and vanishes acting on right-handed weak chiral states.
The bivectors of the right-handed weak group $\SU(2)_\Rchiral$
comprise 3 bivectors equivalent to $\im \rho_k$
with $\rho_k$ satisfying the Pauli algebra,
\begin{subequations}
\label{SUR2bivectorspauli}
\begin{alignat}{5}
\label{SUR2bivectorspauli1}
  \im \rho_1
  &\equiv
  - \tfrac{1}{2} ( 1 + \varkappa_{yz} ) \bgamma^+_y \bgamma^-_z
  &&=
  - \tfrac{1}{2}
  ( \bgamma^+_y \bgamma^-_z + \bgamma^-_y \bgamma^+_z )
  &&=
  \tfrac{\im}{2}
  ( \bgamma_y \wedgie \bgamma_z - \bgamma_{\bar y} \wedgie \bgamma_{\bar z} )
  \ ,
\\
\label{SUR2bivectorspauli2}
  \im \rho_2
  &\equiv
  \phantom{-} \tfrac{1}{2} ( 1 + \varkappa_{yz} ) \bgamma^+_y \bgamma^+_z
  &&=
  \phantom{-} \tfrac{1}{2}
  ( \bgamma^+_y \bgamma^+_z - \bgamma^-_y \bgamma^-_z )
  &&=
  \tfrac{1}{2}
  ( \bgamma_y \wedgie \bgamma_z + \bgamma_{\bar y} \wedgie \bgamma_{\bar z} )
  \ ,
\\
\label{SUR2bivectorspauli3}
  \im \rho_3
  &\equiv
  \phantom{-} \tfrac{1}{2} ( 1 + \varkappa_{yz} ) \bgamma^+_y \bgamma^-_y
  &&=
  \phantom{-} \tfrac{1}{2}
  ( \bgamma^+_y \bgamma^-_y + \bgamma^+_z \bgamma^-_z )
  &&=
  \tfrac{\im}{2}
  ( \bgamma_y \wedgie \bgamma_{\bar y} + \bgamma_z \wedgie \bgamma_{\bar z} )
  \ .
\end{alignat}
\end{subequations}
The right-handed projection operator $\tfrac{1}{2} ( 1 + \varkappa_{yz} )$
equals 1 acting on right-handed weak chiral states,
and vanishes acting on left-handed weak chiral states.
The expressions for the bivectors $\im \rho_1$ and $\im \rho_2$ are equivalent
to equations~(\ref{notsu5bivectorsabpm}) and~(\ref{notsu5bivectorsabpp})
with indices $k$ and $l$ equal to $y$ and $z$.
The diagonal bivector $\im \rho_3$ measures the total $yz$ charge $y+z$,
\begin{equation}
\label{Rbivector}
  \im \rho_3
  =
  \tfrac{1}{2}
  \sum_{k = y,z} \bgamma^+_k \bgamma^-_k
  =
  \tfrac{\im}{2}
  \sum_{k = y,z} \bgamma_k \wedgie \bgamma_{\bar k}
  \ .
\end{equation}

The 15 bivectors involving the colour charges $r$, $g$, and $b$ generate the
colour group $\Spin(6)_c$.
The subgroup of $\Spin(6)_c$ that preserves the number of $rgb$ up-bits
is the standard-model colour group $\SU(3)_c$.
The bivectors of the colour group $\SU(3)_c$
comprise the $6{+}(3{-}1) = 8$ bivectors~(\ref{su5bivectorsab})
and~(\ref{su5bivectorsa})
with $k$ and $l$ running over $r$, $g$, and $b$.
The 7 other bivectors of $\Spin(6)_c$ consist of
the 6 leptoquark bivectors
given by equations~(\ref{notsu5bivectorsab}) with $k,l$
drawn from pairs of distinct colour indices $r,g,b$,
and the 1 diagonal bivector that measures the total $rgb$ charge $r+g+b$,
\begin{equation}
\label{Sbivector}
  - \im \tfrac{3}{2} ( B - L )
  =
  \tfrac{1}{2}
  \sum_{k = r,g,b} \bgamma^+_k \bgamma^-_k
  =
  \tfrac{\im}{2}
  \sum_{k = r,g,b} \bgamma_k \wedgie \bgamma_{\bar k}
  \ .
\end{equation}

The 1 hypercharge bivector, the generator of $\U(1)_Y$,
is defined to be the bivector whose eigenvalue is $\im Y$
where $Y$ is the hypercharge, equation~(\ref{YICY}),
\begin{equation}
\label{uY1bivectora}
  \im Y
  \equiv
  \tfrac{1}{2} \!
  \sum_{k = y,z} \bgamma^+_k \bgamma^-_k
  -
  \tfrac{1}{3} \!
  \sum_{k = r,g,b} \bgamma^+_k \bgamma^-_k
  =
  \im
  \Biggl(
  \tfrac{1}{2} \!
  \sum_{k = y,z} \bgamma_k \wedgie \bgamma_{\bar k}
  -
  \tfrac{1}{3} \!
  \sum_{k = r,g,b} \bgamma_k \wedgie \bgamma_{\bar k}
  \Biggr)
  \ .
\end{equation}
The 1 electromagnetic charge bivector,
the generator of $\U(1)_Q$,
is defined to be the bivector whose eigenvalue is $\im Q$
where $Q$ is the electric charge, equation~(\ref{QIc}),
\begin{equation}
\label{Qbivectora}
  \im Q
  \equiv
  \tfrac{1}{2}
  \bgamma^+_z \bgamma^-_z
  -
  \tfrac{1}{6} \!
  \sum_{k = r,g,b} \bgamma^+_k \bgamma^-_k
  =
  \im
  \Biggl(
  \tfrac{1}{2}
  \bgamma_z \wedgie \bgamma_{\bar z}
  -
  \tfrac{1}{6} \!
  \sum_{k = r,g,b} \bgamma_k \wedgie \bgamma_{\bar k}
  \Biggr)
  \ .
\end{equation}

\printbibliography

\end{document}